\documentclass[twocolumn]{aastex631}

\usepackage[utf8]{inputenc}
\usepackage{color}
\usepackage{soul}
\usepackage{graphicx}
\usepackage{amsmath}
\usepackage{amssymb}
\usepackage{aas_macros}
\usepackage{mathtools}
\usepackage{acronym}

\usepackage{gensymb}
\usepackage{url}
\usepackage{appendix}

\newcommand\msun{\,\rm M_{\odot}}

\acrodef{GW}{gravitational-wave}
\acrodef{LIGO}{Laser Interferometer Gravitational-wave Observatory}
\acrodef{LVK}{LIGO-Virgo-KAGRA}
\acrodef{BBH}{binary black hole}
\acrodef{IMBH}{intermediate mass black hole}
\acrodef{IMF}{initial mass function}
\acrodef{PISN}{Pair Instability Supernova}
\acrodef{BH}{black hole}
\acrodef{SNR}{signal-to-noise ratio}
\acrodef{HoM}{Higher order Multipole}
\acrodef{BNS}{binary neutron star}
\acrodef{PSD}{power spectral density}
\acrodef{CE}{Cosmic Explorer}
\acrodef{ET}{The Einstein Telescope}
\acrodef{SMBH}{supermassive black hole}
\acrodef{AGN}{active galactic nucleus}

\begin{document}

\title {Identifying heavy stellar black holes at cosmological distances \\
with next generation gravitational-wave observatories}
\date{\today}

\author{Stephen Fairhurst}
\affiliation{Gravity Exploration Institute, School of Physics and Astronomy, Cardiff University, Cardiff, UK, CF24 3AA, UK}

\author{Cameron Mills}
\affiliation{Albert-Einstein-Institut, Max-Planck-Institut for Gravitationsphysik, D-30167 Hannover, Germany}
\affiliation{Leibniz Universitat Hannover, D-30167 Hannover, Germany}

\author{Monica Colpi}
\affiliation{Department of Physics,  University of Milano - Bicocca,  Piazza della Scienza 3,I20126 Milano, Italy}
\affiliation{National Institute of Nuclear Physics INFN, Milano - Bicocca, Piazza della Scienza 3, 20126 Milano, Italy}

\author{Raffaella Schneider}
\affiliation{Dipartimento di Fisica, Universitá di Roma `La Sapienza', P.le Aldo Moro 2, I-00185 Roma, Italy}
\affiliation{INAF-Osservatorio Astronomico di Roma, via di Frascati 33, I-00078 Monteporzio Catone, Italy}
\affiliation{INFN, Sezione di Roma I, P.le Aldo Moro 2, I-00185 Roma, Italy}

\author{Alberto Sesana}
\affiliation{Department of Physics,  University of Milano - Bicocca,  Piazza della Scienza 3,I20126 Milano, Italy}
\affiliation{National Institute of Nuclear Physics INFN, Milano - Bicocca, Piazza della Scienza 3, 20126 Milano, Italy}

\author{Alessandro Trinca}
\affiliation{INAF-Osservatorio Astronomico di Roma, via di Frascati 33, I-00078 Monteporzio Catone, Italy}
\affiliation{INFN, Sezione di Roma I, P.le Aldo Moro 2, I-00185 Roma, Italy}

\author{Rosa Valiante}
\affiliation{INAF-Osservatorio Astronomico di Roma, via di Frascati 33, I-00078 Monteporzio Catone, Italy}
\affiliation{INFN, Sezione di Roma I, P.le Aldo Moro 2, I-00185 Roma, Italy}

\begin{abstract}

We investigate the detectability of single-event coalescing black hole binaries with total mass of $100-600\msun$ at cosmological distances ($5 \lesssim z \lesssim 20$) with the next generation of terrestrial gravitational wave observatories, specifically Einstein Telescope and Cosmic Explorer. 
Our ability to observe these binaries is limited by the low-frequency performance of the detectors.
Higher-order Multipoles of the gravitational wave signal are observable in these  systems, and detection of such multipoles serves to both b the mass range over which black hole binaries are observable and improve the recovery of their individual masses and redshift.
For high redshift systems of $\sim 200 \msun$ we will be able to confidently infer that the redshift is at least $z=12$, and for systems of $\sim 400 \msun$ we can infer a minimum redshift of at least $z=8$. 
We discuss the impact that these observations will have in narrowing uncertainties on  the  existence of the {\it pair-instability mass-gap}, and their implications on the formation of the first stellar black holes that could be {\it seeds} for the growth of  supermassive black holes powering high-$z$ quasars.

\end{abstract}

\section{Introduction}
\label{sec:intro}
In the three observing runs by the \ac{LVK} Collaboration, many tens of \ac{GW} transient signals consistent with the merger of \acp{BBH} have been detected  \citep{GWTC-1, GWTC-2.1, GWTC-3}.
The majority of them, observed at $z\lesssim 1$, are found to have one or both component masses between $20\msun$ and $50\msun$ and total mass $\lesssim 80\msun$.\footnote{All masses in this paper are referred to as measured in the source-frame.} A variety of channels have been proposed for their origin: formation as field binaries primarily in low-metallicity galaxies at high redshifts, formation in dense stellar systems, in \ac{AGN} discs, or after generations of repeated mergers \citep [see] [and references therein] {Graziani2020, Mapelli21-review,Gerosa2021NatAs...5..749G,Mandel2022PhR...955....1M}.

The next-generation of ground-based \ac{GW} observatories, specifically the Einstein Telescope  \citep[ET,][]{Punturo2014, 2023JCAP...07..068B} and Cosmic Explorer ~\citep[CE,][]{Evans:2021gyd}, 
will open the prospect of detecting the \ac{GW} signatures of merging \acp{BBH} over a wider mass range and deeper redshifts, extending the realm of observations to \acp{BBH} 
out to $z\sim 30$, when the first stars began to shine, and into the intermediate-mass range ${\cal O}(100-1000)\msun$ \citep{Kalogera2019-BH,Maggiore2020JCAP...03..050M}.  Beyond  redshift $z\sim 30-40$, merging primordial black holes of ${\cal O}(10)\msun$, formed by quantum processes in the early Universe \citep{Carr2021RPPh...84k6902C,Ng2022}, may also be detected and studied \citep{Ng2022,Ng2023PhRvD.107b4041N}.

In this paper we study systems
with individual masses extending from 60 to 480$\msun$, and total 
masses of $180\msun$ to $600\msun$, covering a mass range that is relevant for several reasons, as we explore below.
Measuring such systems is most interesting at cosmological distances, which is only possible due to the enhanced sensitivity of ET and CE at frequencies below $10$ Hz.

The formation of such heavy stellar-mass \acp{BH} requires the presence of  star forming regions of zero or extremely low metallicity, 
where fragmentation and cooling of the parent gas cloud, and mass loss from stellar winds are strongly suppressed \citep [see] [and references therein]  {Bromm2013, Spera17}. These are conditions that occur in the high-redshift Universe, and are expected to result in a top-heavy mass function where stars heavier than $150-300\msun$ are more common than in the conventionally adopted,  Kroupa-like,  stellar \ac{IMF} (see \citealt{Klessen2023} for a recent review on the first stars).
At the highest redshifts, these heavy BHs may represent systems not yet affected by accretion of surrounding gas \citep{Kazanas2019A&A...632L...8R,Zoltan2020ApJ...903L..21S,vanSon2020}, and hence their masses reflect their birth conditions.  Detecting \acp{GW} from these heavy stellar-mass binaries 
will let us constrain their merger rate which is intimately related to the rate of formation of massive stellar binary systems in pristine star forming galaxies \citep{Stacy2013, Chon2019, Sugimura2023}.

Some of the \ac{BBH} masses we investigate reside within the so called {\it pair-instability mass-gap} (often referred to as upper-mass gap or Pair Instability Supernova (PISN) gap. This gap is between about $65\msun$ and $135\msun$ 
where no BH is expected to form in evolution models of isolated stars. This mass-gap is attributed to a pair-instability, arising in metal poor, massive stars between  about $130\msun$ and $250\msun,$ which leads to a supernova explosion due to uncontrolled  $^{12}{\rm C}(\alpha,\gamma)^{16}{\rm O}$ nuclear burning, leaving no remnant \citep{Farmer2020ApJ...902L..36F,Woosley2021}. 
During the third observing run of \ac{LVK} a short duration signal, GW190521, was detected and estimated to be consistent with the merger of two \acp{BH} with component masses of about $85\msun$ and $66\msun$ and total mass of $142\msun$ \citep{GW190521,GW190521-int}.
This is the heaviest BH observed in GWs to date, with an \textit{intermediate-mass} remnant and a primary component residing within the {\it pair-instability mass-gap}.\footnote{Alternative analyses, \citep [e.g.][] {Nitz2021} find that GW190521 could instead be $\sim170\msun$ black hole with a companion of $\sim20\msun$, suggesting that the primary is already an intermediate-mass BH, with a mass beyond the mass gap \citep{Fishbach2020ApJ...904L..26F}.}

Detecting the \ac{GW} signal from high-redshift heavy stellar \ac{BBH} mergers, where one or both components are in the upper-mass gap or straddling it, would be highly informative. Various mechanisms could lead to the formation and coalescence of such  binaries, and among them, evolution in isolated binaries, dynamical encounters in star clusters or a chain of Nth generation mergers.\footnote{There are several proposed channels for the origin of the components of GW190521-like systems observed at low redshift. For instance: mergers from the relics of the first stars (known as Population III stars) \citep{Bromm2020ApJ...903L..40L,Tanikawa2021MNRAS.505.2170T}, isolated binary evolution with no-gap \citep{Belczy2020ApJ...905L..15B}, stellar collisions in star clusters \citep [see] [and references therein] {Costa2022, ArcaSedda2021ApJ...920..128A, DiCarlo2020mnras, Renzo2020ApJ...904L..13R}, and 
hierarchical mergers \citep{Gerosa2021NatAs...5..749G}.}
But at redshifts as high as $\sim 10$ the contributions from the two dynamical channels appear to be negligible \citep{Mapelli2022MNRAS.511.5797M}. Consequently, observing these systems at high $z$ would allow us to better probe the physics of the isolated binary channel, the potential existence of an upper-mass gap and its imprint on the mass function of the earliest stellar BHs.

Estimates of the location and width of the upper-mass gap are at best approximate. Current uncertainties on the reaction rate, on rotation, and on the presence of rich hydrogen envelopes may shift the instability interval for the explosion to higher masses, and narrow further the gap  \citep{Farmer2020ApJ...902L..36F,Belczy2020ApJ...905L..15B, Woosley2021, Marchant-2020A&A...640L..18M,Vink2021MNRAS.504..146V} or even fill it entirely \citep{Costa2021}. 
Testing the existence of this upper-mass gap and inferring it properties from \ac{GW} observations depends critically upon the accuracy with which the masses of the individual \acp{BH}  are measured from the merger signal. Here and for this purpose we carry out a parameter estimation on \acp{BBH} with component masses which touch the {\it edges} of the upper-mass gap, recognizing that
all of the above arguments become compelling if the redshift of the observed systems is $z\gtrsim 10.$  

Determining the {\it lowest} redshift one can claim the source to be {\it beyond}, and inferring posteriors for the distribution of the component masses is of paramount importance \citep{Mancarella2023PhRvD.107j1302M}. 
A key challenge, then, is to accurately infer both the masses and redshift of the binary.  There is a well-known degeneracy between measurements of the distance to and inclination of a binary from \ac{GW} observations \citep{Usman:2018imj}.  At high redshifts, this degeneracy further impacts our ability to infer the masses of the binary. In \ac{GW} observations, masses and redshift are degenerate, and only the redshifted masses, $m_{1, 2} (1 + z)$, can be inferred from the signal.  Given a cosmological model, the measured distance can be used to obtain the redshift and hence the source masses.  However, if the distance is poorly constrained, this leads to significant uncertainties on the mass. For example, it is not unusual to have an uncertainty of $\sim 50\%$ in the distance measurement of \ac{BBH} signals \citep{GWTC-3}.  At a $z=10$ this translates to a redshift uncertainty of $\pm 4$ and consequently an uncertainty in the masses of $40\%$ due to redshift effects alone.  The ability to accurately infer redshifts and masses is improved by a detector network, which can provide more accurate localization and distance measurements \citep{Fairhurst:2010is, mills_localization_2018, Singer:2014qca}, as well as the observation of \acp{HoM} in the \ac{GW} signal which help break the distance-inclination degeneracy \citep{Mills:2020thr, Fairhurst:2023idl}.

The paper is organized as follows.  In Section \ref{sec:hm_obs}, we discuss the observability of high-mass, high-redshift binaries with a focus on the \acp{HoM}.  In Section \ref{sec:pe}, we provide detailed parameter estimation results for a number of astrophysically interesting simulated \ac{BBH} merger signals and in Section \ref{sec:discussion} we summarize our results.  We include two appendices. Appendix \ref{app:sensitivity} provides additional figures showing detector sensitivity for binaries of varying mass ratio and Appendix \ref{app:low_snr} gives parameter estimation accuracy for low \ac{SNR} systems.  
\section{The Importance of Higher order Multipoles}
\label{sec:hm_obs}

\begin{figure}[tb]
 \centering
 \includegraphics[width=0.49\textwidth]{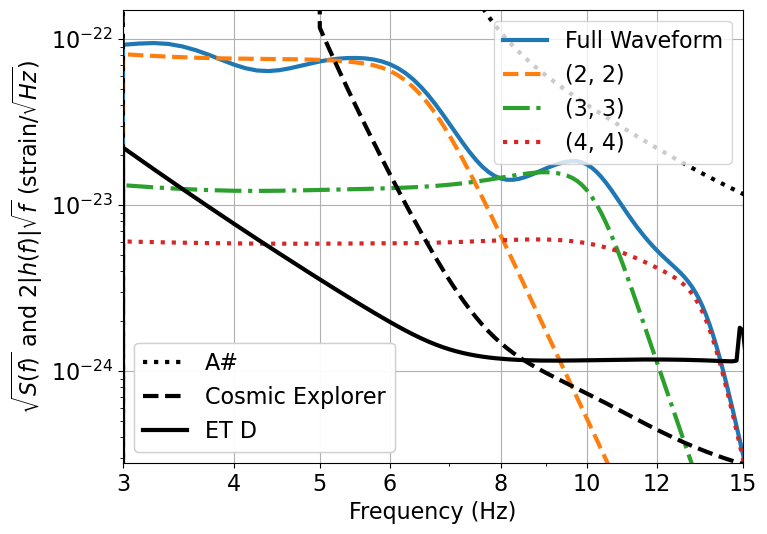}
 \caption{Waveform amplitude from a (120,60) $\msun$ binary at $z=14$, inclined at an angle of $60\degree$ to the line of sight.  
 The figure shows the frequency evolution of the waveform, and also the individual contributions from the three leading multipoles (2, 2), (3, 3) and (4, 4) using the IMRPhenomXPHM waveform \citep{Pratten:2020ceb}. The oscillations in the full waveform are due to constructive/destructive interference between the multipoles. The projected noise curves for three future detectors: Cosmic Explorer, Einstein Telescope and A$^\#$ are also shown. The signal is represented in such a way that area between the wave amplitude and noise is indicative of the \ac{SNR} \citep{O1BBH}. 
 }
 \label{fig:wave}
\end{figure}

High-mass, high-$z$ \ac{BBH} coalesences  are intrinsically low-frequency \ac{GW} sources. This is illustrated in Fig.\ref{fig:wave}, where we show the frequency evolution of the \ac{GW} strain amplitude for a \ac{BBH} of (120-60)$\msun$ placed at redshift $z=14$, with an inclination $\iota$ of $60\degree$ between the orbital angular momentum and theof  line of sight. 
The gravitational waveform for this signal only extends to $15$ Hz and is therefore outside the sensitive frequency range of current \ac{GW} observatories.  The leading \ac{GW} emission from the source, emitted in the (2, 2) multipole at twice the orbital frequency, extends to only $7$ Hz in the detector, making discovery challenging.\footnote{We recall that for all \ac{BBH} observed to date, the (2, 2) multipole, which is emitted at twice the orbital frequency, has been the dominant multipole detected in the \ac{GW} signal.  Additional multipoles of the \ac{GW} signal have been observed for a handful of events \citep{LIGOScientific:2020stg, Abbott:2020khf} but, as their amplitudes are lower, they are generally not identified for the majority of sources.}
Although the (3, 3) and (4, 4) multipoles are intrinsically lower amplitude, they extend to higher frequencies ($\sim$1.5 and 2 times the frequency of the (2, 2)) and can therefore contribute significantly to the observed \ac{SNR}. This improves the prospects of detecting such a system. Furthermore, the identification of these higher-order (higher-frequency) multipoles in the signal can significantly improve the ability to infer the parameters of the system, as they enable us to break measurement degeneracies that exist with observation of only a single multipole. 

There are several well-known degeneracies in the emitted gravitational-waveform, leading to some parameters being very well measured while others being not.  For our purposes, we are most concerned with a degeneracy between the observed distance to  and inclination of a binary, as discussed in \cite{Usman:2018imj}.  When only  the (2, 2) multipole is observed, the amplitude gives a good measurement of $\cos \iota/d_{L}$ where $\iota$ is the binary inclination and $d_{L}$ is the luminosity distance. However, in many cases, the binary inclination is only restricted to the range $\iota \in [0\degree, 60\degree]$, leading to a factor of two uncertainty in distance due to this degeneracy alone.  When the binary is observed at a high redshift, the measurement of the masses also becomes degenerate with  distance and inclination, and a factor of two uncertainty in distance can lead to a similar uncertainty on the masses. The observation of a second \ac{GW} multipole can serve to break this degeneracy \citep{LIGOScientific:2020stg, Mills:2020thr} as the relative amplitude of the different multipoles depends upon the orientation of the binary\footnote{The ratio of the amplitude of the (3, 3) multipole to the (2, 2) scales as $\sin \iota$ while the (4, 4) multipole scales as $\sin^2 \iota$ relative to the (2, 2).}.

\begin{figure*}
    \includegraphics[width=0.49\textwidth]{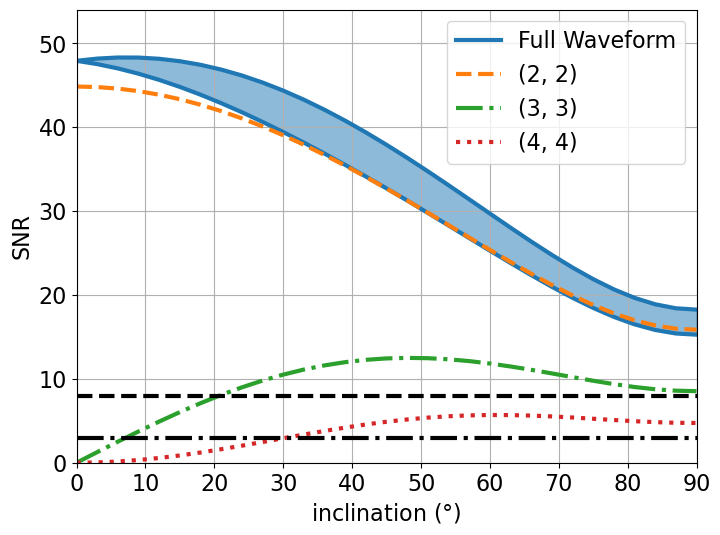}
    \includegraphics[width=0.49\textwidth]{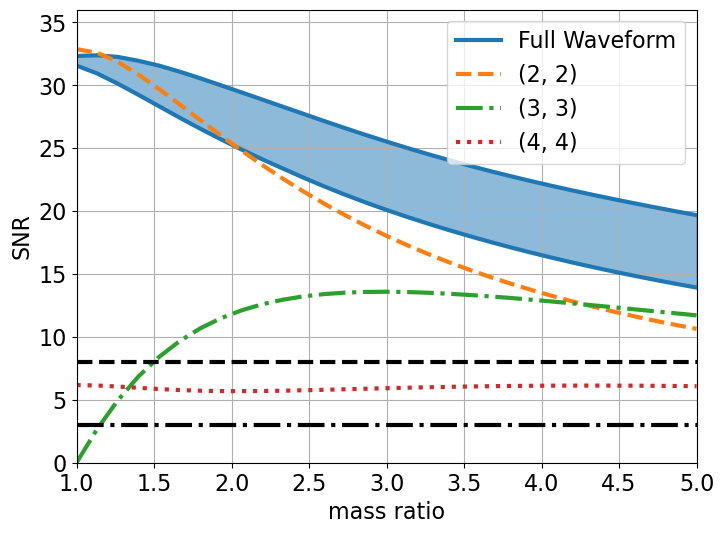}
    \caption{The total \ac{SNR} and \ac{SNR} in each multipole in \ac{ET} for a binary of total mass of $180\msun$, at redshift $z=14$  and overhead the detector. 
    \textit{Left}: Variation of \ac{SNR} with inclination of the binary, for a BBH of ($120,60)\msun$ and the fixed redshift.
    \textit{Right}: Variation of \ac{SNR} with mass ratio, for a $180\msun$ binary with inclination of $60\degree$ at the fixed ratio. 
    The total \ac{SNR} for the system is computed using the (2, 2), (2, 1), (3, 3), (3, 2) and (4, 4) multipoles and individual \acp{SNR} for the (2, 2), (3, 3) and (4, 4) multipoles are shown.  At $\iota=0$ in the left panel the full waveform \ac{SNR} is higher than the (2,2) mode due to the contribution of the (2, 1) and (3, 2) modes, which are not shown. The total \ac{SNR} depends upon the phase of the signal at merger, due to non-zero overlap between multipoles, and is therefore shown as a shaded blue area.
    In both panels, horizontal lines indicate \ac{SNR} = 8, an approximate threshold for detection, and \ac{SNR} = 3, an approximate threshold for observation of a second multipole, given that another multipole has been observed \citep{Mills:2020thr}.
    }
    \label{fig:snr_vs_inc_q_et}
\end{figure*}

In Fig.~\ref{fig:snr_vs_inc_q_et}, we show the variation of the \ac{SNR} with binary mass ratio $q = m_{1}/m_{2}$ (assuming an inclination $\iota = 60^{\circ}$) and inclination (assuming $q=2$) in each of the multipoles for a binary of total mass of $180 \msun$ at $z = 14$ observed by \ac{ET}. The \ac{SNR} of the (2, 2) multipole is greatest for face-on signals ($\iota = 0^{\circ}$) with equal mass components ($q = 1$).  For a face-on signal, the (2, 2) multipole is circularly polarized and, as the inclination increases, the amplitude of both polarizations decreases to a minimum for edge-on systems, $\iota = 90^{\circ}$, whose emission is linearly polarized.  For the other multipoles considered, the \ac{SNR} vanishes at face-on and peaks at $\sim 50\degree$ for the (3, 3) multipole and $\sim60\degree$ for the (4, 4) multipole.  The binary would be observable in \ac{ET} at any orientation.  For inclinations $\iota \gtrsim 10\degree$ or $30\degree$ the (3, 3) and (4, 4) multipoles would be identifiable, respectively.  Since this waveform lasts only a few cycles in the detector band, the contributions from the different multipoles are not orthogonal. Consequently, the total \ac{SNR} varies with the merger phase of the binary.  

The \ac{SNR} of each different multipole, and the full signal, also varies with mass ratio.  The (2, 2) multipole is largest for equal mass systems and decreases by a factor of two by mass ratio $q=5$, while the (3, 3) vanishes for equal mass and peaks around $q=3$.  For this signal, the \ac{SNR} in the (4, 4) multipole does not vary significantly with mass ratio.  The (2, 2) and (4, 4) multipoles would be identifiable at any mass ratio, and the (3, 3) for binaries with mass ratio above $\sim$1.2.  Identification of more than one multipole enables an improved measurement of mass ratio, as well as binary orientation.

\begin{figure*}
     \includegraphics[width=0.49\textwidth]{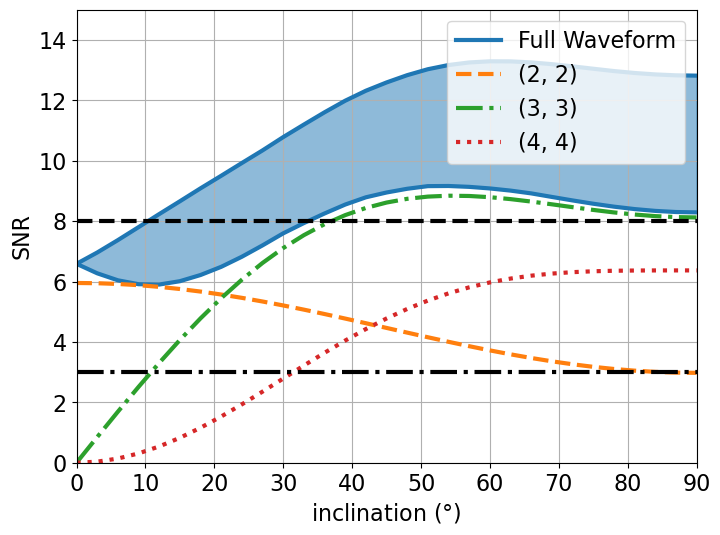}
    \includegraphics[width=0.49\textwidth]{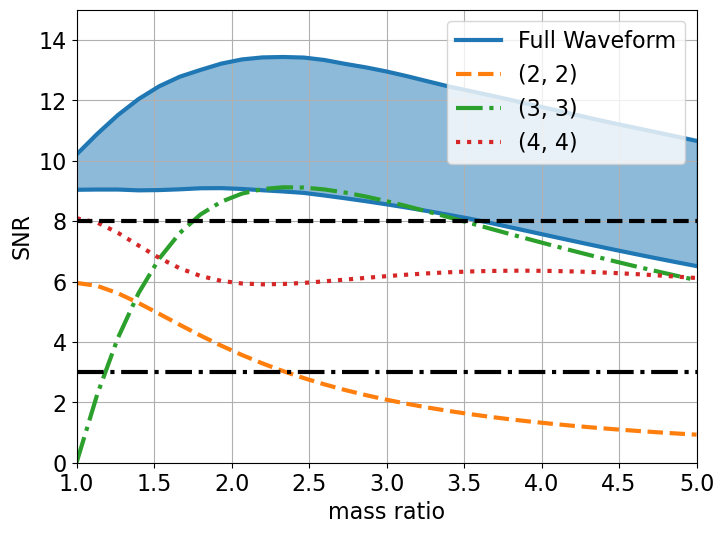}
    \caption{The same as Fig.~\ref{fig:snr_vs_inc_q_et} but for \ac{CE}.
    }
    \label{fig:snr_vs_inc_q_ce}
\end{figure*}

In Fig.~\ref{fig:snr_vs_inc_q_ce}, we show the same dependence of \ac{SNR} with inclination and mass ratio for the \ac{CE} detector.  Since \ac{CE} has sensitivity to the signal above 5 Hz, rather than 3 Hz for \ac{ET}, the overall \ac{SNR} is lower and the signal would be marginally observable.  
Furthermore, a broad range both in inclination and mass ratio, where the (3,3) multiple gives the dominant contribution to the SNR, becomes accessible. This provides a clear example of a signal where the \acp{HoM} enable detection as well improved parameter recovery.

\begin{figure*}
    \centering
    \includegraphics[width=0.49\textwidth]{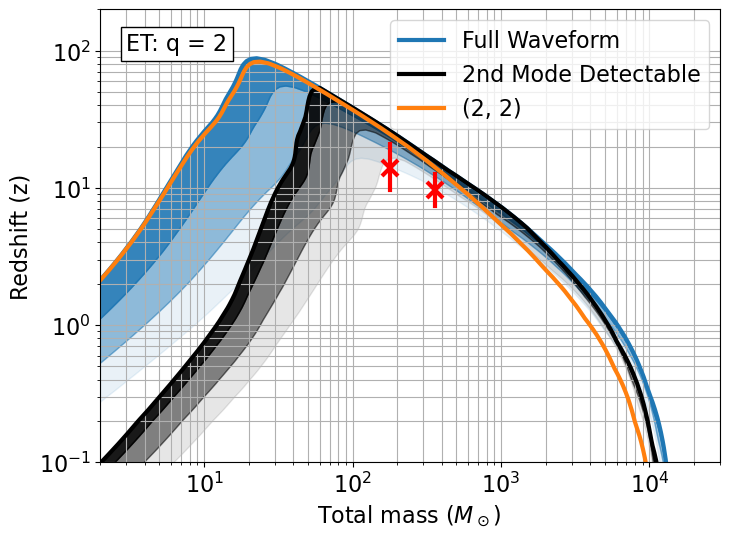}
    \includegraphics[width=0.49\textwidth]{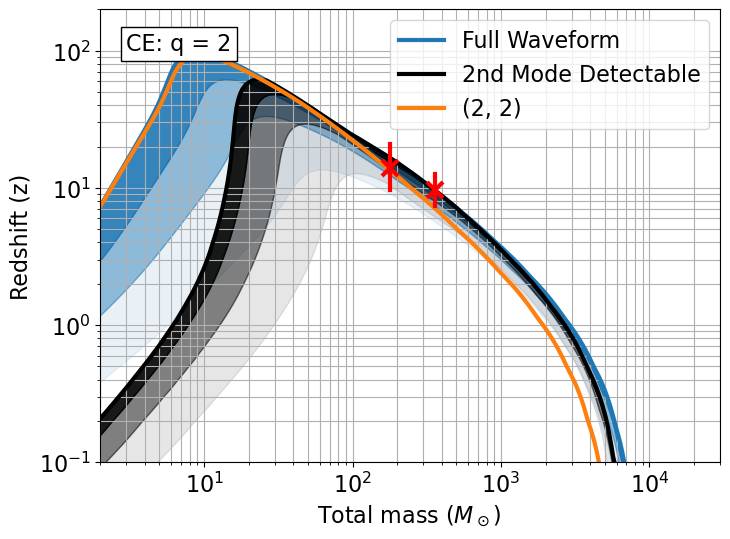}
    \caption{Maximum observable redshift of binaries with mass ratio $q = 2$ with next-generation observatories.
    \textit{Left}: \ac{ET}, \textit{Right}: \ac{CE}.  Shaded regions show redshifts at which 10\%, 50\% and 90\% of sources are observable.  The maximum observable redshift of the (2, 2) multipole at \ac{SNR} = 8 is shown in orange, the observability of the full waveform at \ac{SNR} = 8 is shown in blue, and the observability of the second most significant multipole above \ac{SNR} = 3 is shown in black.  The red asterisks represent the masses and redshifts of the $q = 2$ systems we study in detail in Section \ref{sec:pe}.
    }
    \label{fig:et_ce_horizon}
\end{figure*}

Given the above discussion, we are interested in identifying the regions of the mass space where \acp{HoM} can contribute to either the observability or parameter measurement accuracy of high-mass, high-redshift binaries. In Fig.~\ref{fig:et_ce_horizon} we show the sensitivity of the proposed \ac{ET} and \ac{CE} observatories to \ac{BBH} mergers with mass ratio of 2 as a function of redshift. We show the maximum redshift at which a binary can be observed, at an \ac{SNR} of 8, and also the redshifts at which 10\%, 50\% and 90\% of binaries, averaged over sky location and orientation, will be observed.  The detector sensitivity is shown for both the (2, 2) multipole, in orange, and the full waveform, in blue. At low masses, the (2, 2) multipole dominates the observable signal and therefore the distance to which the full waveform can be observed is essentially equal to that of the (2, 2) multipole.  However, at high masses, the (3, 3) and (4, 4) multipoles contribute more significantly and incorporating them increases the sensitivity of the detectors to these systems. When a system has been observed, the identification of a second multipole, at \ac{SNR} above 3, can greatly improve parameter recovery by breaking degeneracies between distance and inclination and improving mass ratio measurement. The range of masses and redshifts for which the binary would be observed with \ac{SNR} above 8, and with \ac{SNR} above 3 in at least two multipoles, is shown in black in Fig.~\ref{fig:et_ce_horizon}. For example, in \ac{ET} a 4,000$\msun$ system is visible at $z\approx 1$ with the (2, 2) multipole but up to $z \approx 2$ with the full waveform.  

Remarkably, for the majority of binaries with $M \gtrsim 100 \msun$ observed by \ac{ET}, and $M \gtrsim 30 \msun$ observed by \ac{CE}, a second multipole will be observable.  At lower masses, it is the (2, 2) and (3, 3) multipoles which contain most power, while at high masses it is the (3, 3) and (4, 4) multipoles that are observed, with the (2, 2) multipole power occurring at frequencies below the instrumental sensitivity.  The picture is similar at different mass ratios, and figures showing the sensitivity to binaries with $q = 1, 2, 4$ and $10$ are provided in Appendix \ref{app:sensitivity}, for \ac{ET} in Fig.~\ref{fig:et_horizons} and \ac{CE} in Fig.~\ref{fig:ce_horizons}.  The most significant difference occurs for equal mass binaries, where the (3, 3) multipole vanishes and we therefore require both (2, 2) and (4, 4) multipoles to be observable.  This limits the range for which two multipoles can be seen and increases the minimum mass at which we expect to observe two multipoles to $\sim 200\msun$ in \ac{ET} and $\sim50\msun$ for \ac{CE}.  Nonetheless, for the majority of high-mass, high-redshift binaries, we expect to observe multipole multipoles, and therefore obtain good estimates of both the masses and redshift of the system.  In the next section, we investigate those expectations in detail through parameter recovery of a series of systems.
 
\section{Parameter recovery for high-mass, high-redshift binaries}
\label{sec:pe}

Observation in a single \ac{GW} observatory leads to large uncertainties in the sky location of the binary \citep{Singer:2014qca}, and this is again degenerate with the inferred distance and redshift.  A network of detectors with comparable sensitivity can provide accurate localization \citep{mills_localization_2018} and therefore improved redshift and mass accuracy.
Binaries with black hole spins misaligned with the orbital angular momentum will precess.  In principle, the observation of precession can further improve parameter estimates. However, given that so few cycles of the waveform are visible in the detectors, the prospects for observing precession are slim \citep{Green:2020ptm}. Therefore, in what follows we neglect precession effects.

To illustrate the expected performance of a next-generation GW network in observing and measuring these binaries, we perform a number of simulations and obtain parameter estimates with the LALInference \citep{Veitch:2014wba} package and a uniform in comoving volume distance prior. We simulate four different binary mass combinations, denoting (in the source frame) with $m_1$ ($m_2$) the primary (secondary) mass and with $M$ the total mass. We consider $(120, 60)\msun$ and $(90, 90)\msun$ binaries, chosen so that component \acp{BH} lie in, or close to the upper mass-gap, and $(240, 120)\msun$ and $(480, 120)\msun$ binaries chosen to probe observability of high-redshift \ac{IMBH} in binaries.  In all cases, we simulate quasi-circular non-spinning \acp{BBH}, but allow for non-zero, aligned spins when performing parameter estimation\footnote{The restriction to non-spinning \ac{BBH} is solely to simplify presentation --- all results presented here could be easily extended to aligned-spin \acp{BH}.}.
This is important as the degeneracy between the binary mass ratio and \ac{BH} spins \citep{Hannam:2013uu, Fairhurst:2023idl} greatly impacts the accuracy with which mass ratio can be measured.

The simulated signals are added to data from a three-detector network of observatories with sensitivity matching ET \citep{Punturo2014} and CE \citep{Evans:2021gyd}.  Specifically, we use a single, triangular \ac{ET} detector located in Europe and two 40 km \ac{CE} observatories, one in the US and one in India.  The simulations are performed at the optimal sky location for the network.  Given the greater low-frequency sensitivity of \ac{ET}, this leads to the binaries being essentially overhead \ac{ET}.  The signals are generated at varying inclination angle, to enable us to investigate the importance of \acp{HoM}.  We choose the redshift of the sources to ensure a fixed \ac{SNR} for all signals.  In the main text, we use an \ac{SNR} of 30, while in Appendix \ref{app:low_snr} we investigate quieter signals with an \ac{SNR} of 15.

\subsection{Observing mass-gap objects}
\label{ssec:mass_gap}

A mass gap in the \ac{BH} mass distribution is expected due to the presence of the PISN.
\cite{Farmer2020ApJ...902L..36F} investigated the location of this pair-instability region as a function of the temperature-dependent uncertainty in the $^{12}\rm C(\alpha,\gamma)^{16}O$ reaction rate. Determining the value of the $^{12}\rm C(\alpha,\gamma)^{16}O$ reaction rate is extremely important for tracing the evolution of massive stars. Thus, restricting this rate through \ac{GW} observations would be of considerable astrophysical interest. According to \cite{Farmer2020ApJ...902L..36F}, the width of the mass-gap remains roughly constant as a function of the (unknown) reaction rate, but the mass-range where no black hole can form varies.  At the lowest rate relative to the median, the mass-gap extends from $\sim 90 \msun$ to $\sim 175 \msun$. At the highest rate, the location of the mass-gap is between $\sim 60\msun$ and $\sim 120\msun$.  Interestingly, there exists a region of \ac{BH} masses between $90\msun$ and $120\msun$ where we should not expect any black hole to form, {\it for any rate}. We refer to this region as the {\it forbidden strip}. Consequently, we choose to investigate systems which host at least one member with mass touching this narrow strip. Then, if their masses were to be determined with sufficient accuracy, their detection could constrain the $^{12}\rm C(\alpha,\gamma)^{16}O$ reaction rate to be at the extreme of the allowed range \citep [see Fig. 5 of] [] {Farmer2020ApJ...902L..36F}.  In particular, we focus on $(120,60)\msun$ and $(90,90)\msun$ binaries which have components at the lower and upper range of the forbidden strip.
  
As seen in Fig.~\ref{fig:et_ce_horizon}, \acp{BBH} with masses $(120,60)\msun$ will be detectable at a maximum redshift of $z\sim 25$, for an optimally located and oriented system, with 50\% of mergers at $z \sim  17$ and the vast majority of events at $z \sim 10$ being detectable. The sensitivity to $(90, 90)\msun$ systems is comparable. If $(90,90)\msun$ systems were to be observed, this will allow us to constrain the strength of the uncertain $^{12}\rm C(\alpha,\gamma)^{16}O$ reaction rate. In particular, black holes with such masses would imply the rate to be at the lower end of the explored range. A binary with mass $(120,60)\msun$ would be challenging to form through stellar evolution.  Specifically, allowing for variation in the reaction rate, the mass of the primary would require a very high reaction rate for $^{12}\rm C(\alpha,\gamma)^{16}O$, while the mass of the secondary would be compatible with a value below the median. Therefore this would be a system where only one of the two black holes could originate from stellar evolution and the other would require a different formation channel.

\begin{figure*}
 \centering
 \includegraphics[width=0.48\textwidth]{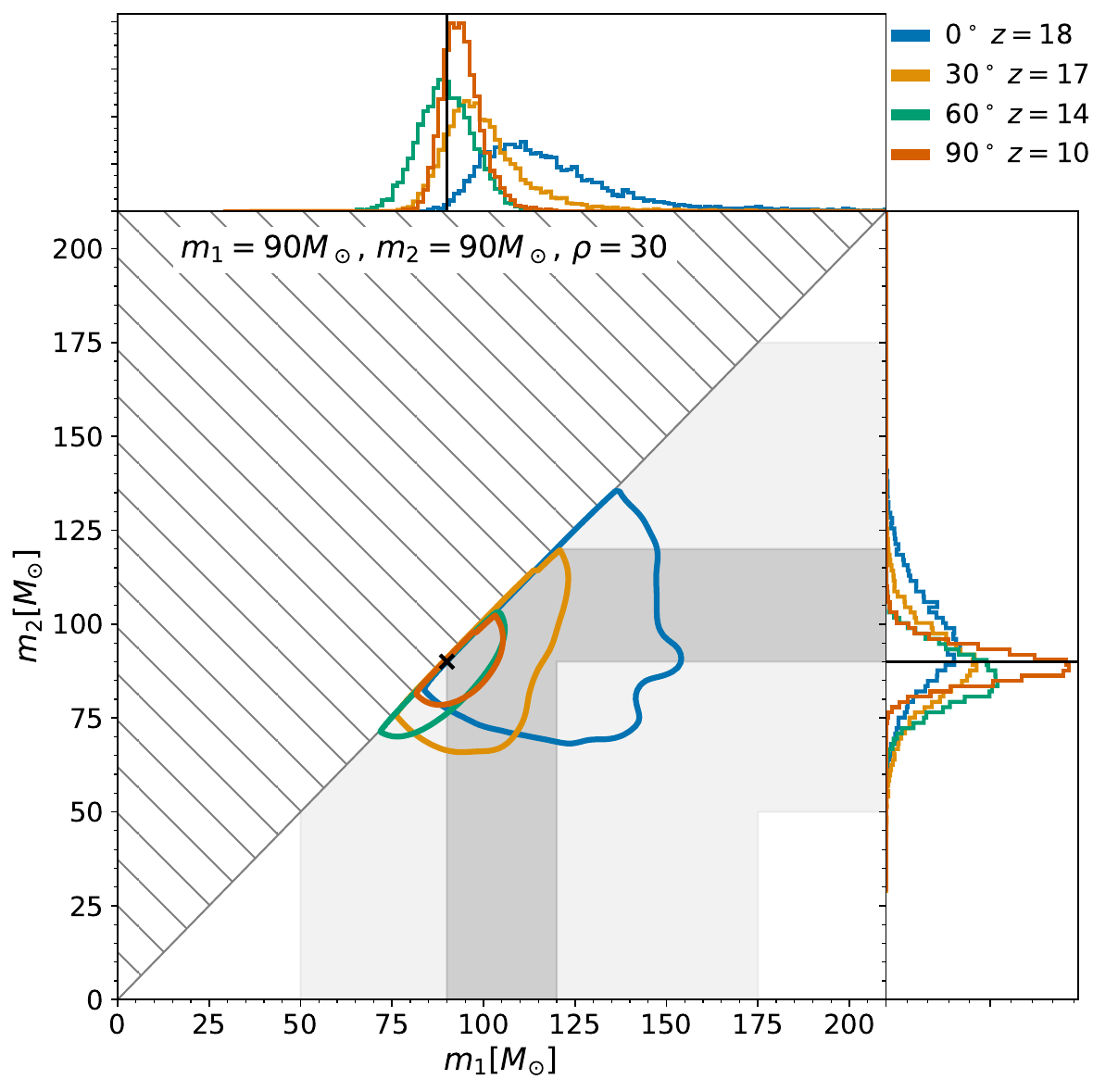}
\includegraphics[width=0.48\textwidth]{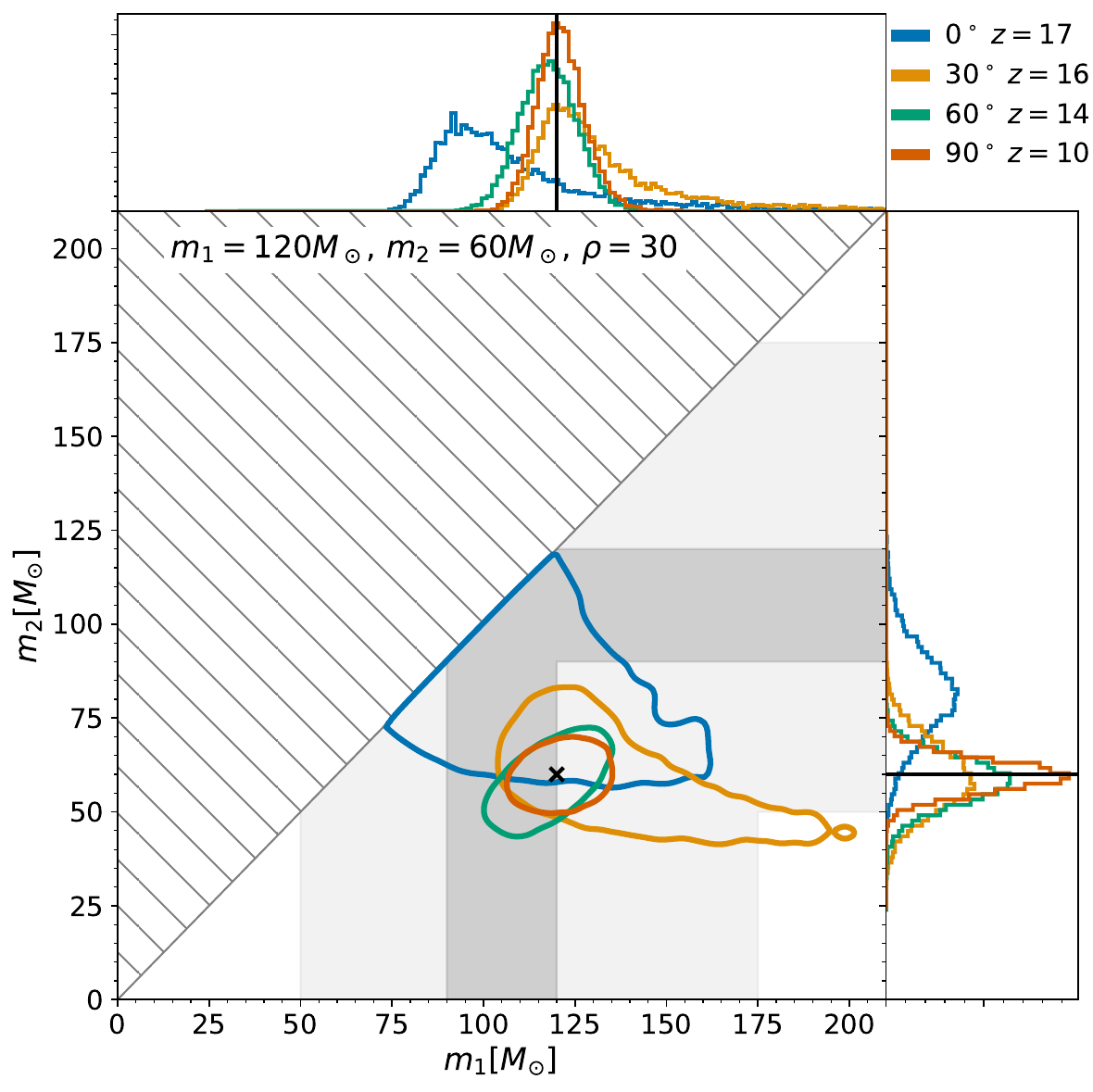}
 \caption{
 Inferred posteriors of component masses for binaries of total mass of $180\msun$ which straddle, or sit within the upper mass gap.  Results are shown for inclinations of $\iota = 0\degree, 30\degree, 60\degree$ and $90\degree$, with the redshift of the system varied, as indicated in the labels, so that the signals are all simulated with an \ac{SNR} of 30 in the \ac{ET}--\ac{CE} network, at the most sensitive sky location for the network. The grey region denotes the pair-instability  mass-gap and the darker grey denotes the forbidden strip  where no black hole is expected to form for any value of the $^{12}\rm C(\alpha,\gamma)^{16}O$ reaction rate \citep{Farmer2020ApJ...902L..36F}.  Simulated values are denoted by a black cross and contours show the 90\% credible region. \textit{Left}: Binary with masses just below the mass gap: $m_1=m_2=90 \msun$. \textit{Right}: Binary with masses that straddle the mass gap: $m_1=120 \msun$ and $m_2=60 \msun$.  Greek letter $\rho$ indicates the \ac{SNR}, throughout the paper.
 }
 \label{fig:mass_gap_pe}
\end{figure*}

\begin{figure*}
 \includegraphics[width=0.48\textwidth]{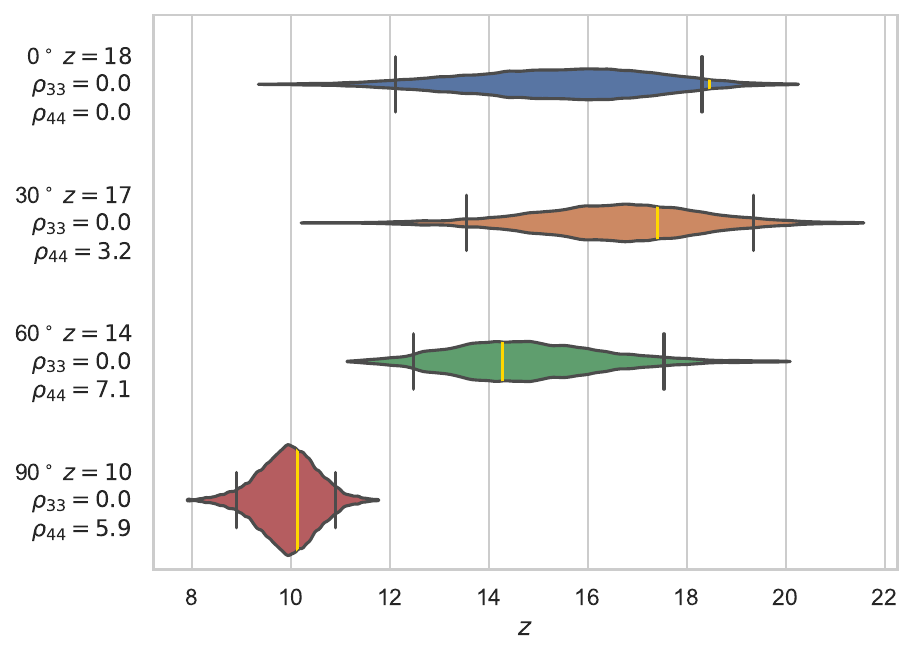}
 \includegraphics[width=0.48\textwidth]{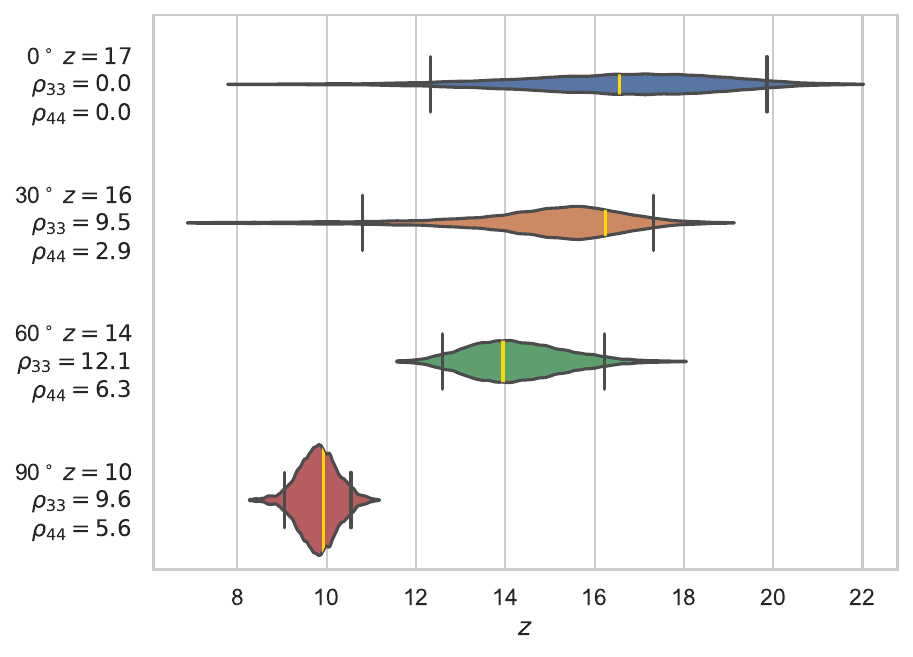}
 \caption{Inferred posteriors for the redshift for binaries of total mass of $180\msun$ which straddle, or sit within the upper mass gap. Signals are simulated with varying inclination (in degrees)  and redshift ($z$), as indicated in the labels, with a fixed \ac{SNR} of 30 in the \ac{ET}--\ac{CE} network, at the most sensitive sky location for the network. Symbols $\rho_{\ell m}$ indicate the \ac{SNR} in the $(\ell, m)$ multipole \acp{HoM}.  Black vertical lines indicate the 90\% credible intervals, violins show the 99.5\% range and yellow vertical lines show simulated values. \textit{Left}: Binary with masses just below the mass gap: $m_1=m_2=90 \msun$. \textit{Right}: Binary with masses that straddle the mass gap: $m_1=120 \msun$ and $m_2=60 \msun$. 
 }
 \label{fig:mass_gap_redshift}
\end{figure*}

In Fig. \ref{fig:mass_gap_pe} and \ref{fig:mass_gap_redshift}, we show the recovered parameter accuracies for both mass and redshift for $(90, 90)\msun$ and $(120, 60)\msun$ binaries observed with \ac{SNR} = 30.  The first thing to note is that these high-mass, high-redshift systems could be identified with good accuracy by the next-generation \ac{GW} network, as would be expected due to the relatively large \acp{SNR}.  For all events, there is, at most, a factor of two uncertainty in the mass of the systems and a $50\%$ uncertainty in the redshift, with both numbers quoted at the 90\% confidence interval.  However, we also notice a substantial variation in the accuracy of parameter measurement between the systems.  The parameters of systems close to face-on ($\iota = 0\degree$ or $30\degree$) are recovered with significantly larger uncertainties than those which are more highly inclined ($\iota = 60\degree$ or $90\degree$).  When the binary is close to face-on, the uncertainty in  the component mass posterior is much greater and, for $\iota = 0\degree$, the 90\% mass region for the $(120,60)\msun$  binary includes equal masses.  As the inclination of the binary is increased, parameter accuracy improves significantly: already at $\iota = 30\degree$ the posterior for the $(120,60)\msun$  binary is inconsistent with equal masses, although large uncertainty in the mass ratio remains.  For binaries inclined at $60\degree$ or $90\degree$ the parameter accuracy is excellent.  In both cases, the mass ratio is very well constrained and uncertainty in total mass and redshift is $\pm (10-20)\%$.  

\begin{figure}[t]
 \includegraphics[width=0.48\textwidth]{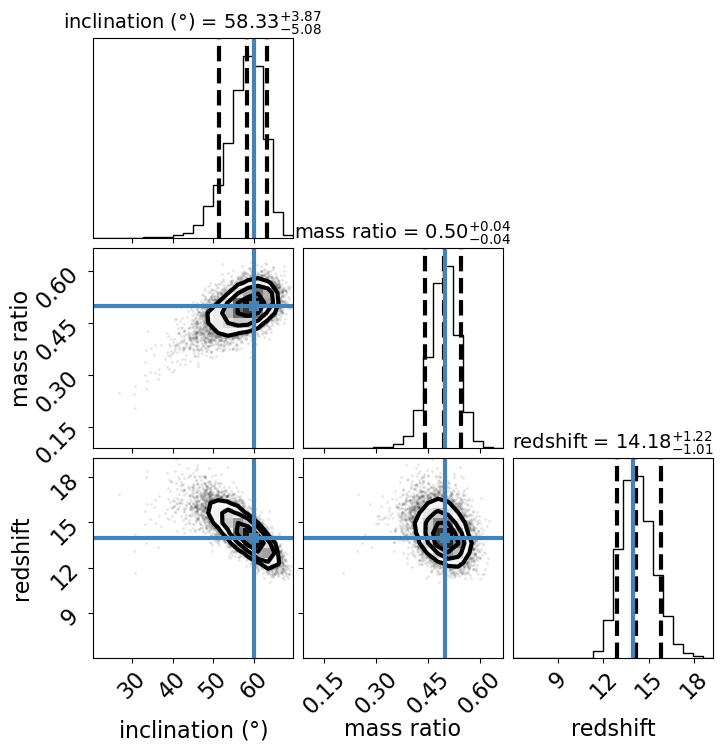}
 \includegraphics[width=0.48\textwidth]{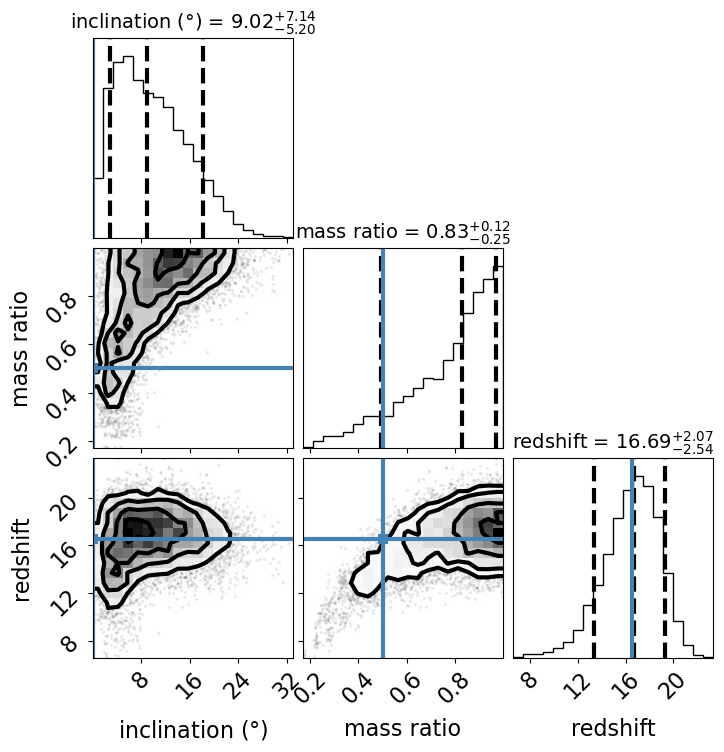}
 \caption{Inferred posteriors for redshift, mass ratio ($1/q=m_2/m_1$) and inclination for $(120,60) \msun$ binary observed at SNR=30.  \textit{Top}: Parameter recovery for a system at $z=14$ and $\iota = 60\degree$. \textit{Bottom}: Parameter recovery for a system at $z=17$ and $\iota = 0\degree$.
 }
 \label{fig:corner}
\end{figure}

 We can explain \textit{why} the results of our parameter accuracy display the features they do, using our understanding of the importance of \acp{HoM} discussed in Section \ref{sec:hm_obs}.  Let us begin by considering the $(120, 60)\msun$ system at $z=14$, inclined at $\iota=60\degree$, that we have plotted in Figs.~\ref{fig:wave}, \ref{fig:snr_vs_inc_q_et} and \ref{fig:snr_vs_inc_q_ce}.  The signal is observed with high \ac{SNR} in \ac{ET} and the (3, 3) multipole is also clearly observed.  In \ac{CE}, the total \ac{SNR} of the system is sufficient for it to be observed, with the (3, 3) multipole providing the dominant contribution.  Since the event is observed in three detectors, it is relatively well localized,\footnote{Since these are very low-frequency systems, the localization is poorer than for events in GWTC-3 \citep{GWTC-3} as the localization depends upon the frequency bandwidth of the signal \citep{Fairhurst:2009tc}.}
with a 90\% localization area of $300\deg^{2}$. The observation of \acp{HoM} in the waveform enables the accurate inference of both the binary inclination and mass ratio.  Since the sky location, mass ratio and inclination are well measured, this enables accurate inference of the distance to the binary and consequently the redshift of the source.  In the top panel of Fig. \ref{fig:corner} we show the recovered values for the redshift, binary inclination and mass ratio.  All three are recovered with an accuracy better than 10\%.

Next, we consider a comparable system observed face-on ($\iota = 0\degree$) at the same \ac{SNR}, and at a redshift of $17$.  In that case, the power in both the (3, 3) and (4, 4) multipoles vanishes.  Consequently, the binary is no longer observable in \ac{CE} as it has an \ac{SNR} of 1.\footnote{The waveform shown in Fig. \ref{fig:snr_vs_inc_q_ce} corresponds to a binary at $z=14$ and we are now considering the same mass binary at $z=17$.  The larger redshift reduces the \ac{SNR} primarily through redshifting the signal which lowers the frequency by 20\%. This leads to an \ac{SNR} which is lower than that shown in the figure.}
At this SNR, the CE observatories are unable to provide localization of the source, with a 90\% localization area of $10,000\deg^{2}$.  Furthermore, the vanishing \acp{HoM} means that only the (2, 2) multipole is observed in \ac{ET}.  Therefore, it is not possible to break the degeneracy between binary orientation and distance, nor to place a tight constraint upon the mass ratio.  The bottom panel of Fig. \ref{fig:corner} shows a the recovered redshift, mass ratio and inclination for this system.  The mass ratio \textit{is not} accurately recovered and, indeed, the binary is inferred most likely be (close-to) equal mass, although the distribution does extend to $1/q=0.5$.  In addition, the binary orientation is not accurately recovered, with a broad distribution of $\iota \lesssim 25\degree$ --- more inclined systems are excluded as they would have observable power in the (4, 4) multipole.  The mass ratio--inclination distribution does show a secondary peak close to the simulated value ($1/q \approx 0.5$ and $\iota < 10 \degree$), however, the preference is for an equal mass system.  
Despite both mass ratio and inclination being offset from the true values, the inferred redshift matches well with the simulated value.  However, due to the uncertainties in other parameters, the redshift uncertainty is now close to $25\%$.  The comparison  of parameter accuracy for these two systems highlights the importance of both a network of detectors and also observability of the \acp{HoM} in accurate inference of binary properties.

It is worth noting that our intuition from current GW observations that the majority of sources are close to face-on (or face-off) no longer holds in the next-generation network \citep{Vitale2016PhRvD..94l1501V}. In the nearby Universe, where sources are approximately uniform in volume, a signal observed with a given \ac{SNR} is most likely to originate from a distant binary which is close to face-on (or face off) as the number of sources increases as $d_{L}^{2}$.  For a high-redshift source, whose redshift is past the peak of the redshift distribution  --- likely around a redshift of $z \approx 2$ at which  star formation peaks --- this is no longer the case.  Now, the most likely origin is from a binary which is at lower redshift, where the intrinsic rate is higher, and is either poorly oriented or from a region of the sky where the detectors have lower sensitivity.  Thus, the results from sources inclined at $60\degree$ and $90\degree$ are more typical of the observed population.

Let us return to the implications for probing the location of the 
\ac{PISN} mass-gap.  For both mass pairs, binaries inclined at $60\degree$ or $90\degree$ are those which provide the best mass measurements.  For the $(90, 90) \msun$ system, we have $m_{1}$ and $m_{2}$ measured in the interval $\in [70, 100] \msun$.  So, this system is consistent with both components lying below the mass-gap provided the $^{12}\rm C(\alpha,\gamma)^{16}O$ rate is low.  For the $(120, 60) \msun$ system, we have $m_{1} \in [100, 140] \msun$ and $m_{2}\in [40, 80] \msun$.  The masses are consistent with one above and one below the gap, provided the reaction rate is high.  If \textit{both} signals were observed, this would be inconsistent with our current understanding of the \ac{PISN} mass gap.

To investigate the observability at even higher redshifts, we have simulated a second set of signals, with the same masses and inclinations but with a lower \ac{SNR} fixed at 15.  For these systems, the redshifts range from $z \gtrsim 20$ for face-on systems to $z \approx 15$ for edge on systems.  Broadly, the results are consistent with those in Fig.  \ref{fig:mass_gap_pe} and \ref{fig:mass_gap_redshift}, but with larger uncertainties due to the lower signal amplitude.  In particular, for all but the face-on systems, we are able to clearly identify that the $(120, 60) \msun$ binary is of unequal mass, due to the observed power in \acp{HoM}.  For the inclined systems, the uncertainty in total mass and redshift is around a factor of two (from 150 to $300 \msun$ and $z=12$ to 25).  Thus, while it is possible to identify these systems as unambiguously high-mass and high-redshift sources, the uncertainties in masses and redshifts make it difficult to perform precision astrophysics.  For the $(90, 90) \msun$ system, it is only at $\iota = 60\degree$ or $90\degree$ that the parameters are well recovered.  These results are shown in Fig. \ref{fig:mass_gap_pe_snr15} and \ref{fig:mass_gap_redshift_snr15} in Appendix \ref{app:low_snr}.

\begin{figure}[t]
\includegraphics[width=0.48\textwidth]{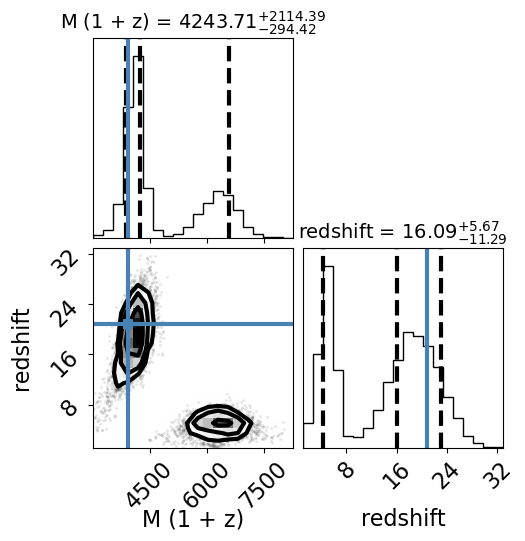}
 \caption{Inferred posteriors for redshift and redshifted mass for a simulated $(120,60) \msun$ \ac{BBH} merger with $\iota = 0\degree$ observed at $z=21$.}
 \label{fig:m_z_corner}
\end{figure}

For the face-on systems, we see an interesting feature whereby the binary can be mistaken for a different system with very different properties.  Fig. \ref{fig:m_z_corner} shows the inferred redshift and redshifted mass --$M(1 + z)$ -- distributions for a $(120, 60)\msun$ system at $z=21$.  The primary peak is at $z = 21$ and $M (1 + z) \approx 4,000 \msun$ corresponding to the simulated value.  However, there is a secondary peak around $z=5$, with a redshifted mass around $6,000 \msun$ corresponding to a binary with mass of $1,000\msun$.  For this system, it is the (3, 3) multipole which is consistent with the simulated signal.  This provides another example of the challenges which arise when identifying high-mass, high-redshift binaries.  The signal would be observed only in \ac{ET}, and have only one observable multipole.  Not only does this lead to poor parameter recovery, but also the inability to distinguish between a $180\msun$ binary at $z=21$ and a $1,000\msun$ binary at $z=5$.  Given the \ac{GW} data alone, it would not be possible to distinguish between the two scenarios.  The relative significance of the two will depend upon astrophysical knowledge of the mass and redshift distributions of \acp{BBH}.  Here, we have used priors which are uniform in comoving volume and component masses.  Other choices might lead to different conclusions about the mass and redshift of the binary. 

In summary, for these two representative sources, and for high \acp{SNR} we would confidently identify the systems as high-mass \acp{BBH} at high redshift. These could be potential seeds for the growth of \acp{SMBH}. The first system of $(90,90)\msun$ would be marginally consistent with being a binary formed at the lower edge of the mass gap, in correspondence of the lowest value  of the $^{12}\rm C(\alpha,\gamma)^{16}O$ reaction rate. Inconsistent otherwise.  The second  system of  $(120,60)\msun$  would be consistent with one BH (the lightest) originating from the core-collapse of a massive star (provided the $^{12}\rm C(\alpha,\gamma)^{16}O$ rate is low) and the second, in the midst of the pair-instability gap, would have a different origin. For a large fraction of the computed rates, the discovery of the latter system  would be inconsistent with the explosion scenario implied by the pair instability that would predict no BHs, and therefore a different channel has to be called for, for both components.  

\subsection{Observing intermediate mass black holes in binaries}  

Next, let us consider higher mass systems, containing an \ac{IMBH}.  For concreteness, we consider binaries with masses $(240,120)\msun$ and $(480,120)\msun$ which would be observable up to $z\sim 10$.  While the $120\msun$ \ac{BH} would be at the border of the pair-instability gap, both \acp{BH} are well above the gap. Their observation would either require a sufficiently high upper-mass end of a Kroupa-like IMF (extending to at least $300\msun$) or a top-heavy IMF \citep{Klessen2023}. Alternatively the primary \ac{BH} (particularly the most massive one of $480\msun$) could have had time to increase its original mass due to accretion of gas from its surroundings.  Thus for these systems, we are interested in determining whether the mass of the primary and the redshift can be accurately inferred in order to identify these early \acp{IMBH}.
   
Assuming the mass gap to be that predicted by the median of the  $^{12}\rm C(\alpha,\gamma)^{16}O$ reaction rate, between $\sim 50 \msun$ and $130\msun$, and a standard Kroupa-like IMF, in the interval between $0.1\msun$ and $150\msun$, all our systems are expected to have dynamical origin in dense star clusters  (see, eg. \citealt{Mapelli2021}).\footnote{We warn that black holes with these masses could be of primordial origin or outcome of post-formation accretion, as mentioned in the Introduction.} Among these, Nuclear Star Clusters (NSCs) could be the sites for 3rd- or 4th-generation \acp{BBH} with observed individual masses up to $\sim 600\msun$, merging  on timescales  smaller than 500 Myr  compatible with the redshift of observation.  Formation in Globular Clusters (GCs) would be marginally compatible with our lightest systems.  According to \citet{Mapelli2022}, above $z\sim 10$ there are two available BBH formation channels, the isolated and the dynamical formation in young star clusters.   However, the maximum masses for these channels are $\sim 50\msun$ and $\sim 100 \msun$ respectively.  This implies that detecting \acp{BBH} with individual masses larger than $\sim 100\msun$ at $z>10$ could point to a top-heavy IMF, as predicted for the first stellar generation. 

\begin{figure*}
 \includegraphics[width=0.48\textwidth]{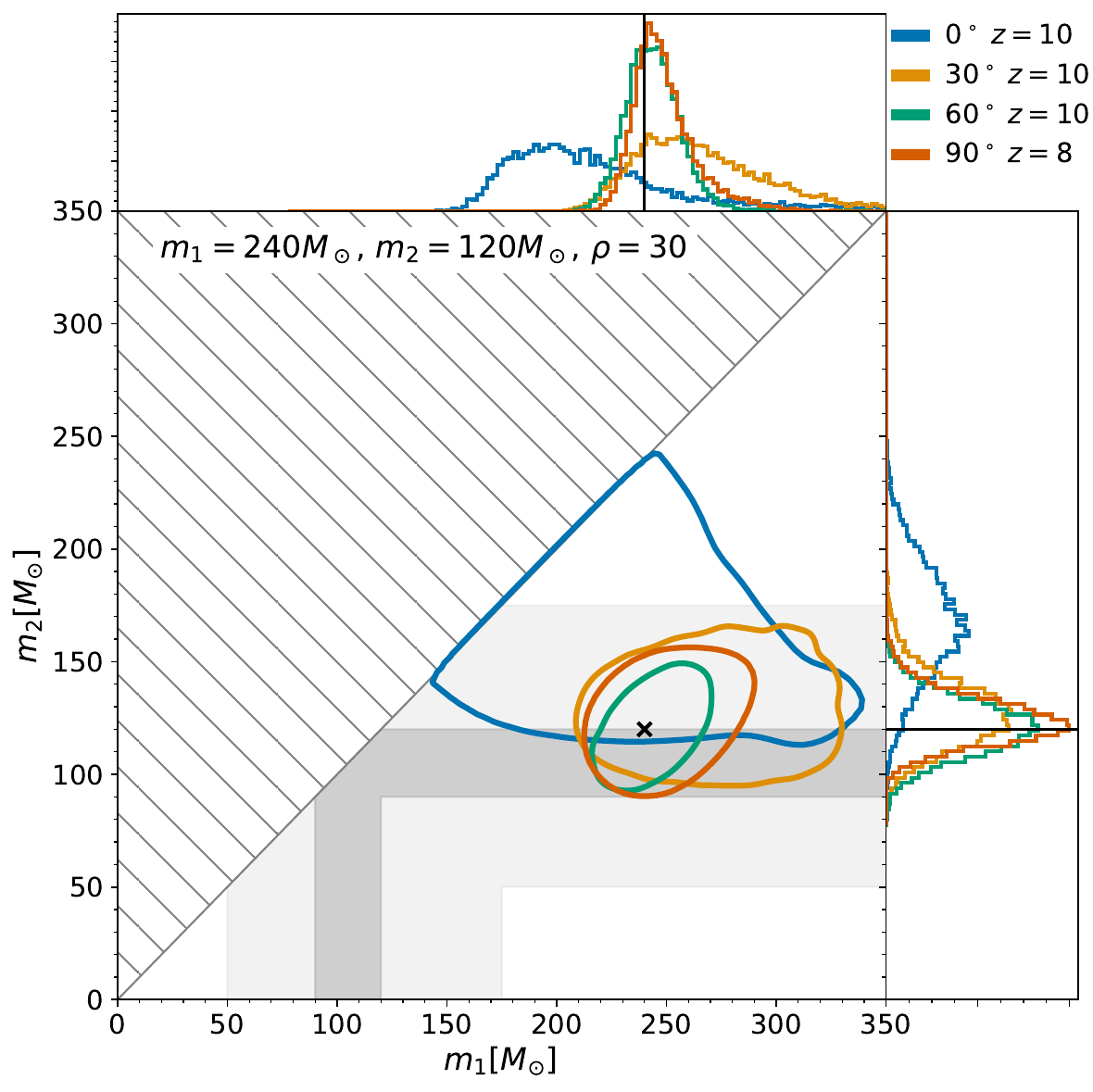}
\includegraphics[width=0.48\textwidth]{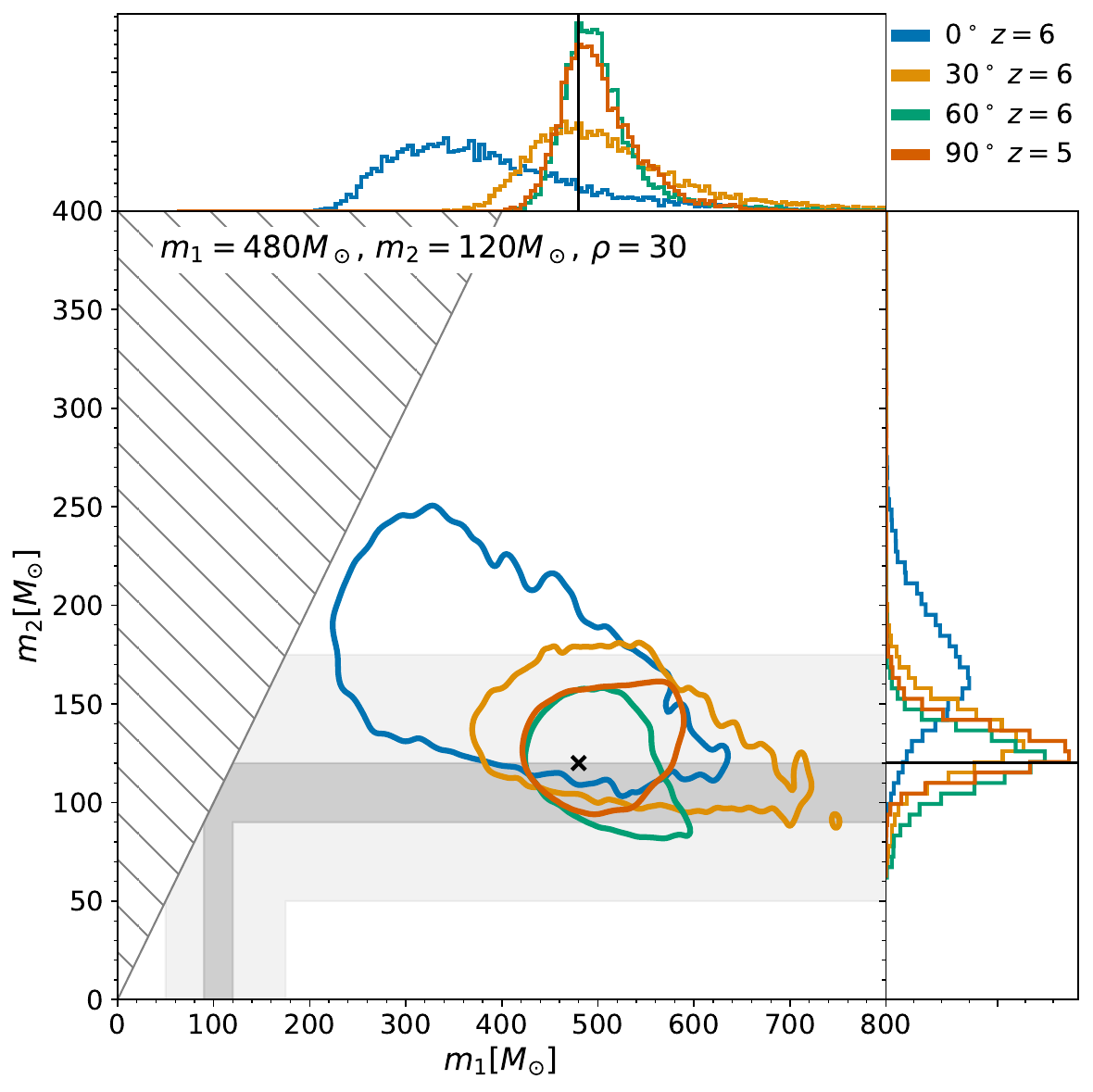}
\caption{
    Inferred posteriors for the component masses for heavy binaries.  Results are shown for inclinations of $\iota = 0\degree, 30\degree, 60\degree$ and $90\degree$, with the redshift of the system varied, as indicated in the labels, so that the signals are all simulated with an \ac{SNR} of 30 in the \ac{ET}-\ac{CE} network,
    at the most sensitive sky location for the network. The grey region denotes the pair-instability  mass-gap and the darker grey denotes the forbidden strip  where no black hole is expected to form for any value of the $^{12}\rm C(\alpha,\gamma)^{16}O$ reaction rate \citep{Farmer2020ApJ...902L..36F}.  
     Simulated values are denoted by a black cross and contours show the 90\% credible region. \textit{Left}:  $m_1=240 \msun$ and $m_2=120 \msun$. \textit{Right}: $m_1=480 \msun$ and $m_2=120 \msun$. 
 }
 \label{fig:intermediate_mass_pe}
\end{figure*} 

\begin{figure*}
 \includegraphics[width=0.48\textwidth]{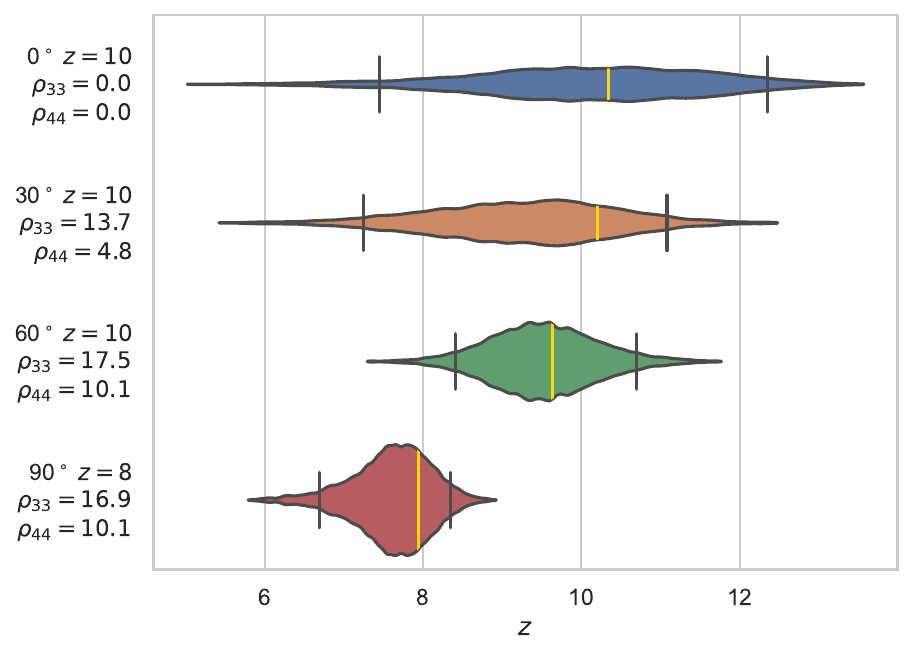}
\includegraphics[width=0.48\textwidth]{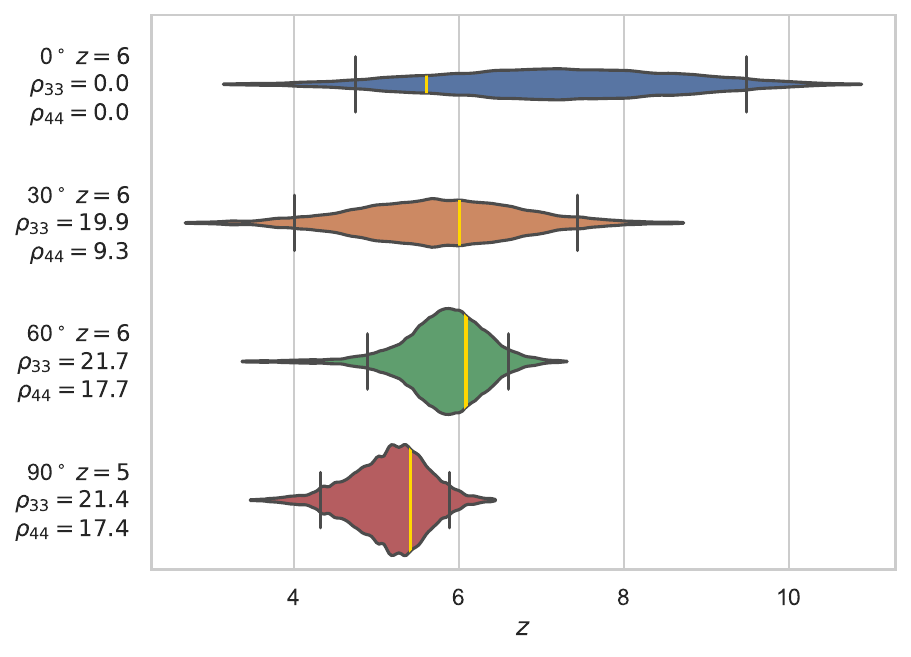}
 \caption{Inferred posteriors for the redshift for heavy binaries. Signals are simulated with varying inclination and redshift, as indicated in the labels, with a fixed \ac{SNR} of 30 in the \ac{ET}--\ac{CE} network, at the most sensitive sky location for the network.  Black vertical lines indicate the 90\% credible intervals, violins show the 99.5\% range and yellow vertical lines show simulated values. Symbols $\rho_{\ell m}$ indicate the \ac{SNR} in the $(\ell, m)$ multipole \acp{HoM}.  \textit{Left}:  $m_1=240 \msun$ and $m_2=120 \msun$. \textit{Right}: $m_1=480 \msun$ and $m_2=120 \msun$.  
 }
 \label{fig:intermediate_mass_redshift}
\end{figure*}

Figures \ref{fig:intermediate_mass_pe} and \ref{fig:intermediate_mass_redshift} show the accuracy with which we can measure the masses and redshifts of the events.
The broad features are similar to what we have already observed for the lower mass systems, namely that the parameter recovery is significantly worse for face-on systems, due to the vanishing \acp{HoM}.  Even though the mass ratio of the systems is 2 or 4, both are inferred to be consistent with equal mass (or nearly equal mass) binaries when viewed face-on.  Furthermore, the uncertainty in redshift and total mass is about a factor of two.  For the inclined systems, the recovery of masses and redshifts improves significantly, particularly for $\iota \ge 60\degree$.  In that case, component masses and redshifts are recovered with a $\sim 20\%$ accuracy.  In particular, the mass of the $120\msun$ \ac{BH} will be constrained to be between $90\msun$ and $\sim 150\msun$ for all except for the face-on system.  This is consistent with a black hole above the mass-gap and, due to uncertainties in the mass measurement, will not significantly restrict the $^{12}\rm C(\alpha,\gamma)^{16}O$ reaction-rate.  In Appendix \ref{app:low_snr}, we also show results for events simulated at higher redshifts and at a lower \ac{SNR} of 15.  The results are comparable to those discussed above, with the masses and redshifts for inclined systems better measured, and masses constrained to be unequal.  For face-on systems, the parameter recovery is significantly worse and we again see multiple peaks in the mass-redshift distributions corresponding to different multipoles matching with the signal.

Remarkably, the next-generation GW observatories have the capability to detect and accurately identify mergers involving $240 \msun$ \ac{BH} at a redshift of 10, and confidently infer a minimum redshift of 7, and mergers involving a $480 \msun$ \ac{BH} at a redshift of 6, and infer a redshift of at least 4.  These systems will be interesting to observe because we do not know if \acp{BH} of those masses exist, and we can hope to shed light on their formation routes, either by accretion from lower-mass \acp{BH} or by direct collapse of very massive stars. 
\section{Discussion}
\label{sec:discussion}
The next-generation of \ac{GW} detectors provide a unique way to probe the existence of heavy stellar black holes in the high-redshift Universe. 
Future \ac{GW} observations of \acp{BH} with masses above $\sim 50\msun$ at redshift $z\sim 10-15$  will enable us to probe the properties of the first stars forming  in the Universe  and their initial mass function.
 If \acp{BH} in the mass range explored here exist, they can contribute, as {\it seeds} \footnote{the so called ``light seeds'' explored in the literature \citep{Valiante17review, Inayoshi2020, SeedNAT2021}.}, to the rapid growth of 
the population of quasars observed close to the recombination epoch,  at $z\approx 7.5$, and housing accreting \acp{BH} of ${\cal O}(10^8- 10^9)\msun$ \citep[e.g.][]{Volonteri05, Madau14, Pezzulli16, Volonteri2015c, Ricarte2018,Valiante-colpi2021,Trinca2022}.
Whether and how the bridge between stellar and supermassive black holes was established when the first galaxies were forming is currently unknown \citep{SeedNAT2021}.
The revolutionary data coming from JWST, with the recent discovery of more than 40 new faint accreting supermassive \acp{BH} of ${\cal O} (10^5-10^7\msun)$
at $4 < z \leq 10.6$ \citep [see] [and references therein] {Maiolino2023}, is an outstanding confirmation of the rich \ac{BH} landscape at cosmic dawn predicted by theoretical models \citep{schneider2023}. In the future, with the Laser Interferometer Space Antenna (LISA) in operation \citep{LISA17}, we will detect low-frequency \acp{GW} from merging massive \acp{BH} of ${\cal O}(10^4-10^6\msun)$ out to $z\sim 10-15$. 
By combining and confronting statistically  all observations of both merging and accreting \acp{BH},  we will be able to shed light into the origin and evolution of the BH populations, from the stellar to the supermassive through the intermediate-mass ones,  across the cosmic epochs \citep{Fragione2023,Valiante-colpi2021}.

In this paper, we focused on the observability of high-redshift stellar \acp{BBH} with high masses, and, equally importantly, on the accuracy with which their masses and redshifts can be inferred.  We have shown that both the observation of systems and the accurate measurement of their parameters depend critically on the inclusion of \acp{HoM} in the \ac{GW} waveform.  At the highest masses and redshifts, \acp{HoM}, which extend the signal to higher frequencies than the (2, 2) multipole, can significantly increase the sensitive range of the detectors.  Across a broad range of masses and redshifts, we expect to see multiple \ac{GW} multipoles in signals observed by \ac{CE} and \ac{ET}.  Observation of more than one multipole, typically a \ac{HoM} in addition to the (2, 2) multipole, enables the breaking of the degeneracy between binary inclination and distance, as well as a more accurate determination of the mass ratio.  Additionally, a network of observatories is required for source's localizations.  When a signal is seen in only a single detector, the sky location is poorly measured and, since the detector response varies significantly over the sky, this leads to large uncertainties in the distance.  For very high-redshift sources, accurate distance/redshift measurement is vital for the measurement of the \ac{BH} individual  masses, as the observed signal depends upon $M(1+z)$.  By performing full parameter estimation on a set of representative systems, we demonstrated that it will be possible to measure masses and redshifts with an accuracy of $10-20\%$, for signals at redshifts up to at least 15.  Those systems which can be observed and accurately measured are typically seen in both CE and ET detectors, so they are well localized and also tend to be viewed away from face-on (or face-off) so that more than one \ac{GW} multipole is observed.

We examined systems with masses $(120, 60) \msun$ and $(90, 90) \msun$, which lie in, or around the {\it pair-instability mass-gap}. 
For the best-measured of the examples we investigated, at a redshift of 10, we could measure redshift and component masses with 10\% uncertainly.  This would, unambiguously, place these sources in the high-redshift Universe and serve to constrain the, currently unknown, location of the pair-instability mass-gap.  We also investigated mergers of $(240, 120)\msun$ and $(480, 120)\msun$ binaries, which enable us to probe the observability of early \ac{IMBH}. It will be possible to observe  these \ac{IMBH} at redshifts up to $z=10$ and constrain the redshift to be at least $z=7$.

The results in this paper complement those in \cite{Ng2022, Ng2023PhRvD.107b4041N} which investigate lower-mass \ac{BH} mergers in the next-generation \ac{GW} network, and \cite{Mancarella2023PhRvD.107j1302M} who  introduced the concept of 
an “inference horizon” which maps the redshift a
source can confidently be placed beyond.  In all cases, it is shown that next-generation \ac{GW} network provides a unique capability to probe high-redshift black hole formation.

The most critical feature of detector sensitivity for observing these systems is the low-frequency sensitivity of the detectors.  
In our study, we used a low-frequency limit of $3$ Hz for \ac{ET} and $5$ Hz for \ac{CE}.  Even the relatively small change from $3$ Hz to $5$ Hz can have a profound impact on sensitivity to high-mass, high-redshift sources as shown in Fig. \ref{fig:snr_vs_inc_q_et} and \ref{fig:snr_vs_inc_q_ce}.  Achieving the instrumental design sensitivity at low frequencies has been challenging in the current LIGO and Virgo observatories.  As the detailed technical designs of the next-generation observatories are finalised, the desire to probe the remnants of high-mass stars in the early Universe should be considered as a motivation to optimize sensitivity at low frequencies. 

\section*{Acknowledgements}

We thank Riccardo Buscicchio for his careful reading and valuable comments. SF acknowledges the support of STFC grant ST/V005618/1 and a Leverhulme Trust International Fellowship.
RS acknowledges support from the Amaldi Research Centre funded by the MIUR programme
``Dipartimento di Eccellenza'' (CUP:B81I18001170001). MC, RS, AT, RV acknowledge the INFN
TEONGRAV specific initiative.  MC acknowledges support by the 2017-NAZ- 0418/PER grant, and
by the Italian Ministry for Universities and Research (MUR) program ``Dipartimenti di Eccellenza 2023-2027'', within the framework of the activities of the ``Centro Bicocca di Cosmologia Quantitativa (BiCoQ)''.
 AS acknowledges the financial support provided under the European Union’s H2020 ERC Consolidator Grant ``Binary Massive Black Hole Astrophysics'' (B Massive, Grant Agreement: 818691). MC and RS thank the Institut d'Astrophysique de Paris for kind
hospitality.

\appendix

\section{Detector Sensitivity}
\label{app:sensitivity}
\begin{figure*}
  \centering
  \includegraphics[width=0.48\textwidth]{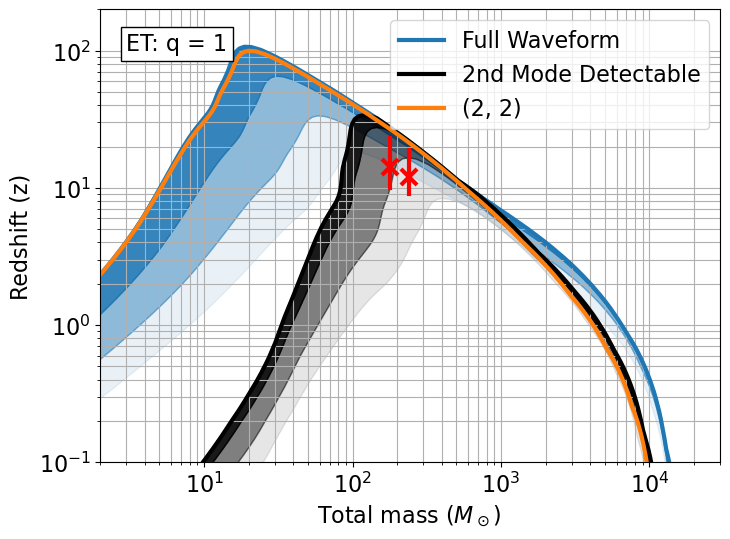}
  \includegraphics[width=0.48\textwidth]{horizon_events_ET_q_2}
  \includegraphics[width=0.48\textwidth]{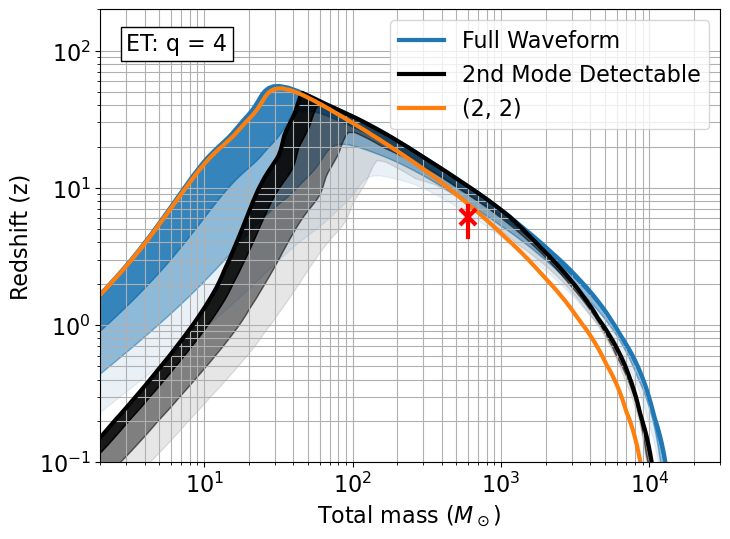}
  \includegraphics[width=0.48\textwidth]{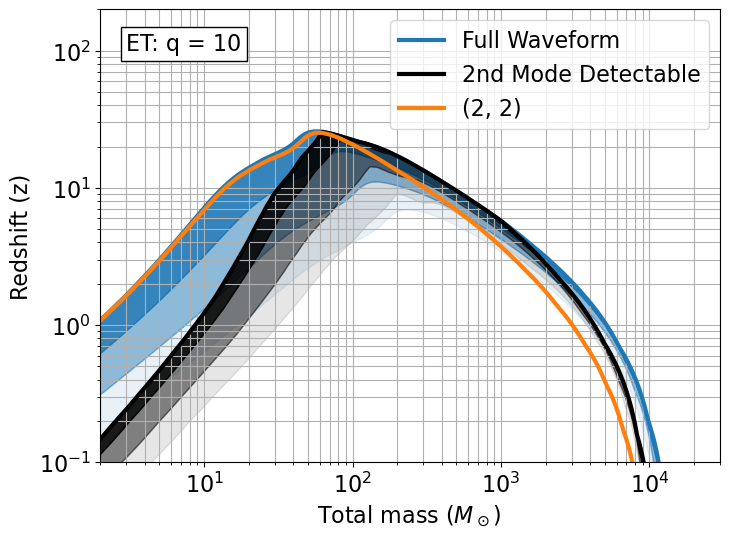}
    \caption{Maximum observable redshift of \ac{BBH} mergers with \ac{ET} for mass ratios $q = 1, 2, 4$ and $10$.  Shaded regions show redshifts at which 10\%, 50\% and 90\% of sources are observable.   The maximum observable redshift of the (2, 2) multipole at \ac{SNR} = 8 is shown in orange, the observability of the full waveform at \ac{SNR} = 8 is shown in blue, and the observability of the second most significant mode above \ac{SNR} = 3 is shown in black.  The red asterisks represent the masses and redshifts of the ($q = 2$) systems we study in detail in Section \ref{sec:pe}.
    }
  \label{fig:et_horizons}
\end{figure*}

\begin{figure*}
  \centering
  \includegraphics[width=0.48\textwidth]{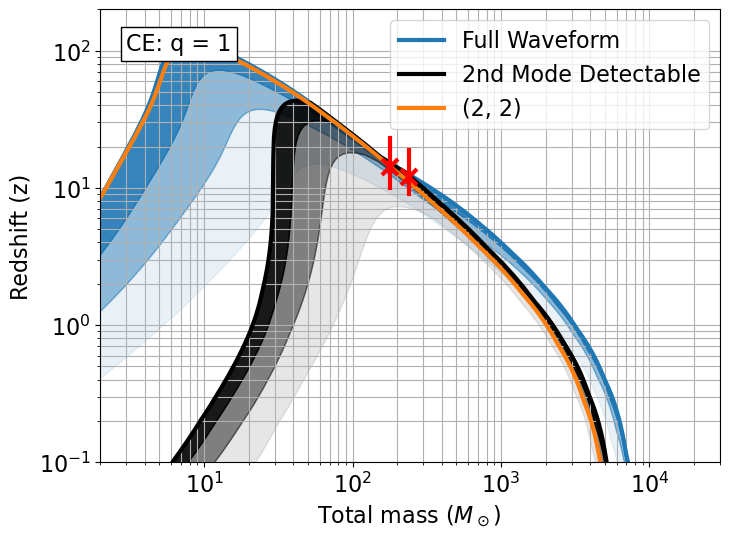}
  \includegraphics[width=0.48\textwidth]{horizon_events_CE_q_2}
  \includegraphics[width=0.48\textwidth]{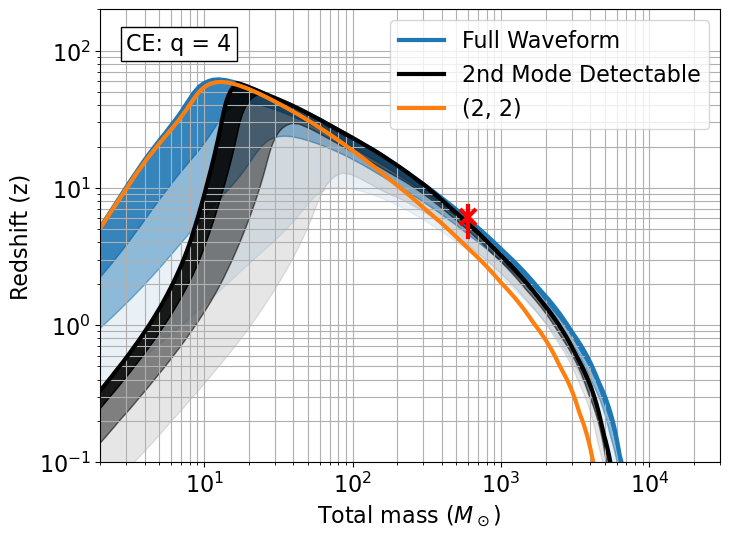}
  \includegraphics[width=0.48\textwidth]{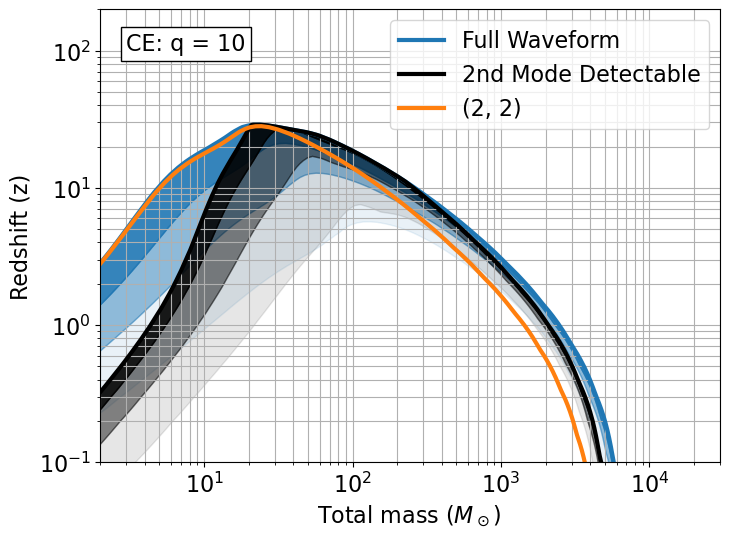}
  \caption{Same as \ref{fig:et_horizons} but for CE.
    }
  \label{fig:ce_horizons}
\end{figure*}

In this Appendix, we show the sensitivity of \ac{ET} and \ac{CE} to binary mergers as a function of mass and redshift.  Fig. \ref{fig:et_ce_horizon} shows the sensitivity to mergers with mass ratio of two.  In Fig. \ref{fig:et_horizons} we show the \ac{ET} sensitivity for binaries with mass ratios 1, 2, 4 and 10.  In Fig. \ref{fig:ce_horizons}, we show the same for \ac{CE}.
The maximum reach of the detectors is for equal mass binaries.  However, at equal mass the (3, 3) multipole vanishes so there is a larger range for which only one multipole is visible.  As we increase the mass ratio, the maximum sensitive redshift decreases, as the amplitude of the emitted \ac{GW} also decreases.  However, the relative significance of the \acp{HoM} increases so that an increasing fraction of sources will be observed with at least two multipoles.
 
\section{Low SNR signals}
\label{app:low_snr}
In this Appendix we present parameter estimation results for systems that would have a network \ac{SNR} of 15 in the ET-CE network described in Section \ref{sec:pe}, to complement the results for \ac{SNR} 30 presented in that section of the paper.  Since these events are at a lower \ac{SNR}, they are also at a higher redshift, with the $(120, 60)\msun$ and $(90, 90)\msun$ binaries at redshifts between 23 and 15 (depending upon inclination).  At these redshifts, the signal is shifted to such low frequencies that it lies essentially outside of the sensitive band of \ac{CE} --- the \ac{SNR} of these events in \ac{CE} is less than 2 in all cases.  Consequently, all sources are poorly localized on the sky, with a typical 90\% localization of thousands of square degrees.

\begin{figure*}
 \centering
 \includegraphics[width=0.48\textwidth]{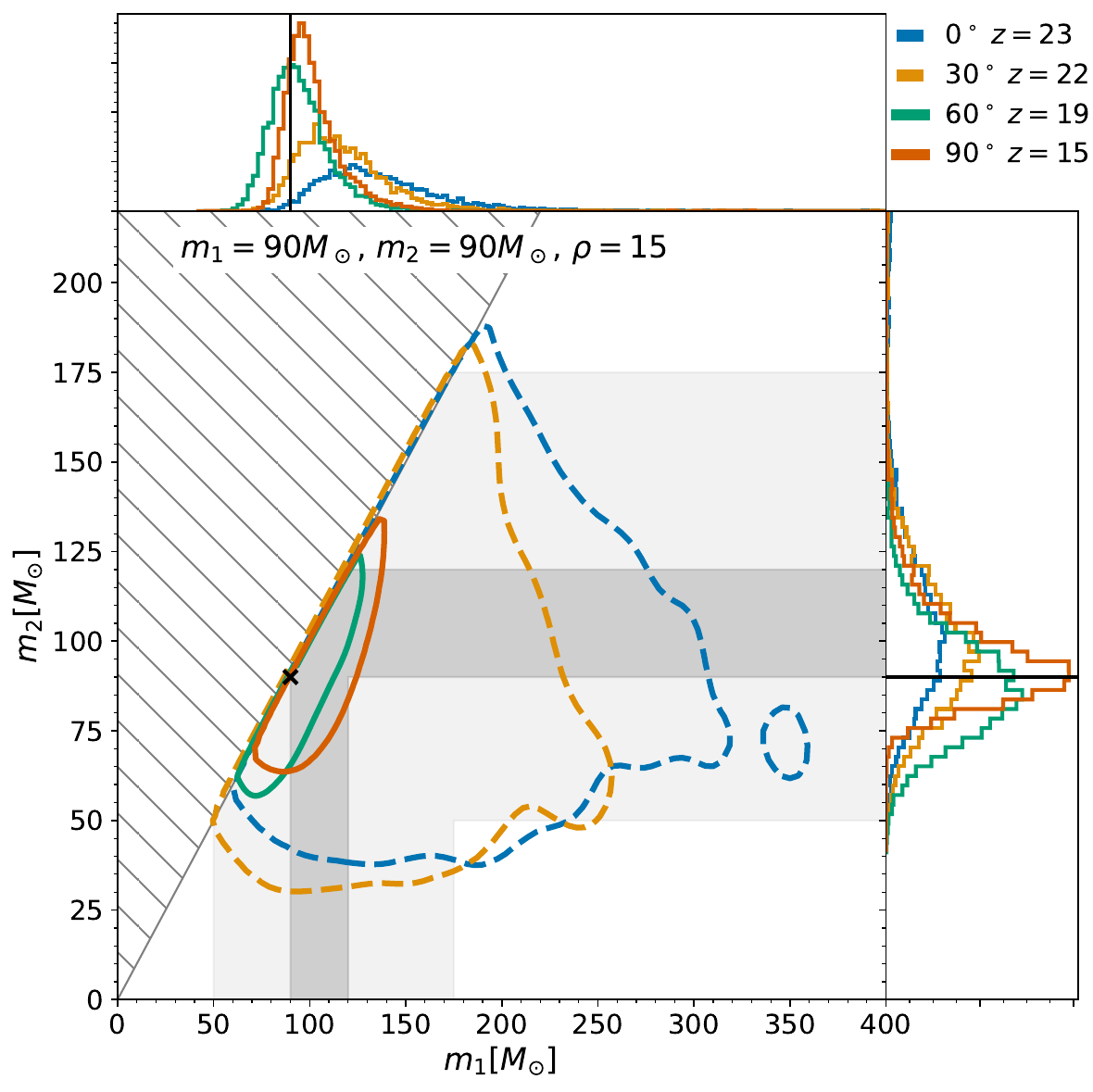}
 \includegraphics[width=0.48\textwidth]{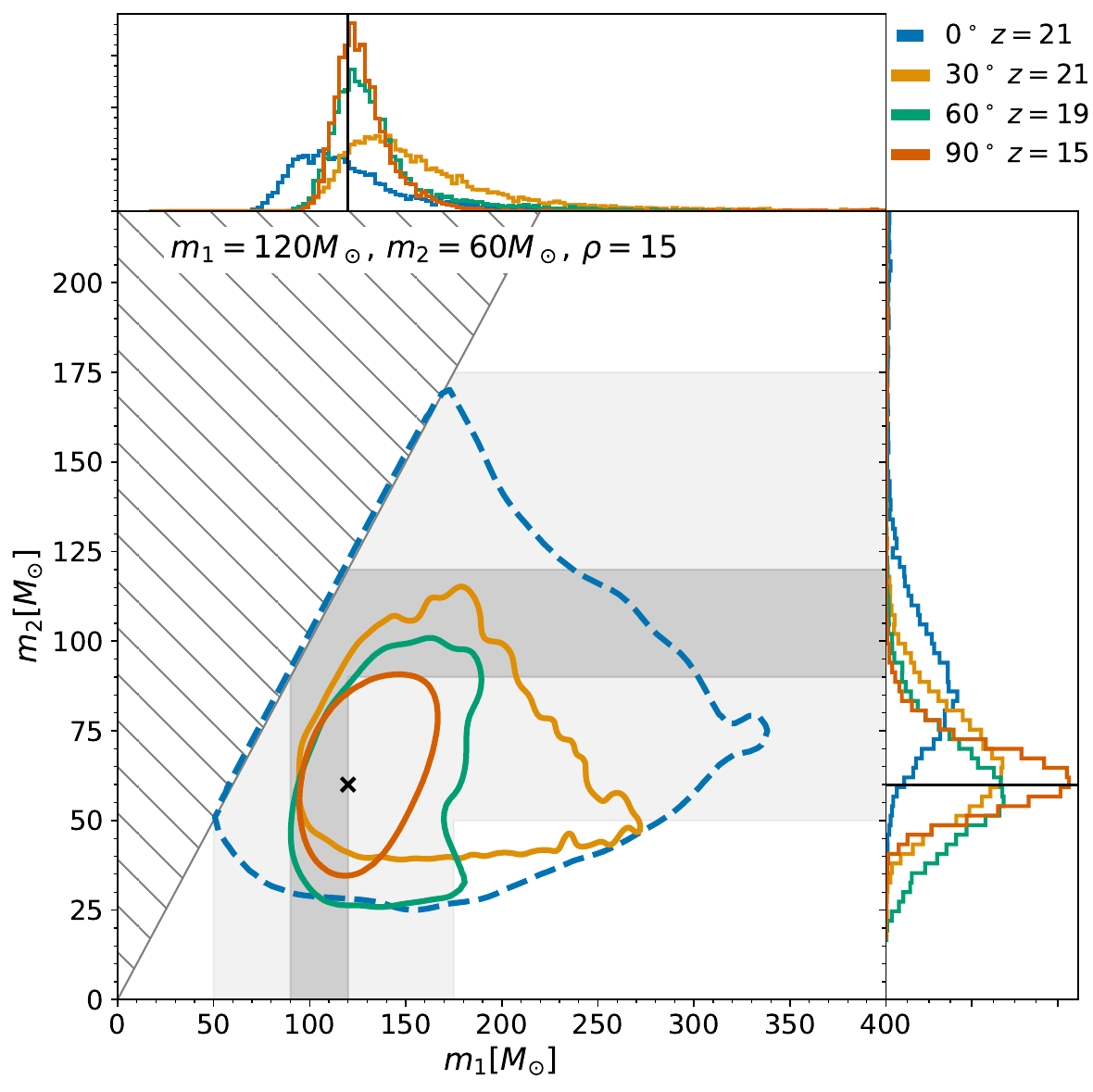}
 \caption{
  Inferred posteriors of component masses for binaries of total mass of $180\msun$ which straddle, or sit within the upper mass gap.  Results are shown for inclinations of $\iota = 0\degree, 30\degree, 60\degree$ and $90\degree$, with the redshift of the system varied, as indicated in the labels, so that the signals are all simulated with an \ac{SNR} of 15 in the \ac{ET}-\ac{CE} network, at the most sensitive sky location for the network. 
   The grey region denotes the mass gap, whose location varies with the $^{12}\rm C(\alpha,\gamma)^{16}O$ reaction rate, and the darker grey denotes the forbidden strip  \citep{Farmer2020ApJ...902L..36F}. Simulated values are denoted by a black cross and contours show the 90\% credible region. \textit{Left}: Binary with masses just below the mass gap: $m_1=m_2=90 \msun$. \textit{Right}: Binary with masses that straddle the mass gap: $m_1=120 \msun$ and $m_2=60 \msun$.
 }
 \label{fig:mass_gap_pe_snr15}
\end{figure*}

\begin{figure*}
 \includegraphics[width=0.48\textwidth]{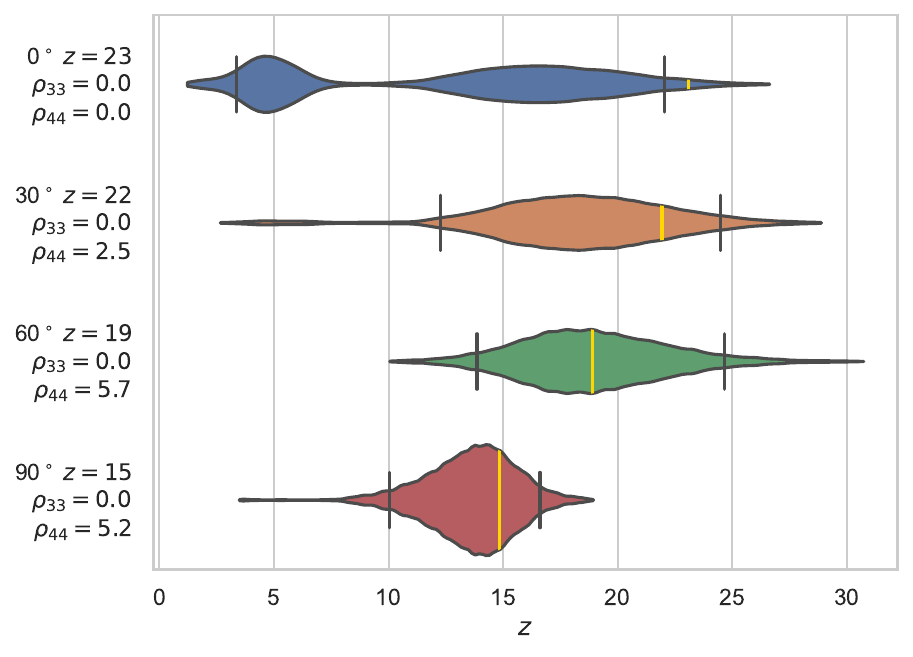}
 \includegraphics[width=0.48\textwidth]{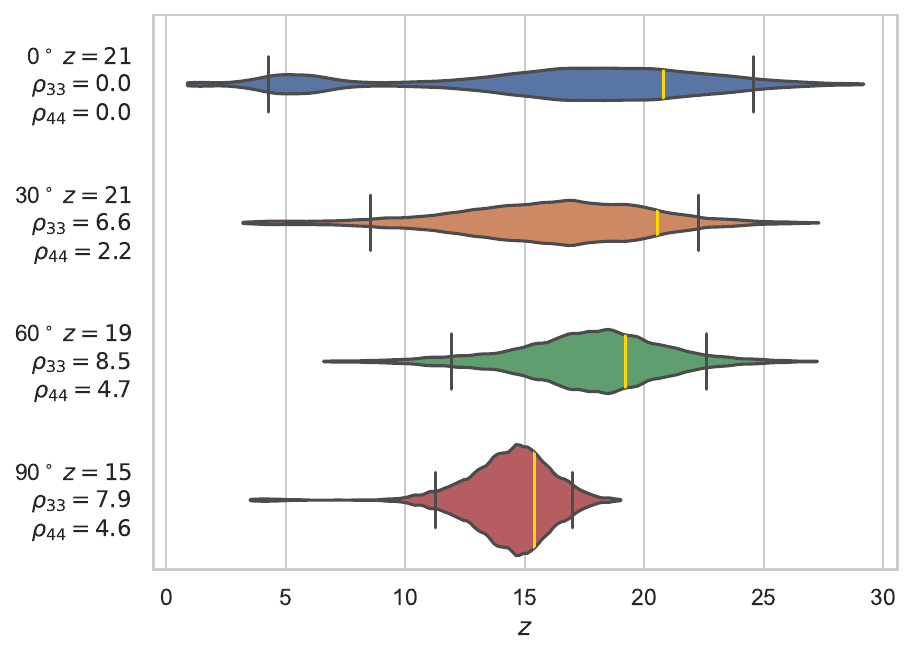}
 \caption{Inferred posteriors for the redshift for binaries of total mass of $180\msun$ which straddle, or sit within the upper mass gap. Signals are simulated with varying inclination and redshift, as indicated in the labels, with a fixed \ac{SNR} of 15 in the \ac{ET}--\ac{CE} network, at the most sensitive sky location for the network.  Black vertical lines indicate the 90\% credible intervals, violins show the 99.5\% range and yellow vertical lines show simulated values.  Symbols $\rho_{\ell m}$ indicate the \ac{SNR} in the $(\ell, m)$ multipole \acp{HoM}. \textit{Left}: Binary with masses just below the mass gap: $m_1=m_2=90 \msun$. \textit{Right}: Binary with masses that straddle the mass gap: $m_1=120 \msun$ and $m_2=60 \msun$.
 }
 \label{fig:mass_gap_redshift_snr15}
\end{figure*}

In Fig. \ref{fig:mass_gap_pe_snr15} and \ref{fig:mass_gap_redshift_snr15}, we show the accuracy with which the masses and redshift are recovered for the $(120, 60)\msun$ and $(90, 90)\msun$ binaries.  The qualitative results are similar to those for the \ac{SNR} 30 signals presented in Section \ref{sec:pe}, with broader posteriors as expected due to the lower \ac{SNR}.  Specifically, the masses and redshifts are poorly measured for face-on systems, and measurement accuracy improves for inclined systems (particularly $\iota = 60\degree, 90\degree$) where there is observable power in the \acp{HoM}.  In the best-case scenarios, masses and redshifts are measured with $\sim25\%$ accuracy.  For all systems other than $\iota=0\degree$ the $(120, 60)\msun$ system is clearly identified as having unequal masses.  However, the mass distributions are broad enough that limited information about the location of the pair-insatability mass-gap can be extracted.  

For the $\iota=0\degree$ systems, and $\iota=30\degree$ for the $(90, 90)\msun$ binary, there is a bimodality in the recovered redshift.  In addition, the inferred mass distribution is broader than that shown in Fig.\ref{fig:mass_gap_pe_snr15} and extends to $\sim 1000 \msun$.  For these events, there is zero (or limited) power in the \acp{HoM} so only a single \ac{GW} multipole is observable.  The secondary peak at high masses and $z\approx 5$ corresponds to a binary configuration where the (3, 3) multipole has the correct amplitude and frequency content to match the simulated signal.  This is discussed in more detail in Section \ref{sec:pe}, around Fig.  \ref{fig:m_z_corner}.

\begin{figure*}
 \centering
 \includegraphics[width=0.48\textwidth]{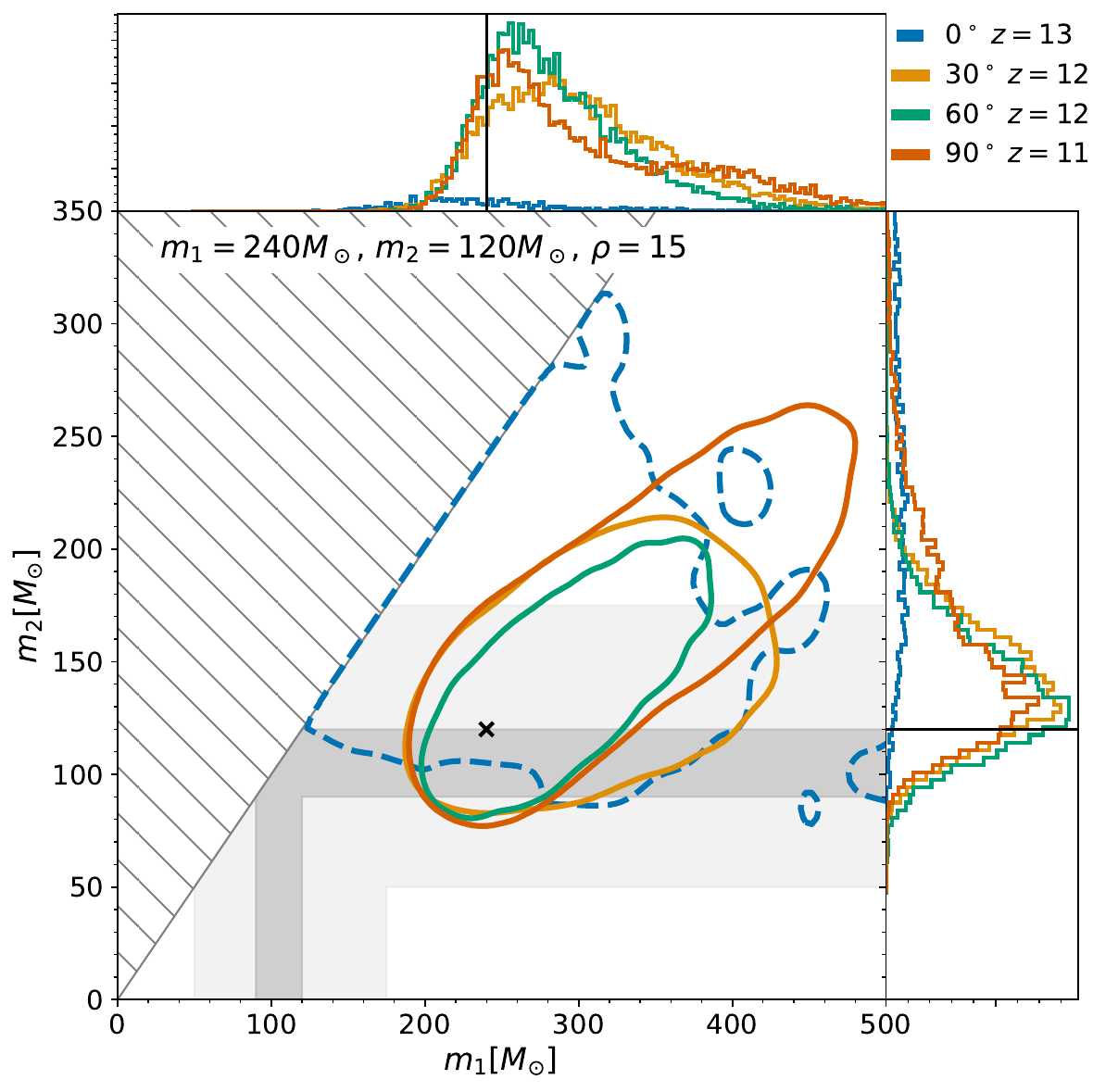}
 \includegraphics[width=0.48\textwidth]{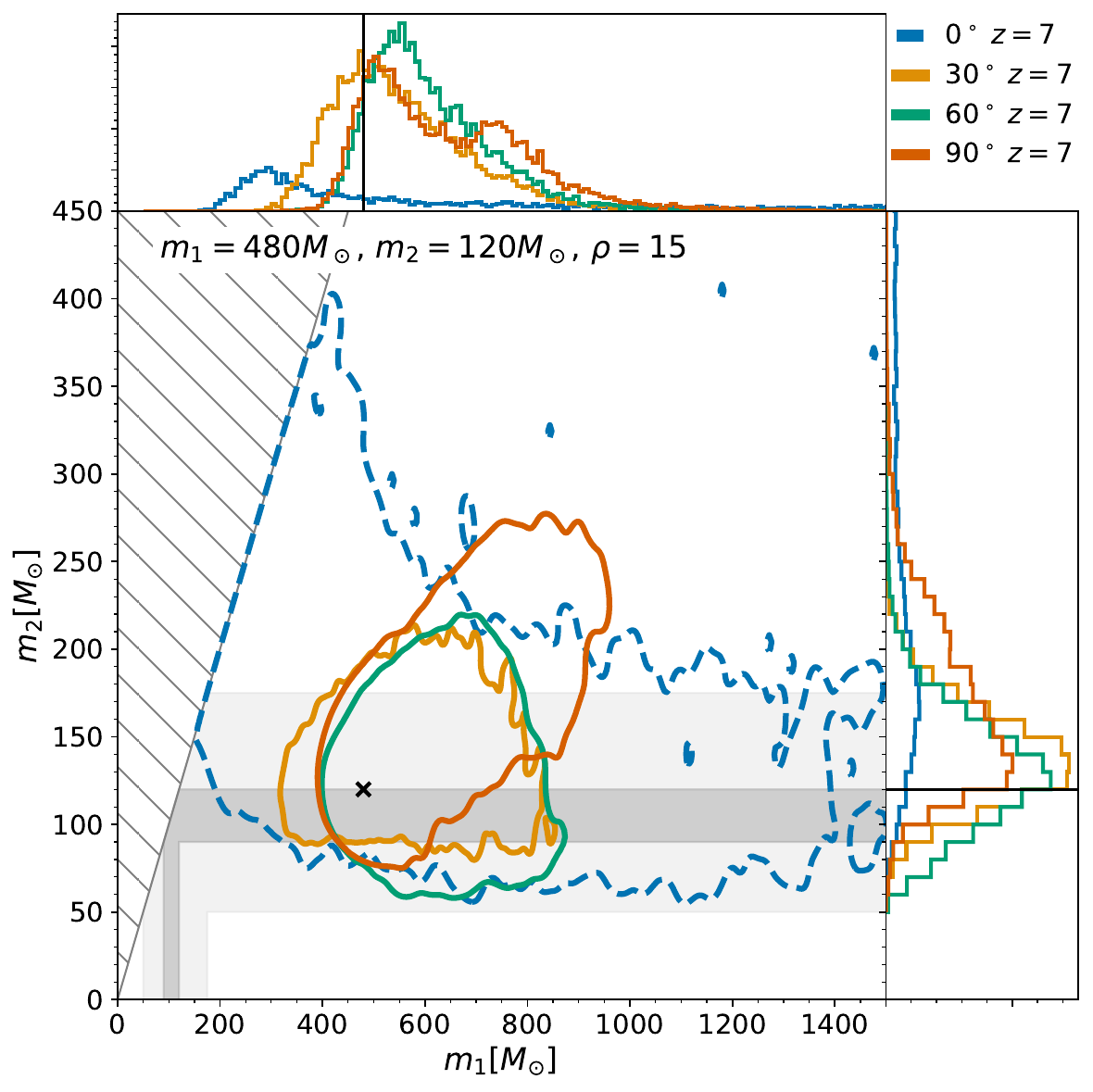}
 \caption{
     Inferred posteriors for the component masses for heavy binaries.  Results are shown for inclinations of $\iota = 0\degree, 30\degree, 60\degree$ and $90\degree$, with the redshift of the system varied, as indicated in the labels, so that the signals are all simulated with an \ac{SNR} of 15 in the \ac{ET}-\ac{CE} network, at the most sensitive sky location for the network. The grey region denotes the pair-instability  mass-gap and the darker grey denotes the forbidden strip  where no black hole is expected to form for any value of the $^{12}\rm C(\alpha,\gamma)^{16}O$ reaction rate \citep{Farmer2020ApJ...902L..36F}.  Simulated values are denoted by a black cross and contours show the 90\% credible region. \textit{Left}:  $m_1=240 \msun$ and $m_2=120 \msun$. \textit{Right}: $m_1=480 \msun$ and $m_2=120 \msun$. 
 }
 \label{fig:intermediate_mass_pe_snr15}
\end{figure*} 

\begin{figure*}
 \centering
 \includegraphics[width=0.48\textwidth]{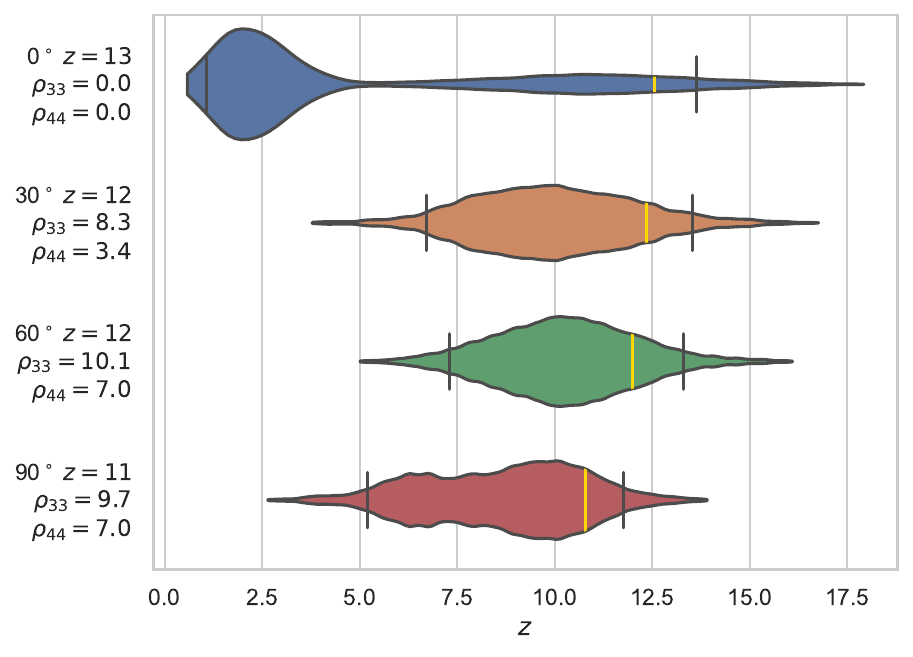}
 \includegraphics[width=0.48\textwidth]{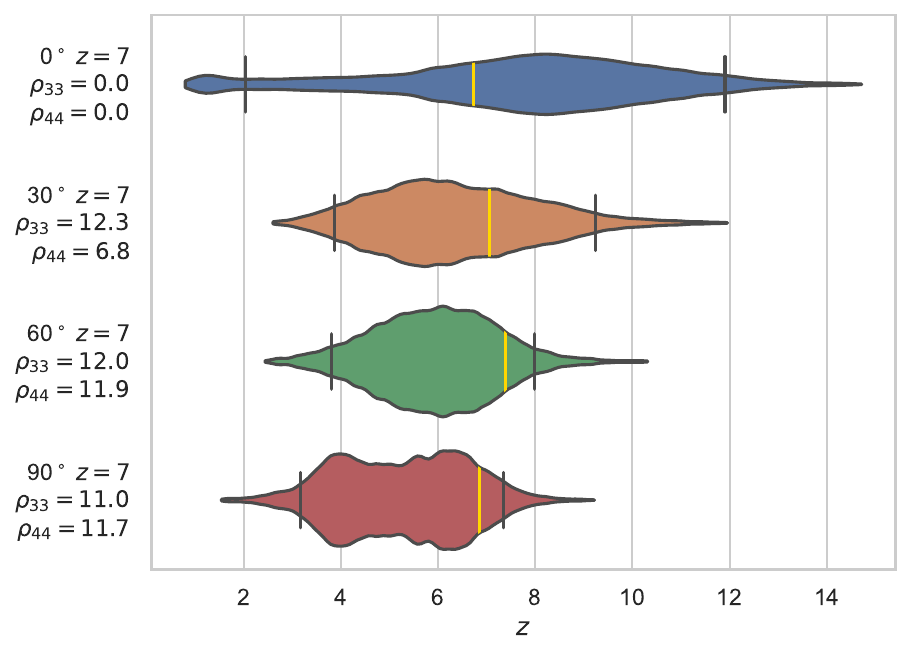}
 \caption{
 Inferred posteriors for the redshift for heavy binaries. Signals are simulated with varying inclination and redshift, as indicated in the labels, with a fixed \ac{SNR} of 15 in the \ac{ET}--\ac{CE} network, at the most sensitive sky location for the network.  Black vertical lines indicate the 90\% credible intervals, violins show the 99.5\% range and yellow vertical lines show simulated values. Symbols $\rho_{\ell m}$ indicate the \ac{SNR} in the $(\ell, m)$ multipole \acp{HoM}.  \textit{Left}:  $m_1=240 \msun$ and $m_2=120 \msun$. \textit{Right}: $m_1=480 \msun$ and $m_2=120 \msun$.  
 }
 \label{fig:intermediate_mass_redshift_snr15}
\end{figure*}

In Fig. \ref{fig:intermediate_mass_pe_snr15} and \ref{fig:intermediate_mass_redshift_snr15}, we show the accuracy with which the masses and redshift are recovered for the $(240, 120)\msun$ and $(480, 120)\msun$ binaries.  As for the lower-mass systems, the qualitative results are similar to those for the \ac{SNR} 30 signals presented in Section \ref{sec:pe}, with broader posteriors as expected due to the lower \ac{SNR}.  Nonetheless, other than the face-on ($\iota=0\degree$) systems, the binaries are clearly identified as unequal mass systems containing an \ac{IMBH} with minimum mass $200\msun$/$400\msun$ for the two system.  Redshifts are generally underestimated, likely due to the poor sky localization, and lower bounds on the redshift are \textit{no better} than for the higher \ac{SNR} systems shown in Fig. \ref{fig:intermediate_mass_redshift}. 

Again, the face-on systems show significant bimodality with a second peak at much lower redshifts and higher masses.  As before, this corresponds to a system where the \acp{HoM}, rather that the (2, 2) multipole, are associated with the observed waveform. 

\bibliography{monica_refs, thesis_refs}

\begin{thebibliography}{}
\expandafter\ifx\csname natexlab\endcsname\relax\def\natexlab#1{#1}\fi
\providecommand{\url}[1]{\href{#1}{#1}}
\providecommand{\dodoi}[1]{doi:~\href{http://doi.org/#1}{\nolinkurl{#1}}}
\providecommand{\doeprint}[1]{\href{http://ascl.net/#1}{\nolinkurl{http://ascl.net/#1}}}
\providecommand{\doarXiv}[1]{\href{https://arxiv.org/abs/#1}{\nolinkurl{https://arxiv.org/abs/#1}}}

\bibitem[{{Abbott} {et~al.}(2016){Abbott}, {Abbott}, {Abbott}, {Abernathy},
  {Acernese}, {Ackley}, {Adams}, {Adams}, {Addesso}, {Adhikari},
  {et~al.}}]{O1BBH}
{Abbott}, B.~P., {Abbott}, R., {Abbott}, T.~D., {et~al.} 2016, \prx, 6, 041015,
  \dodoi{10.1103/PhysRevX.6.041015}

\bibitem[{{Abbott} {et~al.}(2019){Abbott}, {Abbott}, {Abbott}, {Abraham},
  {Acernese}, {Ackley}, {Adams}, {Adhikari}, {Adya}, {Affeldt}, {Agathos},
  {Agatsuma}, {Aggarwal}, {Aguiar}, {Aiello}, {Ain}, {Ajith}, {Allen},
  {Allocca}, {Aloy}, {Altin}, {Amato}, {Ananyeva}, {Anderson}, {Anderson},
  {Angelova}, {Antier}, {Appert}, {Arai}, {Araya}, {Areeda}, {Ar{\`e}ne},
  {Arnaud}, {Arun}, {Ascenzi}, {Ashton}, {Aston}, {Astone}, {Aubin}, {Aufmuth},
  {AultONeal}, {Austin}, {Avendano}, {Avila-Alvarez}, {Babak}, {Bacon},
  {Badaracco}, {Bader}, {Bae}, {Baker}, {Baldaccini}, {Ballardin}, {Ballmer},
  {Banagiri}, {Barayoga}, {Barclay}, {Barish}, {Barker}, {Barkett}, {Barnum},
  {Barone}, {Barr}, {Barsotti}, {Barsuglia}, {Barta}, {Bartlett}, {Bartos},
  {Bassiri}, {Basti}, {Bawaj}, {Bayley}, {Bazzan}, {B{\'e}csy}, {Bejger},
  {Belahcene}, {Bell}, {Beniwal}, {Berger}, {Bergmann}, {Bernuzzi}, {Bero},
  {Berry}, {Bersanetti}, {Bertolini}, {Betzwieser}, {Bhandare}, {Bidler},
  {Bilenko}, {Bilgili}, {Billingsley}, {Birch}, {Birney}, {Birnholtz},
  {Biscans}, {Biscoveanu}, {Bisht}, {Bitossi}, {Bizouard}, {Blackburn},
  {Blackman}, {Blair}, {Blair}, {Blair}, {Bloemen}, {Bode}, {Boer}, {Boetzel},
  {Bogaert}, {Bondu}, {Bonilla}, {Bonnand}, {Booker}, {Boom}, {Booth}, {Bork},
  {Boschi}, {Bose}, {Bossie}, {Bossilkov}, {Bosveld}, {Bouffanais}, {Bozzi},
  {Bradaschia}, {Brady}, {Bramley}, {Branchesi}, {Brau}, {Briant}, {Briggs},
  {Brighenti}, {Brillet}, {Brinkmann}, {Brisson}, {Brockill}, {Brooks},
  {Brown}, {Brunett}, {Buikema}, {Bulik}, {Bulten}, {Buonanno}, {Buskulic},
  {Bustamante Rosell}, {Buy}, {Byer}, {Cabero}, {Cadonati}, {Cagnoli},
  {Cahillane}, {Calder{\'o}n Bustillo}, {Callister}, {Calloni}, {Camp},
  {Campbell}, {Canepa}, {Cannon}, {Cao}, {Cao}, {Capocasa}, {Carbognani},
  {Caride}, {Carney}, {Carullo}, {Casanueva Diaz}, {Casentini}, {Caudill},
  {Cavagli{\`a}}, {Cavalier}, {Cavalieri}, {Cella}, {Cerd{\'a}-Dur{\'a}n},
  {Cerretani}, {Cesarini}, {Chaibi}, {Chakravarti}, {Chamberlin}, {Chan},
  {Chao}, {Charlton}, {Chase}, {Chassande-Mottin}, {Chatterjee}, {Chaturvedi},
  {Chatziioannou}, {Cheeseboro}, {Chen}, {Chen}, {Chen}, {Cheng}, {Cheong},
  {Chia}, {Chincarini}, {Chiummo}, {Cho}, {Cho}, {Cho}, {Christensen}, {Chu},
  {Chua}, {Chung}, {Chung}, {Ciani}, {Ciobanu}, {Ciolfi}, {Cipriano}, {Cirone},
  {Clara}, {Clark}, {Clearwater}, {Cleva}, {Cocchieri}, {Coccia}, {Cohadon},
  {Cohen}, {Colgan}, {Colleoni}, {Collette}, {Collins}, {Cominsky},
  {Constancio}, {Conti}, {Cooper}, {Corban}, {Corbitt}, {Cordero-Carri{\'o}n},
  {Corley}, {Cornish}, {Corsi}, {Cortese}, {Costa}, {Cotesta}, {Coughlin},
  {Coughlin}, {Coulon}, {Countryman}, {Couvares}, {Covas}, {Cowan}, {Coward},
  {Cowart}, {Coyne}, {Coyne}, {Creighton}, {Creighton}, {Cripe}, {Croquette},
  {Crowder}, {Cullen}, {Cumming}, {Cunningham}, {Cuoco}, {Canton}, {D{\'a}lya},
  {Danilishin}, {D'Antonio}, {Danzmann}, {Dasgupta}, {Da Silva Costa},
  {Datrier}, {Dattilo}, {Dave}, {Davier}, {Davis}, {Daw}, {DeBra},
  {Deenadayalan}, {Degallaix}, {De Laurentis}, {Del{\'e}glise}, {Del Pozzo},
  {DeMarchi}, {Demos}, {Dent}, {De Pietri}, {Derby}, {De Rosa}, {De Rossi},
  {DeSalvo}, {de Varona}, {Dhurandhar}, {D{\'\i}az}, {Dietrich}, {Di Fiore},
  {Di Giovanni}, {Di Girolamo}, {Di Lieto}, {Ding}, {Di Pace}, {Di Palma}, {Di
  Renzo}, {Dmitriev}, {Doctor}, {Donovan}, {Dooley}, {Doravari}, {Dorrington},
  {Downes}, {Drago}, {Driggers}, {Du}, {Ducoin}, {Dupej}, {Dwyer}, {Easter},
  {Edo}, {Edwards}, {Effler}, {Ehrens}, {Eichholz}, {Eikenberry}, {Eisenmann},
  {Eisenstein}, {Essick}, {Estelles}, {Estevez}, {Etienne}, {Etzel}, {Evans},
  {Evans}, {Fafone}, {Fair}, {Fairhurst}, {Fan}, {Farinon}, {Farr}, {Farr},
  {Fauchon-Jones}, {Favata}, {Fays}, {Fazio}, {Fee}, {Feicht}, {Fejer}, {Feng},
  {Fernandez-Galiana}, {Ferrante}, {Ferreira}, {Ferreira}, {Ferrini},
  {Fidecaro}, {Fiori}, {Fiorucci}, {Fishbach}, {Fisher}, {Fishner},
  {Fitz-Axen}, {Flaminio}, {Fletcher}, {Flynn}, {Fong}, {Font}, {Forsyth},
  {Fournier}, {Frasca}, {Frasconi}, {Frei}, {Freise}, {Frey}, {Frey},
  {Fritschel}, {Frolov}, {Fulda}, {Fyffe}, {Gabbard}, {Gadre}, {Gaebel},
  {Gair}, {Gammaitoni}, {Ganija}, {Gaonkar}, {Garcia},
  {Garc{\'\i}a-Quir{\'o}s}, {Garufi}, {Gateley}, {Gaudio}, {Gaur}, {Gayathri},
  {Gemme}, {Genin}, {Gennai}, {George}, {George}, {Gergely}, {Germain},
  {Ghonge}, {Ghosh}, {Ghosh}, {Ghosh}, {Giacomazzo}, {Giaime}, {Giardina},
  {Giazotto}, {Gill}, {Giordano}, {Glover}, {Godwin}, {Goetz}, {Goetz},
  {Goncharov}, {Gonz{\'a}lez}, {Gonzalez Castro}, {Gopakumar}, {Gorodetsky},
  {Gossan}, {Gosselin}, {Gouaty}, {Grado}, {Graef}, {Granata}, {Grant}, {Gras},
  {Grassia}, {Gray}, {Gray}, {Greco}, {Green}, {Green}, {Gretarsson}, {Groot},
  {Grote}, {Grunewald}, {Gruning}, {Guidi}, {Gulati}, {Guo}, {Gupta}, {Gupta},
  {Gustafson}, {Gustafson}, {Haegel}, {Halim}, {Hall}, {Hall}, {Hamilton},
  {Hammond}, {Haney}, {Hanke}, {Hanks}, {Hanna}, {Hannam}, {Hannuksela},
  {Hanson}, {Hardwick}, {Haris}, {Harms}, {Harry}, {Harry}, {Haster},
  {Haughian}, {Hayes}, {Healy}, {Heidmann}, {Heintze}, {Heitmann}, {Hello},
  {Hemming}, {Hendry}, {Heng}, {Hennig}, {Heptonstall}, {Hernandez Vivanco},
  {Heurs}, {Hild}, {Hinderer}, {Hoak}, {Hochheim}, {Hofman}, {Holgado},
  {Holland}, {Holt}, {Holz}, {Hopkins}, {Horst}, {Hough}, {Howell}, {Hoy},
  {Hreibi}, {Huang}, {Huerta}, {Huet}, {Hughey}, {Hulko}, {Husa}, {Huttner},
  {Huynh-Dinh}, {Idzkowski}, {Iess}, {Ingram}, {Inta}, {Intini}, {Irwin},
  {Isa}, {Isac}, {Isi}, {Iyer}, {Izumi}, {Jacqmin}, {Jadhav}, {Jani},
  {Janthalur}, {Jaranowski}, {Jenkins}, {Jiang}, {Johnson}, {Johnson-McDaniel},
  {Jones}, {Jones}, {Jones}, {Jonker}, {Ju}, {Junker}, {Kalaghatgi},
  {Kalogera}, {Kamai}, {Kandhasamy}, {Kang}, {Kanner}, {Kapadia}, {Karki},
  {Karvinen}, {Kashyap}, {Kasprzack}, {Katsanevas}, {Katsavounidis}, {Katzman},
  {Kaufer}, {Kawabe}, {Keerthana}, {K{\'e}f{\'e}lian}, {Keitel}, {Kennedy},
  {Key}, {Khalili}, {Khan}, {Khan}, {Khan}, {Khan}, {Khazanov}, {Khursheed},
  {Kijbunchoo}, {Kim}, {Kim}, {Kim}, {Kim}, {Kim}, {Kim}, {Kimball}, {King},
  {King}, {Kinley-Hanlon}, {Kirchhoff}, {Kissel}, {Kleybolte}, {Klika},
  {Klimenko}, {Knowles}, {Koch}, {Koehlenbeck}, {Koekoek}, {Koley},
  {Kondrashov}, {Kontos}, {Koper}, {Korobko}, {Korth}, {Kowalska}, {Kozak},
  {Kringel}, {Krishnendu}, {Kr{\'o}lak}, {Kuehn}, {Kumar}, {Kumar}, {Kumar},
  {Kumar}, {Kuo}, {Kutynia}, {Kwang}, {Lackey}, {Lai}, {Lam}, {Landry}, {Lane},
  {Lang}, {Lange}, {Lantz}, {Lanza}, {Lartaux-Vollard}, {Lasky}, {Laxen},
  {Lazzarini}, {Lazzaro}, {Leaci}, {Leavey}, {Lecoeuche}, {Lee}, {Lee}, {Lee},
  {Lee}, {Lee}, {Lee}, {Lehmann}, {Lenon}, {Leroy}, {Letendre}, {Levin}, {Li},
  {Li}, {Li}, {Li}, {Lin}, {Linde}, {Linker}, {Littenberg}, {Liu}, {Liu}, {Lo},
  {Lockerbie}, {London}, {Longo}, {Lorenzini}, {Loriette}, {Lormand},
  {Losurdo}, {Lough}, {Lousto}, {Lovelace}, {Lower}, {L{\"u}ck}, {Lumaca},
  {Lundgren}, {Lynch}, {Ma}, {Macas}, {Macfoy}, {MacInnis}, {Macleod},
  {Macquet}, {Maga{\~n}a-Sandoval}, {Maga{\~n}a Zertuche}, {Magee}, {Majorana},
  {Maksimovic}, {Malik}, {Man}, {Mandic}, {Mangano}, {Mansell}, {Manske},
  {Mantovani}, {Marchesoni}, {Marion}, {M{\'a}rka}, {M{\'a}rka}, {Markakis},
  {Markosyan}, {Markowitz}, {Maros}, {Marquina}, {Marsat}, {Martelli},
  {Martin}, {Martin}, {Martynov}, {Mason}, {Massera}, {Masserot}, {Massinger},
  {Masso-Reid}, {Mastrogiovanni}, {Matas}, {Matichard}, {Matone}, {Mavalvala},
  {Mazumder}, {McCann}, {McCarthy}, {McClelland}, {McCormick}, {McCuller},
  {McGuire}, {McIver}, {McManus}, {McRae}, {McWilliams}, {Meacher}, {Meadors},
  {Mehmet}, {Mehta}, {Meidam}, {Melatos}, {Mendell}, {Mercer}, {Mereni},
  {Merilh}, {Merzougui}, {Meshkov}, {Messenger}, {Messick}, {Metzdorff},
  {Meyers}, {Miao}, {Michel}, {Middleton}, {Mikhailov}, {Milano}, {Miller},
  {Miller}, {Millhouse}, {Mills}, {Milovich-Goff}, {Minazzoli}, {Minenkov},
  {Mishkin}, {Mishra}, {Mistry}, {Mitra}, {Mitrofanov}, {Mitselmakher},
  {Mittleman}, {Mo}, {Moffa}, {Mogushi}, {Mohapatra}, {Montani}, {Moore},
  {Moraru}, {Moreno}, {Morisaki}, {Mours}, {Mow-Lowry}, {Mukherjee},
  {Mukherjee}, {Mukherjee}, {Mukund}, {Mullavey}, {Munch}, {Mu{\~n}iz},
  {Muratore}, {Murray}, {Nagar}, {Nardecchia}, {Naticchioni}, {Nayak},
  {Neilson}, {Nelemans}, {Nelson}, {Nery}, {Neunzert}, {Ng}, {Ng}, {Nguyen},
  {Nichols}, {Nielsen}, {Nissanke}, {Nitz}, {Nocera}, {North}, {Nuttall},
  {Obergaulinger}, {Oberling}, {O'Brien}, {O'Dea}, {Ogin}, {Oh}, {Oh}, {Ohme},
  {Ohta}, {Okada}, {Oliver}, {Oppermann}, {Oram}, {O'Reilly}, {Ormiston},
  {Ortega}, {O'Shaughnessy}, {Ossokine}, {Ottaway}, {Overmier}, {Owen}, {Pace},
  {Pagano}, {Page}, {Pai}, {Pai}, {Palamos}, {Palashov}, {Palomba},
  {Pal-Singh}, {Pan}, {Pang}, {Pang}, {Pankow}, {Pannarale}, {Pant},
  {Paoletti}, {Paoli}, {Papa}, {Parida}, {Parker}, {Pascucci}, {Pasqualetti},
  {Passaquieti}, {Passuello}, {Patil}, {Patricelli}, {Pearlstone}, {Pedersen},
  {Pedraza}, {Pedurand}, {Pele}, {Penn}, {Perego}, {Perez}, {Perreca},
  {Pfeiffer}, {Phelps}, {Phukon}, {Piccinni}, {Pichot}, {Piergiovanni},
  {Pillant}, {Pinard}, {Pirello}, {Pitkin}, {Poggiani}, {Pong}, {Ponrathnam},
  {Popolizio}, {Porter}, {Powell}, {Prajapati}, {Prasad}, {Prasai}, {Prasanna},
  {Pratten}, {Prestegard}, {Privitera}, {Prodi}, {Prokhorov}, {Puncken},
  {Punturo}, {Puppo}, {P{\"u}rrer}, {Qi}, {Quetschke}, {Quinonez}, {Quintero},
  {Quitzow-James}, {Raab}, {Radkins}, {Radulescu}, {Raffai}, {Raja}, {Rajan},
  {Rajbhandari}, {Rakhmanov}, {Ramirez}, {Ramos-Buades}, {Rana}, {Rao},
  {Rapagnani}, {Raymond}, {Razzano}, {Read}, {Regimbau}, {Rei}, {Reid},
  {Reitze}, {Ren}, {Ricci}, {Richardson}, {Richardson}, {Ricker},
  {Riemenschneider}, {Riles}, {Rizzo}, {Robertson}, {Robie}, {Robinet},
  {Rocchi}, {Rolland}, {Rollins}, {Roma}, {Romanelli}, {Romano}, {Romel},
  {Romie}, {Rose}, {Rosi{\'n}ska}, {Rosofsky}, {Ross}, {Rowan}, {R{\"u}diger},
  {Ruggi}, {Rutins}, {Ryan}, {Sachdev}, {Sadecki}, {Sakellariadou}, {Salafia},
  {Salconi}, {Saleem}, {Salemi}, {Samajdar}, {Sammut}, {Sanchez}, {Sanchez},
  {Sanchis-Gual}, {Sandberg}, {Sanders}, {Santiago}, {Sarin}, {Sassolas},
  {Sathyaprakash}, {Saulson}, {Sauter}, {Savage}, {Schale}, {Scheel},
  {Scheuer}, {Schmidt}, {Schnabel}, {Schofield}, {Sch{\"o}nbeck}, {Schreiber},
  {Schulte}, {Schutz}, {Schwalbe}, {Scott}, {Scott}, {Seidel}, {Sellers},
  {Sengupta}, {Sennett}, {Sentenac}, {Sequino}, {Sergeev}, {Setyawati},
  {Shaddock}, {Shaffer}, {Shahriar}, {Shaner}, {Shao}, {Sharma}, {Shawhan},
  {Shen}, {Shink}, {Shoemaker}, {Shoemaker}, {ShyamSundar}, {Siellez},
  {Sieniawska}, {Sigg}, {Silva}, {Singer}, {Singh}, {Singhal}, {Sintes},
  {Sitmukhambetov}, {Skliris}, {Slagmolen}, {Slaven-Blair}, {Smith}, {Smith},
  {Somala}, {Son}, {Sorazu}, {Sorrentino}, {Souradeep}, {Sowell}, {Spencer},
  {Srivastava}, {Srivastava}, {Staats}, {Stachie}, {Standke}, {Steer},
  {Steinke}, {Steinlechner}, {Steinlechner}, {Steinmeyer}, {Stevenson},
  {Stocks}, {Stone}, {Stops}, {Strain}, {Stratta}, {Strigin}, {Strunk},
  {Sturani}, {Stuver}, {Sudhir}, {Summerscales}, {Sun}, {Sunil}, {Suresh},
  {Sutton}, {Swinkels}, {Szczepa{\'n}czyk}, {Tacca}, {Tait}, {Talbot},
  {Talukder}, {Tanner}, {T{\'a}pai}, {Taracchini}, {Tasson}, {Taylor}, {Thies},
  {Thomas}, {Thomas}, {Thondapu}, {Thorne}, {Thrane}, {Tiwari}, {Tiwari},
  {Tiwari}, {Toland}, {Tonelli}, {Tornasi}, {Torres-Forn{\'e}}, {Torrie},
  {T{\"o}yr{\"a}}, {Travasso}, {Traylor}, {Tringali}, {Trovato}, {Trozzo},
  {Trudeau}, {Tsang}, {Tse}, {Tso}, {Tsukada}, {Tsuna}, {Tuyenbayev}, {Ueno},
  {Ugolini}, {Unnikrishnan}, {Urban}, {Usman}, {Vahlbruch}, {Vajente},
  {Valdes}, {van Bakel}, {van Beuzekom}, {van den Brand}, {Van Den Broeck},
  {Vander-Hyde}, {van Heijningen}, {van der Schaaf}, {van Veggel}, {Vardaro},
  {Varma}, {Vass}, {Vas{\'u}th}, {Vecchio}, {Vedovato}, {Veitch}, {Veitch},
  {Venkateswara}, {Venugopalan}, {Verkindt}, {Vetrano}, {Vicer{\'e}}, {Viets},
  {Vine}, {Vinet}, {Vitale}, {Vo}, {Vocca}, {Vorvick}, {Vyatchanin}, {Wade},
  {Wade}, {Wade}, {Walet}, {Walker}, {Wallace}, {Walsh}, {Wang}, {Wang},
  {Wang}, {Wang}, {Wang}, {Ward}, {Warden}, {Warner}, {Was}, {Watchi},
  {Weaver}, {Wei}, {Weinert}, {Weinstein}, {Weiss}, {Wellmann}, {Wen},
  {Wessel}, {We{\ss}els}, {Westhouse}, {Wette}, {Whelan}, {White}, {Whiting},
  {Whittle}, {Wilken}, {Williams}, {Williamson}, {Willis}, {Willke}, {Wimmer},
  {Winkler}, {Wipf}, {Wittel}, {Woan}, {Woehler}, {Wofford}, {Worden},
  {Wright}, {Wu}, {Wysocki}, {Xiao}, {Yamamoto}, {Yancey}, {Yang}, {Yap},
  {Yazback}, {Yeeles}, {Yu}, {Yu}, {Yuen}, {Yvert}, {Zadro{\.Z}ny}, {Zanolin},
  {Zappa}, {Zelenova}, {Zendri}, {Zevin}, {Zhang}, {Zhang}, {Zhang}, {Zhao},
  {Zhou}, {Zhou}, {Zhu}, {Zimmerman}, {Zlochower}, {Zucker}, {Zweizig}, {LIGO
  Scientific Collaboration}, \& {Virgo Collaboration}}]{GWTC-1}
---. 2019, Physical Review X, 9, 031040, \dodoi{10.1103/PhysRevX.9.031040}

\bibitem[{{Abbott} {et~al.}(2020{\natexlab{a}}){Abbott}, {Abbott}, {Abraham},
  {Acernese}, {Ackley}, {Adams}, {Adhikari}, {Adya}, {Affeldt}, {Agathos},
  {Agatsuma}, {Aggarwal}, {Aguiar}, {Aich}, {Aiello}, {Ain}, {Ajith}, {Akcay},
  {Allen}, {Allocca}, {Altin}, {Amato}, {Anand}, {Ananyeva}, {Anderson},
  {Anderson}, {Angelova}, {Ansoldi}, {Antier}, {Appert}, {Arai}, {Araya},
  {Areeda}, {Ar{\`e}ne}, {Arnaud}, {Aronson}, {Arun}, {Asali}, {Ascenzi},
  {Ashton}, {Aston}, {Astone}, {Aubin}, {Aufmuth}, {AultONeal}, {Austin},
  {Avendano}, {Babak}, {Bacon}, {Badaracco}, {Bader}, {Bae}, {Baer}, {Baird},
  {Baldaccini}, {Ballardin}, {Ballmer}, {Bals}, {Balsamo}, {Baltus},
  {Banagiri}, {Bankar}, {Bankar}, {Barayoga}, {Barbieri}, {Barish}, {Barker},
  {Barkett}, {Barneo}, {Barone}, {Barr}, {Barsotti}, {Barsuglia}, {Barta},
  {Bartlett}, {Bartos}, {Bassiri}, {Basti}, {Bawaj}, {Bayley}, {Bazzan},
  {B{\'e}csy}, {Bejger}, {Belahcene}, {Bell}, {Beniwal}, {Benjamin}, {Bentley},
  {Bergamin}, {Berger}, {Bergmann}, {Bernuzzi}, {Berry}, {Bersanetti},
  {Bertolini}, {Betzwieser}, {Bhandare}, {Bhandari}, {Bidler}, {Biggs},
  {Bilenko}, {Billingsley}, {Birney}, {Birnholtz}, {Biscans}, {Bischi},
  {Biscoveanu}, {Bisht}, {Bissenbayeva}, {Bitossi}, {Bizouard}, {Blackburn},
  {Blackman}, {Blair}, {Blair}, {Blair}, {Bobba}, {Bode}, {Boer}, {Boetzel},
  {Bogaert}, {Bondu}, {Bonilla}, {Bonnand}, {Booker}, {Boom}, {Bork}, {Boschi},
  {Bose}, {Bossilkov}, {Bosveld}, {Bouffanais}, {Bozzi}, {Bradaschia}, {Brady},
  {Bramley}, {Branchesi}, {Brau}, {Breschi}, {Briant}, {Briggs}, {Brighenti},
  {Brillet}, {Brinkmann}, {Brockill}, {Brooks}, {Brooks}, {Brown}, {Brunett},
  {Bruno}, {Bruntz}, {Buikema}, {Bulik}, {Bulten}, {Buonanno}, {Buscicchio},
  {Buskulic}, {Byer}, {Cabero}, {Cadonati}, {Cagnoli}, {Cahillane},
  {Calder{\'o}n Bustillo}, {Callaghan}, {Callister}, {Calloni}, {Camp},
  {Canepa}, {Cannon}, {Cao}, {Cao}, {Carapella}, {Carbognani}, {Caride},
  {Carney}, {Carullo}, {Casanueva Diaz}, {Casentini}, {Casta{\~n}eda},
  {Caudill}, {Cavagli{\`a}}, {Cavalier}, {Cavalieri}, {Cella},
  {Cerd{\'a}-Dur{\'a}n}, {Cesarini}, {Chaibi}, {Chakravarti}, {Chan}, {Chan},
  {Chandra}, {Chao}, {Charlton}, {Chase}, {Chassande-Mottin}, {Chatterjee},
  {Chaturvedi}, {Chatziioannou}, {Chen}, {Chen}, {Chen}, {Cheng}, {Cheong},
  {Chia}, {Chiadini}, {Chierici}, {Chincarini}, {Chiummo}, {Cho}, {Cho}, {Cho},
  {Christensen}, {Chu}, {Chua}, {Chung}, {Chung}, {Ciani}, {Ciecielag},
  {Cie{\'s}lar}, {Ciobanu}, {Ciolfi}, {Cipriano}, {Cirone}, {Clara}, {Clark},
  {Clearwater}, {Clesse}, {Cleva}, {Coccia}, {Cohadon}, {Cohen}, {Colleoni},
  {Collette}, {Collins}, {Colpi}, {Constancio}, {Conti}, {Cooper}, {Corban},
  {Corbitt}, {Cordero-Carri{\'o}n}, {Corezzi}, {Corley}, {Cornish}, {Corre},
  {Corsi}, {Cortese}, {Costa}, {Cotesta}, {Coughlin}, {Coughlin}, {Coulon},
  {Countryman}, {Couvares}, {Covas}, {Coward}, {Cowart}, {Coyne}, {Coyne},
  {Creighton}, {Creighton}, {Cripe}, {Croquette}, {Crowder}, {Cudell},
  {Cullen}, {Cumming}, {Cummings}, {Cunningham}, {Cuoco}, {Curylo}, {Canton},
  {D{\'a}lya}, {Dana}, {Daneshgaran-Bajastani}, {D'Angelo}, {Danilishin},
  {D'Antonio}, {Danzmann}, {Darsow-Fromm}, {Dasgupta}, {Datrier}, {Dattilo},
  {Dave}, {Davier}, {Davies}, {Davis}, {Daw}, {DeBra}, {Deenadayalan},
  {Degallaix}, {De Laurentis}, {Del{\'e}glise}, {Delfavero}, {De Lillo}, {Del
  Pozzo}, {DeMarchi}, {D'Emilio}, {Demos}, {Dent}, {De Pietri}, {De Rosa}, {De
  Rossi}, {DeSalvo}, {de Varona}, {Dhurandhar}, {D{\'\i}az}, {Diaz-Ortiz},
  {Dietrich}, {Di Fiore}, {Di Fronzo}, {Di Giorgio}, {Di Giovanni}, {Di
  Giovanni}, {Di Girolamo}, {Di Lieto}, {Ding}, {Di Pace}, {Di Palma}, {Di
  Renzo}, {Divakarla}, {Dmitriev}, {Doctor}, {Donovan}, {Dooley}, {Doravari},
  {Dorrington}, {Downes}, {Drago}, {Driggers}, {Du}, {Ducoin}, {Dupej},
  {Durante}, {D'Urso}, {Dwyer}, {Easter}, {Eddolls}, {Edelman}, {Edo}, {Edy},
  {Effler}, {Ehrens}, {Eichholz}, {Eikenberry}, {Eisenmann}, {Eisenstein},
  {Ejlli}, {Errico}, {Essick}, {Estelles}, {Estevez}, {Etienne}, {Etzel},
  {Evans}, {Evans}, {Ewing}, {Fafone}, {Fairhurst}, {Fan}, {Farinon}, {Farr},
  {Farr}, {Fauchon-Jones}, {Favata}, {Fays}, {Fazio}, {Feicht}, {Fejer},
  {Feng}, {Fenyvesi}, {Ferguson}, {Fernandez-Galiana}, {Ferrante}, {Ferreira},
  {Ferreira}, {Fidecaro}, {Fiori}, {Fiorucci}, {Fishbach}, {Fisher},
  {Fittipaldi}, {Fitz-Axen}, {Fiumara}, {Flaminio}, {Floden}, {Flynn}, {Fong},
  {Font}, {Forsyth}, {Fournier}, {Frasca}, {Frasconi}, {Frei}, {Freise},
  {Frey}, {Frey}, {Fritschel}, {Frolov}, {Fronz{\`e}}, {Fulda}, {Fyffe},
  {Gabbard}, {Gadre}, {Gaebel}, {Gair}, {Galaudage}, {Ganapathy}, {Ganguly},
  {Gaonkar}, {Garc{\'\i}a-Quir{\'o}s}, {Garufi}, {Gateley}, {Gaudio},
  {Gayathri}, {Gemme}, {Genin}, {Gennai}, {George}, {George}, {Gergely},
  {Ghonge}, {Ghosh}, {Ghosh}, {Ghosh}, {Giacomazzo}, {Giaime}, {Giardina},
  {Gibson}, {Gier}, {Gill}, {Glanzer}, {Gniesmer}, {Godwin}, {Goetz}, {Goetz},
  {Gohlke}, {Goncharov}, {Gonz{\'a}lez}, {Gopakumar}, {Gossan}, {Gosselin},
  {Gouaty}, {Grace}, {Grado}, {Granata}, {Grant}, {Gras}, {Grassia}, {Gray},
  {Gray}, {Greco}, {Green}, {Green}, {Gretarsson}, {Griggs}, {Grignani},
  {Grimaldi}, {Grimm}, {Grote}, {Grunewald}, {Gruning}, {Guidi}, {Guimaraes},
  {Guix{\'e}}, {Gulati}, {Guo}, {Gupta}, {Gupta}, {Gupta}, {Gustafson},
  {Gustafson}, {Haegel}, {Halim}, {Hall}, {Hamilton}, {Hammond}, {Haney},
  {Hanke}, {Hanks}, {Hanna}, {Hannam}, {Hannuksela}, {Hansen}, {Hanson},
  {Harder}, {Hardwick}, {Haris}, {Harms}, {Harry}, {Harry}, {Hasskew},
  {Haster}, {Haughian}, {Hayes}, {Healy}, {Heidmann}, {Heintze}, {Heinze},
  {Heitmann}, {Hellman}, {Hello}, {Hemming}, {Hendry}, {Heng}, {Hennes},
  {Hennig}, {Heurs}, {Hild}, {Hinderer}, {Hoback}, {Hochheim}, {Hofgard},
  {Hofman}, {Holgado}, {Holland}, {Holt}, {Holz}, {Hopkins}, {Horst}, {Hough},
  {Howell}, {Hoy}, {Huang}, {H{\"u}bner}, {Huerta}, {Huet}, {Hughey}, {Hui},
  {Husa}, {Huttner}, {Huxford}, {Huynh-Dinh}, {Idzkowski}, {Iess}, {Inchauspe},
  {Ingram}, {Intini}, {Isac}, {Isi}, {Iyer}, {Jacqmin}, {Jadhav}, {Jadhav},
  {James}, {Jani}, {Janthalur}, {Jaranowski}, {Jariwala}, {Jaume}, {Jenkins},
  {Jiang}, {Johns}, {Johnson-McDaniel}, {Jones}, {Jones}, {Jones}, {Jones},
  {Jones}, {Jonker}, {Ju}, {Junker}, {Kalaghatgi}, {Kalogera}, {Kamai},
  {Kandhasamy}, {Kang}, {Kanner}, {Kapadia}, {Karki}, {Kashyap}, {Kasprzack},
  {Kastaun}, {Katsanevas}, {Katsavounidis}, {Katzman}, {Kaufer}, {Kawabe},
  {K{\'e}f{\'e}lian}, {Keitel}, {Keivani}, {Kennedy}, {Key}, {Khadka},
  {Khalili}, {Khan}, {Khan}, {Khan}, {Khazanov}, {Khetan}, {Khursheed},
  {Kijbunchoo}, {Kim}, {Kim}, {Kim}, {Kim}, {Kim}, {Kim}, {Kim}, {Kimball},
  {King}, {Kinley-Hanlon}, {Kirchhoff}, {Kissel}, {Kleybolte}, {Klimenko},
  {Knowles}, {Knyazev}, {Koch}, {Koehlenbeck}, {Koekoek}, {Koley},
  {Kondrashov}, {Kontos}, {Koper}, {Korobko}, {Korth}, {Kovalam}, {Kozak},
  {Kringel}, {Krishnendu}, {Kr{\'o}lak}, {Krupinski}, {Kuehn}, {Kumar},
  {Kumar}, {Kumar}, {Kumar}, {Kumar}, {Kuo}, {Kutynia}, {Lackey}, {Laghi},
  {Lalande}, {Lam}, {Lamberts}, {Landry}, {Lane}, {Lang}, {Lange}, {Lantz},
  {Lanza}, {La Rosa}, {Lartaux-Vollard}, {Lasky}, {Laxen}, {Lazzarini},
  {Lazzaro}, {Leaci}, {Leavey}, {Lecoeuche}, {Lee}, {Lee}, {Lee}, {Lee}, {Lee},
  {Lehmann}, {Leroy}, {Letendre}, {Levin}, {Li}, {Li}, {li}, {Li}, {Li},
  {Linde}, {Linker}, {Linley}, {Littenberg}, {Liu}, {Liu},
  {Llorens-Monteagudo}, {Lo}, {Lockwood}, {London}, {Longo}, {Lorenzini},
  {Loriette}, {Lormand}, {Losurdo}, {Lough}, {Lousto}, {Lovelace}, {L{\"u}ck},
  {Lumaca}, {Lundgren}, {Ma}, {Macas}, {Macfoy}, {MacInnis}, {Macleod},
  {MacMillan}, {Macquet}, {Maga{\~n}a Hernandez}, {Maga{\~n}a-Sandoval},
  {Magee}, {Majorana}, {Maksimovic}, {Malik}, {Man}, {Mandic}, {Mangano},
  {Mansell}, {Manske}, {Mantovani}, {Mapelli}, {Marchesoni}, {Marion},
  {M{\'a}rka}, {M{\'a}rka}, {Markakis}, {Markosyan}, {Markowitz}, {Maros},
  {Marquina}, {Marsat}, {Martelli}, {Martin}, {Martin}, {Martinez}, {Martynov},
  {Masalehdan}, {Mason}, {Massera}, {Masserot}, {Massinger}, {Masso-Reid},
  {Mastrogiovanni}, {Matas}, {Matichard}, {Mavalvala}, {Maynard}, {McCann},
  {McCarthy}, {McClelland}, {McCormick}, {McCuller}, {McGuire}, {McIsaac},
  {McIver}, {McManus}, {McRae}, {McWilliams}, {Meacher}, {Meadors}, {Mehmet},
  {Mehta}, {Mejuto Villa}, {Melatos}, {Mendell}, {Mercer}, {Mereni}, {Merfeld},
  {Merilh}, {Merritt}, {Merzougui}, {Meshkov}, {Messenger}, {Messick},
  {Metzdorff}, {Meyers}, {Meylahn}, {Mhaske}, {Miani}, {Miao}, {Michaloliakos},
  {Michel}, {Middleton}, {Milano}, {Miller}, {Millhouse}, {Mills}, {Milotti},
  {Milovich-Goff}, {Minazzoli}, {Minenkov}, {Mishkin}, {Mishra}, {Mistry},
  {Mitra}, {Mitrofanov}, {Mitselmakher}, {Mittleman}, {Mo}, {Mogushi},
  {Mohapatra}, {Mohite}, {Molina-Ruiz}, {Mondin}, {Montani}, {Moore}, {Moraru},
  {Morawski}, {Moreno}, {Morisaki}, {Mours}, {Mow-Lowry}, {Mozzon},
  {Muciaccia}, {Mukherjee}, {Mukherjee}, {Mukherjee}, {Mukherjee}, {Mukund},
  {Mullavey}, {Munch}, {Mu{\~n}iz}, {Murray}, {Nagar}, {Nardecchia},
  {Naticchioni}, {Nayak}, {Neil}, {Neilson}, {Nelemans}, {Nelson}, {Nery},
  {Neunzert}, {Ng}, {Ng}, {Nguyen}, {Nguyen}, {Nichols}, {Nichols}, {Nissanke},
  {Nitz}, {Nocera}, {Noh}, {North}, {Nothard}, {Nuttall}, {Oberling},
  {O'Brien}, {Oganesyan}, {Ogin}, {Oh}, {Oh}, {Ohme}, {Ohta}, {Okada},
  {Oliver}, {Olivetto}, {Oppermann}, {Oram}, {O'Reilly}, {Ormiston}, {Ortega},
  {O'Shaughnessy}, {Ossokine}, {Osthelder}, {Ottaway}, {Overmier}, {Owen},
  {Pace}, {Pagano}, {Page}, {Pagliaroli}, {Pai}, {Pai}, {Palamos}, {Palashov},
  {Palomba}, {Pan}, {Panda}, {Pang}, {Pankow}, {Pannarale}, {Pant}, {Paoletti},
  {Paoli}, {Parida}, {Parker}, {Pascucci}, {Pasqualetti}, {Passaquieti},
  {Passuello}, {Patricelli}, {Payne}, {Pearlstone}, {Pechsiri}, {Pedersen},
  {Pedraza}, {Pele}, {Penn}, {Perego}, {Perez}, {P{\'e}rigois}, {Perreca},
  {Perri{\`e}s}, {Petermann}, {Pfeiffer}, {Phelps}, {Phukon}, {Piccinni},
  {Pichot}, {Piendibene}, {Piergiovanni}, {Pierro}, {Pillant}, {Pinard},
  {Pinto}, {Piotrzkowski}, {Pirello}, {Pitkin}, {Plastino}, {Poggiani}, {Pong},
  {Ponrathnam}, {Popolizio}, {Porter}, {Powell}, {Prajapati}, {Prasai},
  {Prasanna}, {Pratten}, {Prestegard}, {Principe}, {Prodi}, {Prokhorov},
  {Punturo}, {Puppo}, {P{\"u}rrer}, {Qi}, {Quetschke}, {Quinonez}, {Raab},
  {Raaijmakers}, {Radkins}, {Radulesco}, {Raffai}, {Rafferty}, {Raja}, {Rajan},
  {Rajbhandari}, {Rakhmanov}, {Ramirez}, {Ramos-Buades}, {Rana}, {Rao},
  {Rapagnani}, {Raymond}, {Razzano}, {Read}, {Regimbau}, {Rei}, {Reid},
  {Reitze}, {Rettegno}, {Ricci}, {Richardson}, {Richardson}, {Ricker},
  {Riemenschneider}, {Riles}, {Rizzo}, {Robertson}, {Robinet}, {Rocchi},
  {Rodriguez-Soto}, {Rolland}, {Rollins}, {Roma}, {Romanelli}, {Romano},
  {Romel}, {Romero-Shaw}, {Romie}, {Rose}, {Rose}, {Rose}, {Rosi{\'n}ska},
  {Rosofsky}, {Ross}, {Rowan}, {Rowlinson}, {Roy}, {Roy}, {Roy}, {Ruggi},
  {Rutins}, {Ryan}, {Sachdev}, {Sadecki}, {Sakellariadou}, {Salafia},
  {Salconi}, {Saleem}, {Salemi}, {Samajdar}, {Sanchez}, {Sanchez},
  {Sanchis-Gual}, {Sanders}, {Santiago}, {Santos}, {Sarin}, {Sassolas},
  {Sathyaprakash}, {Sauter}, {Savage}, {Savant}, {Sawant}, {Sayah}, {Schaetzl},
  {Schale}, {Scheel}, {Scheuer}, {Schmidt}, {Schnabel}, {Schofield},
  {Sch{\"o}nbeck}, {Schreiber}, {Schulte}, {Schutz}, {Schwarm}, {Schwartz},
  {Scott}, {Scott}, {Seidel}, {Sellers}, {Sengupta}, {Sennett}, {Sentenac},
  {Sequino}, {Sergeev}, {Setyawati}, {Shaddock}, {Shaffer}, {Sharifi},
  {Shahriar}, {Sharma}, {Sharma}, {Shawhan}, {Shen}, {Shikauchi}, {Shink},
  {Shoemaker}, {Shoemaker}, {Shukla}, {ShyamSundar}, {Siellez}, {Sieniawska},
  {Sigg}, {Singer}, {Singh}, {Singh}, {Singha}, {Singhal}, {Sintes}, {Sipala},
  {Skliris}, {Slagmolen}, {Slaven-Blair}, {Smetana}, {Smith}, {Smith},
  {Somala}, {Son}, {Soni}, {Sorazu}, {Sordini}, {Sorrentino}, {Souradeep},
  {Sowell}, {Spencer}, {Spera}, {Srivastava}, {Srivastava}, {Staats},
  {Stachie}, {Standke}, {Steer}, {Steinke}, {Steinlechner}, {Steinlechner},
  {Steinmeyer}, {Stevenson}, {Stocks}, {Stops}, {Stover}, {Strain}, {Stratta},
  {Strunk}, {Sturani}, {Stuver}, {Sudhagar}, {Sudhir}, {Summerscales}, {Sun},
  {Sunil}, {Sur}, {Suresh}, {Sutton}, {Swinkels}, {Szczepa{\'n}czyk}, {Tacca},
  {Tait}, {Talbot}, {Tanasijczuk}, {Tanner}, {Tao}, {T{\'a}pai}, {Tapia},
  {Tapia San Martin}, {Tasson}, {Taylor}, {Tenorio}, {Terkowski},
  {Thirugnanasambandam}, {Thomas}, {Thomas}, {Thompson}, {Thondapu}, {Thorne},
  {Thrane}, {Tinsman}, {Saravanan}, {Tiwari}, {Tiwari}, {Tiwari}, {Toland},
  {Tonelli}, {Tornasi}, {Torres-Forn{\'e}}, {Torrie}, {Tosta e Melo},
  {T{\"o}yr{\"a}}, {Travasso}, {Traylor}, {Tringali}, {Tripathee}, {Trovato},
  {Trudeau}, {Tsang}, {Tse}, {Tso}, {Tsukada}, {Tsuna}, {Tsutsui}, {Turconi},
  {Ubhi}, {Udall}, {Ueno}, {Ugolini}, {Unnikrishnan}, {Urban}, {Usman},
  {Utina}, {Vahlbruch}, {Vajente}, {Valdes}, {Valentini}, {van Bakel}, {van
  Beuzekom}, {van den Brand}, {Van Den Broeck}, {Vander-Hyde}, {van der
  Schaaf}, {Van Heijningen}, {van Veggel}, {Vardaro}, {Varma}, {Vass},
  {Vas{\'u}th}, {Vecchio}, {Vedovato}, {Veitch}, {Veitch}, {Venkateswara},
  {Venugopalan}, {Verkindt}, {Veske}, {Vetrano}, {Vicer{\'e}}, {Viets},
  {Vinciguerra}, {Vine}, {Vinet}, {Vitale}, {Vivanco}, {Vo}, {Vocca},
  {Vorvick}, {Vyatchanin}, {Wade}, {Wade}, {Wade}, {Walet}, {Walker},
  {Wallace}, {Wallace}, {Walsh}, {Wang}, {Wang}, {Wang}, {Ward}, {Warden},
  {Warner}, {Was}, {Watchi}, {Weaver}, {Wei}, {Weinert}, {Weinstein}, {Weiss},
  {Wellmann}, {Wen}, {We{\ss}els}, {Westhouse}, {Wette}, {Whelan}, {Whiting},
  {Whittle}, {Wilken}, {Williams}, {Willis}, {Willke}, {Winkler}, {Wipf},
  {Wittel}, {Woan}, {Woehler}, {Wofford}, {Wong}, {Wright}, {Wu}, {Wysocki},
  {Xiao}, {Yamamoto}, {Yang}, {Yang}, {Yang}, {Yap}, {Yazback}, {Yeeles}, {Yu},
  {Yu}, {Yuen}, {Zadro{\.Z}ny}, {Zadro{\.Z}ny}, {Zanolin}, {Zelenova},
  {Zendri}, {Zevin}, {Zhang}, {Zhang}, {Zhang}, {Zhao}, {Zhao}, {Zhou}, {Zhou},
  {Zhu}, {Zimmerman}, {Zucker}, {Zweizig}, {LIGO Scientific Collaboration}, \&
  {Virgo Collaboration}}]{GW190521}
{Abbott}, R., {Abbott}, T.~D., {Abraham}, S., {et~al.} 2020{\natexlab{a}},
  \prl, 125, 101102, \dodoi{10.1103/PhysRevLett.125.101102}

\bibitem[{{Abbott} {et~al.}(2020{\natexlab{b}}){Abbott}, {Abbott}, {Abraham},
  {Acernese}, {Ackley}, {Adams}, {Adhikari}, {Adya}, {Affeldt}, {Agathos},
  {Agatsuma}, {Aggarwal}, {Aguiar}, {Aich}, {Aiello}, {Ain}, {Ajith}, {Akcay},
  {Allen}, {Allocca}, {Altin}, {Amato}, {Anand}, {Ananyeva}, {Anderson},
  {Anderson}, {Angelova}, {Ansoldi}, {Antier}, {Appert}, {Arai}, {Araya},
  {Areeda}, {Ar{\`e}ne}, {Arnaud}, {Aronson}, {Arun}, {Asali}, {Ascenzi},
  {Ashton}, {Aston}, {Astone}, {Aubin}, {Aufmuth}, {AultONeal}, {Austin},
  {Avendano}, {Babak}, {Bacon}, {Badaracco}, {Bader}, {Bae}, {Baer}, {Baird},
  {Baldaccini}, {Ballardin}, {Ballmer}, {Bals}, {Balsamo}, {Baltus},
  {Banagiri}, {Bankar}, {Bankar}, {Barayoga}, {Barbieri}, {Barish}, {Barker},
  {Barkett}, {Barneo}, {Barone}, {Barr}, {Barsotti}, {Barsuglia}, {Barta},
  {Bartlett}, {Bartos}, {Bassiri}, {Basti}, {Bawaj}, {Bayley}, {Bazzan},
  {B{\'e}csy}, {Bejger}, {Belahcene}, {Bell}, {Beniwal}, {Benjamin}, {Bentley},
  {Bergamin}, {Berger}, {Bergmann}, {Bernuzzi}, {Berry}, {Bersanetti},
  {Bertolini}, {Betzwieser}, {Bhandare}, {Bhandari}, {Bidler}, {Biggs},
  {Bilenko}, {Billingsley}, {Birney}, {Birnholtz}, {Biscans}, {Bischi},
  {Biscoveanu}, {Bisht}, {Bissenbayeva}, {Bitossi}, {Bizouard}, {Blackburn},
  {Blackman}, {Blair}, {Blair}, {Blair}, {Bobba}, {Bode}, {Boer}, {Boetzel},
  {Bogaert}, {Bondu}, {Bonilla}, {Bonnand}, {Booker}, {Boom}, {Bork}, {Boschi},
  {Bose}, {Bossilkov}, {Bosveld}, {Bouffanais}, {Bozzi}, {Bradaschia}, {Brady},
  {Bramley}, {Branchesi}, {Brau}, {Breschi}, {Briant}, {Briggs}, {Brighenti},
  {Brillet}, {Brinkmann}, {Brockill}, {Brooks}, {Brooks}, {Brown}, {Brunett},
  {Bruno}, {Bruntz}, {Buikema}, {Bulik}, {Bulten}, {Buonanno}, {Buscicchio},
  {Buskulic}, {Byer}, {Cabero}, {Cadonati}, {Cagnoli}, {Cahillane}, {Bustillo},
  {Callaghan}, {Callister}, {Calloni}, {Camp}, {Canepa}, {Cannon}, {Cao},
  {Cao}, {Carapella}, {Carbognani}, {Caride}, {Carney}, {Carullo}, {Diaz},
  {Casentini}, {Casta{\~n}eda}, {Caudill}, {Cavagli{\`a}}, {Cavalier},
  {Cavalieri}, {Cella}, {Cerd{\'a}-Dur{\'a}n}, {Cesarini}, {Chaibi},
  {Chakravarti}, {Chan}, {Chan}, {Chao}, {Charlton}, {Chase},
  {Chassande-Mottin}, {Chatterjee}, {Chaturvedi}, {Chatziioannou}, {Chen},
  {Chen}, {Chen}, {Cheng}, {Cheong}, {Chia}, {Chiadini}, {Chierici},
  {Chincarini}, {Chiummo}, {Cho}, {Cho}, {Cho}, {Christensen}, {Chu}, {Chua},
  {Chung}, {Chung}, {Ciani}, {Ciecielag}, {Cie{\'s}lar}, {Ciobanu}, {Ciolfi},
  {Cipriano}, {Cirone}, {Clara}, {Clark}, {Clearwater}, {Clesse}, {Cleva},
  {Coccia}, {Cohadon}, {Cohen}, {Colleoni}, {Collette}, {Collins}, {Colpi},
  {Constancio}, {Conti}, {Cooper}, {Corban}, {Corbitt}, {Cordero-Carri{\'o}n},
  {Corezzi}, {Corley}, {Cornish}, {Corre}, {Corsi}, {Cortese}, {Costa},
  {Cotesta}, {Coughlin}, {Coughlin}, {Coulon}, {Countryman}, {Couvares},
  {Covas}, {Coward}, {Cowart}, {Coyne}, {Coyne}, {Creighton}, {Creighton},
  {Cripe}, {Croquette}, {Crowder}, {Cudell}, {Cullen}, {Cumming}, {Cummings},
  {Cunningham}, {Cuoco}, {Curylo}, {Canton}, {D{\'a}lya}, {Dana},
  {Daneshgaran-Bajastani}, {D'Angelo}, {Danilishin}, {D'Antonio}, {Danzmann},
  {Darsow-Fromm}, {Dasgupta}, {Datrier}, {Dattilo}, {Dave}, {Davier}, {Davies},
  {Davis}, {Daw}, {DeBra}, {Deenadayalan}, {Degallaix}, {De Laurentis},
  {Del{\'e}glise}, {Delfavero}, {De Lillo}, {Del Pozzo}, {DeMarchi},
  {D'Emilio}, {Demos}, {Dent}, {De Pietri}, {De Rosa}, {De Rossi}, {DeSalvo},
  {de Varona}, {Dhurandhar}, {D{\'\i}az}, {Diaz-Ortiz}, {Dietrich}, {Di Fiore},
  {Di Fronzo}, {Di Giorgio}, {Di Giovanni}, {Di Giovanni}, {Di Girolamo}, {Di
  Lieto}, {Ding}, {Di Pace}, {Di Palma}, {Di Renzo}, {Divakarla}, {Dmitriev},
  {Doctor}, {Donovan}, {Dooley}, {Doravari}, {Dorrington}, {Downes}, {Drago},
  {Driggers}, {Du}, {Ducoin}, {Dupej}, {Durante}, {D'Urso}, {Dwyer}, {Easter},
  {Eddolls}, {Edelman}, {Edo}, {Edy}, {Effler}, {Ehrens}, {Eichholz},
  {Eikenberry}, {Eisenmann}, {Eisenstein}, {Ejlli}, {Errico}, {Essick},
  {Estelles}, {Estevez}, {Etienne}, {Etzel}, {Evans}, {Evans}, {Ewing},
  {Fafone}, {Fairhurst}, {Fan}, {Farinon}, {Farr}, {Farr}, {Fauchon-Jones},
  {Favata}, {Fays}, {Fazio}, {Feicht}, {Fejer}, {Feng}, {Fenyvesi}, {Ferguson},
  {Fernandez-Galiana}, {Ferrante}, {Ferreira}, {Ferreira}, {Fidecaro}, {Fiori},
  {Fiorucci}, {Fishbach}, {Fisher}, {Fittipaldi}, {Fitz-Axen}, {Fiumara},
  {Flaminio}, {Floden}, {Flynn}, {Fong}, {Font}, {Forsyth}, {Fournier},
  {Frasca}, {Frasconi}, {Frei}, {Freise}, {Frey}, {Frey}, {Fritschel},
  {Frolov}, {Fronz{\`e}}, {Fulda}, {Fyffe}, {Gabbard}, {Gadre}, {Gaebel},
  {Gair}, {Galaudage}, {Ganapathy}, {Gaonkar}, {Garc{\'\i}a-Quir{\'o}s},
  {Garufi}, {Gateley}, {Gaudio}, {Gayathri}, {Gemme}, {Genin}, {Gennai},
  {George}, {George}, {Gergely}, {Ghonge}, {Ghosh}, {Ghosh}, {Ghosh},
  {Giacomazzo}, {Giaime}, {Giardina}, {Gibson}, {Gier}, {Gill}, {Glanzer},
  {Gniesmer}, {Godwin}, {Goetz}, {Goetz}, {Gohlke}, {Goncharov},
  {Gonz{\'a}lez}, {Gopakumar}, {Gossan}, {Gosselin}, {Gouaty}, {Grace},
  {Grado}, {Granata}, {Grant}, {Gras}, {Grassia}, {Gray}, {Gray}, {Greco},
  {Green}, {Green}, {Gretarsson}, {Griggs}, {Grignani}, {Grimaldi}, {Grimm},
  {Grote}, {Grunewald}, {Gruning}, {Guidi}, {Guimaraes}, {Guix{\'e}}, {Gulati},
  {Guo}, {Gupta}, {Gupta}, {Gupta}, {Gustafson}, {Gustafson}, {Haegel},
  {Halim}, {Hall}, {Hamilton}, {Hammond}, {Haney}, {Hanke}, {Hanks}, {Hanna},
  {Hannam}, {Hannuksela}, {Hansen}, {Hanson}, {Harder}, {Hardwick}, {Haris},
  {Harms}, {Harry}, {Harry}, {Hasskew}, {Haster}, {Haughian}, {Hayes}, {Healy},
  {Heidmann}, {Heintze}, {Heinze}, {Heitmann}, {Hellman}, {Hello}, {Hemming},
  {Hendry}, {Heng}, {Hennes}, {Hennig}, {Heurs}, {Hild}, {Hinderer}, {Hoback},
  {Hochheim}, {Hofgard}, {Hofman}, {Holgado}, {Holland}, {Holt}, {Holz},
  {Hopkins}, {Horst}, {Hough}, {Howell}, {Hoy}, {Huang}, {H{\"u}bner},
  {Huerta}, {Huet}, {Hughey}, {Hui}, {Husa}, {Huttner}, {Huxford},
  {Huynh-Dinh}, {Idzkowski}, {Iess}, {Inchauspe}, {Ingram}, {Intini}, {Isac},
  {Isi}, {Iyer}, {Jacqmin}, {Jadhav}, {Jadhav}, {James}, {Jani}, {Janthalur},
  {Jaranowski}, {Jariwala}, {Jaume}, {Jenkins}, {Jiang}, {Johns},
  {Johnson-McDaniel}, {Jones}, {Jones}, {Jones}, {Jones}, {Jones}, {Jonker},
  {Ju}, {Junker}, {Kalaghatgi}, {Kalogera}, {Kamai}, {Kandhasamy}, {Kang},
  {Kanner}, {Kapadia}, {Karki}, {Kashyap}, {Kasprzack}, {Kastaun},
  {Katsanevas}, {Katsavounidis}, {Katzman}, {Kaufer}, {Kawabe},
  {K{\'e}f{\'e}lian}, {Keitel}, {Keivani}, {Kennedy}, {Key}, {Khadka},
  {Khalili}, {Khan}, {Khan}, {Khan}, {Khazanov}, {Khetan}, {Khursheed},
  {Kijbunchoo}, {Kim}, {Kim}, {Kim}, {Kim}, {Kim}, {Kim}, {Kim}, {Kimball},
  {King}, {Kinley-Hanlon}, {Kirchhoff}, {Kissel}, {Kleybolte}, {Klimenko},
  {Knowles}, {Knyazev}, {Koch}, {Koehlenbeck}, {Koekoek}, {Koley},
  {Kondrashov}, {Kontos}, {Koper}, {Korobko}, {Korth}, {Kovalam}, {Kozak},
  {Kringel}, {Krishnendu}, {Kr{\'o}lak}, {Krupinski}, {Kuehn}, {Kumar},
  {Kumar}, {Kumar}, {Kumar}, {Kumar}, {Kuo}, {Kutynia}, {Lackey}, {Laghi},
  {Lalande}, {Lam}, {Lamberts}, {Landry}, {Lane}, {Lang}, {Lange}, {Lantz},
  {Lanza}, {La Rosa}, {Lartaux-Vollard}, {Lasky}, {Laxen}, {Lazzarini},
  {Lazzaro}, {Leaci}, {Leavey}, {Lecoeuche}, {Lee}, {Lee}, {Lee}, {Lee}, {Lee},
  {Lehmann}, {Leroy}, {Letendre}, {Levin}, {Li}, {Li}, {li}, {Li}, {Li},
  {Linde}, {Linker}, {Linley}, {Littenberg}, {Liu}, {Liu},
  {Llorens-Monteagudo}, {Lo}, {Lockwood}, {London}, {Longo}, {Lorenzini},
  {Loriette}, {Lormand}, {Losurdo}, {Lough}, {Lousto}, {Lovelace}, {L{\"u}ck},
  {Lumaca}, {Lundgren}, {Ma}, {Macas}, {Macfoy}, {MacInnis}, {Macleod},
  {MacMillan}, {Macquet}, {Hernandez}, {Maga{\~n}a-Sandoval}, {Magee},
  {Majorana}, {Maksimovic}, {Malik}, {Man}, {Mandic}, {Mangano}, {Mansell},
  {Manske}, {Mantovani}, {Mapelli}, {Marchesoni}, {Marion}, {M{\'a}rka},
  {M{\'a}rka}, {Markakis}, {Markosyan}, {Markowitz}, {Maros}, {Marquina},
  {Marsat}, {Martelli}, {Martin}, {Martin}, {Martinez}, {Martynov},
  {Masalehdan}, {Mason}, {Massera}, {Masserot}, {Massinger}, {Masso-Reid},
  {Mastrogiovanni}, {Matas}, {Matichard}, {Mavalvala}, {Maynard}, {McCann},
  {McCarthy}, {McClelland}, {McCormick}, {McCuller}, {McGuire}, {McIsaac},
  {McIver}, {McManus}, {McRae}, {McWilliams}, {Meacher}, {Meadors}, {Mehmet},
  {Mehta}, {Villa}, {Melatos}, {Mendell}, {Mercer}, {Mereni}, {Merfeld},
  {Merilh}, {Merritt}, {Merzougui}, {Meshkov}, {Messenger}, {Messick},
  {Metzdorff}, {Meyers}, {Meylahn}, {Mhaske}, {Miani}, {Miao}, {Michaloliakos},
  {Michel}, {Middleton}, {Milano}, {Miller}, {Millhouse}, {Mills}, {Milotti},
  {Milovich-Goff}, {Minazzoli}, {Minenkov}, {Mishkin}, {Mishra}, {Mistry},
  {Mitra}, {Mitrofanov}, {Mitselmakher}, {Mittleman}, {Mo}, {Mogushi},
  {Mohapatra}, {Mohite}, {Molina-Ruiz}, {Mondin}, {Montani}, {Moore}, {Moraru},
  {Morawski}, {Moreno}, {Morisaki}, {Mours}, {Mow-Lowry}, {Mozzon},
  {Muciaccia}, {Mukherjee}, {Mukherjee}, {Mukherjee}, {Mukherjee}, {Mukund},
  {Mullavey}, {Munch}, {Mu{\~n}iz}, {Murray}, {Nagar}, {Nardecchia},
  {Naticchioni}, {Nayak}, {Neil}, {Neilson}, {Nelemans}, {Nelson}, {Nery},
  {Neunzert}, {Ng}, {Ng}, {Nguyen}, {Nguyen}, {Nichols}, {Nichols}, {Nissanke},
  {Nocera}, {Noh}, {North}, {Nothard}, {Nuttall}, {Oberling}, {O'Brien},
  {Oganesyan}, {Ogin}, {Oh}, {Oh}, {Ohme}, {Ohta}, {Okada}, {Oliver},
  {Olivetto}, {Oppermann}, {Oram}, {O'Reilly}, {Ormiston}, {Ortega},
  {O'Shaughnessy}, {Ossokine}, {Osthelder}, {Ottaway}, {Overmier}, {Owen},
  {Pace}, {Pagano}, {Page}, {Pagliaroli}, {Pai}, {Pai}, {Palamos}, {Palashov},
  {Palomba}, {Pan}, {Panda}, {Pang}, {Pankow}, {Pannarale}, {Pant}, {Paoletti},
  {Paoli}, {Parida}, {Parker}, {Pascucci}, {Pasqualetti}, {Passaquieti},
  {Passuello}, {Patricelli}, {Payne}, {Pearlstone}, {Pechsiri}, {Pedersen},
  {Pedraza}, {Pele}, {Penn}, {Perego}, {Perez}, {P{\'e}rigois}, {Perreca},
  {Perri{\`e}s}, {Petermann}, {Pfeiffer}, {Phelps}, {Phukon}, {Piccinni},
  {Pichot}, {Piendibene}, {Piergiovanni}, {Pierro}, {Pillant}, {Pinard},
  {Pinto}, {Piotrzkowski}, {Pirello}, {Pitkin}, {Plastino}, {Poggiani}, {Pong},
  {Ponrathnam}, {Popolizio}, {Porter}, {Powell}, {Prajapati}, {Prasai},
  {Prasanna}, {Pratten}, {Prestegard}, {Principe}, {Prodi}, {Prokhorov},
  {Punturo}, {Puppo}, {P{\"u}rrer}, {Qi}, {Quetschke}, {Quinonez}, {Raab},
  {Raaijmakers}, {Radkins}, {Radulesco}, {Raffai}, {Rafferty}, {Raja}, {Rajan},
  {Rajbhandari}, {Rakhmanov}, {Ramirez}, {Ramos-Buades}, {Rana}, {Rao},
  {Rapagnani}, {Raymond}, {Razzano}, {Read}, {Regimbau}, {Rei}, {Reid},
  {Reitze}, {Rettegno}, {Ricci}, {Richardson}, {Richardson}, {Ricker},
  {Riemenschneider}, {Riles}, {Rizzo}, {Robertson}, {Robinet}, {Rocchi},
  {Rodriguez-Soto}, {Rolland}, {Rollins}, {Roma}, {Romanelli}, {Romano},
  {Romel}, {Romero-Shaw}, {Romie}, {Rose}, {Rose}, {Rose}, {Rosi{\'n}ska},
  {Rosofsky}, {Ross}, {Rowan}, {Rowlinson}, {Roy}, {Roy}, {Roy}, {Ruggi},
  {Rutins}, {Ryan}, {Sachdev}, {Sadecki}, {Sakellariadou}, {Salafia},
  {Salconi}, {Saleem}, {Samajdar}, {Sanchez}, {Sanchez}, {Sanchis-Gual},
  {Sanders}, {Santiago}, {Santos}, {Sarin}, {Sassolas}, {Sathyaprakash},
  {Sauter}, {Savage}, {Savant}, {Sawant}, {Sayah}, {Schaetzl}, {Schale},
  {Scheel}, {Scheuer}, {Schmidt}, {Schnabel}, {Schofield}, {Sch{\"o}nbeck},
  {Schreiber}, {Schulte}, {Schutz}, {Schwarm}, {Schwartz}, {Scott}, {Scott},
  {Seidel}, {Sellers}, {Sengupta}, {Sennett}, {Sentenac}, {Sequino}, {Sergeev},
  {Setyawati}, {Shaddock}, {Shaffer}, {Shahriar}, {Sharifi}, {Sharma},
  {Sharma}, {Shawhan}, {Shen}, {Shikauchi}, {Shink}, {Shoemaker}, {Shoemaker},
  {Shukla}, {ShyamSundar}, {Siellez}, {Sieniawska}, {Sigg}, {Singer}, {Singh},
  {Singh}, {Singha}, {Singhal}, {Sintes}, {Sipala}, {Skliris}, {Slagmolen},
  {Slaven-Blair}, {Smetana}, {Smith}, {Smith}, {Somala}, {Son}, {Soni},
  {Sorazu}, {Sordini}, {Sorrentino}, {Souradeep}, {Sowell}, {Spencer}, {Spera},
  {Srivastava}, {Srivastava}, {Staats}, {Stachie}, {Standke}, {Steer},
  {Steinke}, {Steinlechner}, {Steinlechner}, {Steinmeyer}, {Stevenson},
  {Stocks}, {Stops}, {Stover}, {Strain}, {Stratta}, {Strunk}, {Sturani},
  {Stuver}, {Sudhagar}, {Sudhir}, {Summerscales}, {Sun}, {Sunil}, {Sur},
  {Suresh}, {Sutton}, {Swinkels}, {Szczepa{\'n}czyk}, {Tacca}, {Tait},
  {Talbot}, {Tanasijczuk}, {Tanner}, {Tao}, {T{\'a}pai}, {Tapia}, {San Martin},
  {Tasson}, {Taylor}, {Tenorio}, {Terkowski}, {Thirugnanasambandam}, {Thomas},
  {Thomas}, {Thompson}, {Thondapu}, {Thorne}, {Thrane}, {Tinsman}, {Saravanan},
  {Tiwari}, {Tiwari}, {Tiwari}, {Toland}, {Tonelli}, {Tornasi},
  {Torres-Forn{\'e}}, {Torrie}, {Tosta e Melo}, {T{\"o}yr{\"a}}, {Trail},
  {Travasso}, {Traylor}, {Tringali}, {Tripathee}, {Trovato}, {Trudeau},
  {Tsang}, {Tse}, {Tso}, {Tsukada}, {Tsuna}, {Tsutsui}, {Turconi}, {Ubhi},
  {Udall}, {Ueno}, {Ugolini}, {Unnikrishnan}, {Urban}, {Usman}, {Utina},
  {Vahlbruch}, {Vajente}, {Valdes}, {Valentini}, {van Bakel}, {van Beuzekom},
  {van den Brand}, {Van Den Broeck}, {Vander-Hyde}, {van der Schaaf}, {Van
  Heijningen}, {van Veggel}, {Vardaro}, {Varma}, {Vass}, {Vas{\'u}th},
  {Vecchio}, {Vedovato}, {Veitch}, {Veitch}, {Venkateswara}, {Venugopalan},
  {Verkindt}, {Veske}, {Vetrano}, {Vicer{\'e}}, {Viets}, {Vinciguerra}, {Vine},
  {Vinet}, {Vitale}, {Vivanco}, {Vo}, {Vocca}, {Vorvick}, {Vyatchanin}, {Wade},
  {Wade}, {Wade}, {Walet}, {Walker}, {Wallace}, {Wallace}, {Walsh}, {Wang},
  {Wang}, {Wang}, {Ward}, {Warden}, {Warner}, {Was}, {Watchi}, {Weaver}, {Wei},
  {Weinert}, {Weinstein}, {Weiss}, {Wellmann}, {Wen}, {We{\ss}els},
  {Westhouse}, {Wette}, {Whelan}, {Whiting}, {Whittle}, {Wilken}, {Williams},
  {Willis}, {Willke}, {Winkler}, {Wipf}, {Wittel}, {Woan}, {Woehler},
  {Wofford}, {Wong}, {Wright}, {Wu}, {Wysocki}, {Xiao}, {Yamamoto}, {Yang},
  {Yang}, {Yang}, {Yap}, {Yazback}, {Yeeles}, {Yu}, {Yu}, {Yuen},
  {Zadro{\.z}ny}, {Zadro{\.z}ny}, {Zanolin}, {Zelenova}, {Zendri}, {Zevin},
  {Zhang}, {Zhang}, {Zhang}, {Zhao}, {Zhao}, {Zhou}, {Zhou}, {Zhu},
  {Zimmerman}, {Zlochower}, {Zucker}, {Zweizig}, {LIGO Scientific
  Collaboration}, \& {Virgo Collaboration}}]{GW190521-int}
---. 2020{\natexlab{b}}, \apjl, 900, L13, \dodoi{10.3847/2041-8213/aba493}

\bibitem[{Abbott {et~al.}(2020{\natexlab{a}})}]{LIGOScientific:2020stg}
Abbott, R., {et~al.} 2020{\natexlab{a}}, Phys. Rev. D, 102, 043015,
  \dodoi{10.1103/PhysRevD.102.043015}

\bibitem[{Abbott {et~al.}(2020{\natexlab{b}})}]{Abbott:2020khf}
---. 2020{\natexlab{b}}, Astrophys. J. Lett., 896, L44,
  \dodoi{10.3847/2041-8213/ab960f}

\bibitem[{{Amaro-Seoane} {et~al.}(2017)}]{LISA17}
{Amaro-Seoane}, P., {et~al.} 2017, ArXiv e-prints.
\newblock \doarXiv{1702.00786}

\bibitem[{{Arca-Sedda} {et~al.}(2021){Arca-Sedda}, {Rizzuto}, {Naab},
  {Ostriker}, {Giersz}, \& {Spurzem}}]{ArcaSedda2021ApJ...920..128A}
{Arca-Sedda}, M., {Rizzuto}, F.~P., {Naab}, T., {et~al.} 2021, \apj, 920, 128,
  \dodoi{10.3847/1538-4357/ac1419}

\bibitem[{{Belczynski}(2020)}]{Belczy2020ApJ...905L..15B}
{Belczynski}, K. 2020, \apjl, 905, L15, \dodoi{10.3847/2041-8213/abcbf1}

\bibitem[{{Branchesi} {et~al.}(2023){Branchesi}, {Maggiore}, {Alonso},
  {Badger}, {Banerjee}, {Beirnaert}, {Belgacem}, {Bhagwat}, {Boileau},
  {Borhanian}, {Brown}, {Leong Chan}, {Cusin}, {Danilishin}, {Degallaix}, {De
  Luca}, {Dhani}, {Dietrich}, {Dupletsa}, {Foffa}, {Franciolini}, {Freise},
  {Gemme}, {Goncharov}, {Ghosh}, {Gulminelli}, {Gupta}, {Kumar Gupta}, {Harms},
  {Hazra}, {Hild}, {Hinderer}, {Siong Heng}, {Iacovelli}, {Janquart},
  {Janssens}, {Jenkins}, {Kalaghatgi}, {Koroveshi}, {Li}, {Li}, {Loffredo},
  {Maggio}, {Mancarella}, {Mapelli}, {Martinovic}, {Maselli}, {Meyers},
  {Miller}, {Mondal}, {Muttoni}, {Narola}, {Oertel}, {Oganesyan}, {Pacilio},
  {Palomba}, {Pani}, {Pasqualetti}, {Perego}, {P{\'e}rigois}, {Pieroni},
  {Piccinni}, {Puecher}, {Puppo}, {Ricciardone}, {Riotto}, {Ronchini},
  {Sakellariadou}, {Samajdar}, {Santoliquido}, {Sathyaprakash}, {Steinlechner},
  {Steinlechner}, {Utina}, {Van Den Broeck}, \& {Zhang}}]{2023JCAP...07..068B}
{Branchesi}, M., {Maggiore}, M., {Alonso}, D., {et~al.} 2023, \jcap, 2023, 068,
  \dodoi{10.1088/1475-7516/2023/07/068}

\bibitem[{{Bromm}(2013)}]{Bromm2013}
{Bromm}, V. 2013, Reports on Progress in Physics, 76, 112901,
  \dodoi{10.1088/0034-4885/76/11/112901}

\bibitem[{{Carr} {et~al.}(2021){Carr}, {Kohri}, {Sendouda}, \&
  {Yokoyama}}]{Carr2021RPPh...84k6902C}
{Carr}, B., {Kohri}, K., {Sendouda}, Y., \& {Yokoyama}, J. 2021, Reports on
  Progress in Physics, 84, 116902, \dodoi{10.1088/1361-6633/ac1e31}

\bibitem[{{Chon} \& {Hosokawa}(2019)}]{Chon2019}
{Chon}, S., \& {Hosokawa}, T. 2019, \mnras, 488, 2658,
  \dodoi{10.1093/mnras/stz1824}

\bibitem[{{Costa} {et~al.}(2022){Costa}, {Ballone}, {Mapelli}, \&
  {Bressan}}]{Costa2022}
{Costa}, G., {Ballone}, A., {Mapelli}, M., \& {Bressan}, A. 2022, \mnras, 516,
  1072, \dodoi{10.1093/mnras/stac2222}

\bibitem[{{Costa} {et~al.}(2021){Costa}, {Bressan}, {Mapelli}, {Marigo},
  {Iorio}, \& {Spera}}]{Costa2021}
{Costa}, G., {Bressan}, A., {Mapelli}, M., {et~al.} 2021, \mnras, 501, 4514,
  \dodoi{10.1093/mnras/staa3916}

\bibitem[{{Di Carlo} {et~al.}(2020){Di Carlo}, {Mapelli}, {Giacobbo}, {Spera},
  {Bouffanais}, {Rastello}, {Santoliquido}, {Pasquato}, {Ballone}, {Trani},
  {Torniamenti}, \& {Haardt}}]{DiCarlo2020mnras}
{Di Carlo}, U.~N., {Mapelli}, M., {Giacobbo}, N., {et~al.} 2020, \mnras, 498,
  495, \dodoi{10.1093/mnras/staa2286}

\bibitem[{Evans {et~al.}(2021)}]{Evans:2021gyd}
Evans, M., {et~al.} 2021.
\newblock \doarXiv{2109.09882}

\bibitem[{Fairhurst(2009)}]{Fairhurst:2009tc}
Fairhurst, S. 2009, New J. Phys., 11, 123006,
  \dodoi{10.1088/1367-2630/11/12/123006}

\bibitem[{Fairhurst(2011)}]{Fairhurst:2010is}
---. 2011, Class.Quant.Grav., 28, 105021,
  \dodoi{10.1088/0264-9381/28/10/105021}

\bibitem[{Fairhurst {et~al.}(2023)Fairhurst, Hoy, Green, Mills, \&
  Usman}]{Fairhurst:2023idl}
Fairhurst, S., Hoy, C., Green, R., Mills, C., \& Usman, S.~A. 2023.
\newblock \doarXiv{2304.03731}

\bibitem[{{Farmer} {et~al.}(2020){Farmer}, {Renzo}, {de Mink}, {Fishbach}, \&
  {Justham}}]{Farmer2020ApJ...902L..36F}
{Farmer}, R., {Renzo}, M., {de Mink}, S.~E., {Fishbach}, M., \& {Justham}, S.
  2020, \apjl, 902, L36, \dodoi{10.3847/2041-8213/abbadd}

\bibitem[{{Fishbach} \& {Holz}(2020)}]{Fishbach2020ApJ...904L..26F}
{Fishbach}, M., \& {Holz}, D.~E. 2020, \apjl, 904, L26,
  \dodoi{10.3847/2041-8213/abc827}

\bibitem[{{Fragione} \& {Loeb}(2023)}]{Fragione2023}
{Fragione}, G., \& {Loeb}, A. 2023, \apj, 944, 81,
  \dodoi{10.3847/1538-4357/acb34e}

\bibitem[{{Gerosa} \& {Fishbach}(2021)}]{Gerosa2021NatAs...5..749G}
{Gerosa}, D., \& {Fishbach}, M. 2021, Nature Astronomy, 5, 749,
  \dodoi{10.1038/s41550-021-01398-w}

\bibitem[{{Graziani} {et~al.}(2020){Graziani}, {Schneider}, {Marassi}, {Del
  Pozzo}, {Mapelli}, \& {Giacobbo}}]{Graziani2020}
{Graziani}, L., {Schneider}, R., {Marassi}, S., {et~al.} 2020, \mnras, 495,
  L81, \dodoi{10.1093/mnrasl/slaa063}

\bibitem[{Green {et~al.}(2020)Green, Hoy, Fairhurst, Hannam, Pannarale, \&
  Thomas}]{Green:2020ptm}
Green, R., Hoy, C., Fairhurst, S., {et~al.} 2020.
\newblock \doarXiv{2010.04131}

\bibitem[{Hannam {et~al.}(2013)Hannam, Brown, Fairhurst, Fryer, \&
  Harry}]{Hannam:2013uu}
Hannam, M., Brown, D.~A., Fairhurst, S., Fryer, C.~L., \& Harry, I.~W. 2013,
  \apj, 766, L14, \dodoi{10.1088/2041-8205/766/1/L14}

\bibitem[{{Inayoshi} {et~al.}(2020){Inayoshi}, {Visbal}, \&
  {Haiman}}]{Inayoshi2020}
{Inayoshi}, K., {Visbal}, E., \& {Haiman}, Z. 2020, \araa, 58, 27,
  \dodoi{10.1146/annurev-astro-120419-014455}

\bibitem[{{Kalogera} {et~al.}(2019){Kalogera}, {Berry}, {Colpi}, {Fairhurst},
  {Justham}, {Mandel}, {Mangiagli}, {Mapelli}, {Mills}, {Sathyaprakash},
  {Schneider}, {Tauris}, \& {Valiante}}]{Kalogera2019-BH}
{Kalogera}, V., {Berry}, C. P.~L., {Colpi}, M., {et~al.} 2019, \baas, 51, 242.
\newblock \doarXiv{1903.09220}

\bibitem[{{Klessen} \& {Glover}(2023)}]{Klessen2023}
{Klessen}, R.~S., \& {Glover}, S. C.~O. 2023, arXiv e-prints, arXiv:2303.12500,
  \dodoi{10.48550/arXiv.2303.12500}

\bibitem[{{Liu} \& {Bromm}(2020)}]{Bromm2020ApJ...903L..40L}
{Liu}, B., \& {Bromm}, V. 2020, \apjl, 903, L40,
  \dodoi{10.3847/2041-8213/abc552}

\bibitem[{{Madau} {et~al.}(2014){Madau}, {Haardt}, \& {Dotti}}]{Madau14}
{Madau}, P., {Haardt}, F., \& {Dotti}, M. 2014, \apj, 784, L38,
  \dodoi{10.1088/2041-8205/784/2/L38}

\bibitem[{{Maggiore} {et~al.}(2020){Maggiore}, {Van Den Broeck}, {Bartolo},
  {Belgacem}, {Bertacca}, {Bizouard}, {Branchesi}, {Clesse}, {Foffa},
  {Garc{\'\i}a-Bellido}, {Grimm}, {Harms}, {Hinderer}, {Matarrese}, {Palomba},
  {Peloso}, {Ricciardone}, \& {Sakellariadou}}]{Maggiore2020JCAP...03..050M}
{Maggiore}, M., {Van Den Broeck}, C., {Bartolo}, N., {et~al.} 2020, \jcap,
  2020, 050, \dodoi{10.1088/1475-7516/2020/03/050}

\bibitem[{{Maiolino} {et~al.}(2023){Maiolino}, {Scholtz}, {Curtis-Lake},
  {Carniani}, {Baker}, {de Graaff}, {Tacchella}, {{\"U}bler}, {D'Eugenio},
  {Witstok}, {Curti}, {Arribas}, {Bunker}, {Charlot}, {Chevallard},
  {Eisenstein}, {Egami}, {Ji}, {Jones}, {Lyu}, {Rawle}, {Robertson},
  {Rujopakarn}, {Perna}, {Sun}, {Venturi}, {Williams}, \&
  {Willott}}]{Maiolino2023}
{Maiolino}, R., {Scholtz}, J., {Curtis-Lake}, E., {et~al.} 2023, arXiv
  e-prints, arXiv:2308.01230, \dodoi{10.48550/arXiv.2308.01230}

\bibitem[{{Mancarella} {et~al.}(2023){Mancarella}, {Iacovelli}, \&
  {Gerosa}}]{Mancarella2023PhRvD.107j1302M}
{Mancarella}, M., {Iacovelli}, F., \& {Gerosa}, D. 2023, \prd, 107, L101302,
  \dodoi{10.1103/PhysRevD.107.L101302}

\bibitem[{{Mandel} \& {Farmer}(2022)}]{Mandel2022PhR...955....1M}
{Mandel}, I., \& {Farmer}, A. 2022, \physrep, 955, 1,
  \dodoi{10.1016/j.physrep.2022.01.003}

\bibitem[{{Mapelli}(2021)}]{Mapelli21-review}
{Mapelli}, M. 2021, in Handbook of Gravitational Wave Astronomy, 16,
  \dodoi{10.1007/978-981-15-4702-7_16-1}

\bibitem[{{Mapelli} {et~al.}(2022{\natexlab{a}}){Mapelli}, {Bouffanais},
  {Santoliquido}, {Arca Sedda}, \& {Artale}}]{Mapelli2022MNRAS.511.5797M}
{Mapelli}, M., {Bouffanais}, Y., {Santoliquido}, F., {Arca Sedda}, M., \&
  {Artale}, M.~C. 2022{\natexlab{a}}, \mnras, 511, 5797,
  \dodoi{10.1093/mnras/stac422}

\bibitem[{{Mapelli} {et~al.}(2022{\natexlab{b}}){Mapelli}, {Bouffanais},
  {Santoliquido}, {Arca Sedda}, \& {Artale}}]{Mapelli2022}
---. 2022{\natexlab{b}}, \mnras, 511, 5797, \dodoi{10.1093/mnras/stac422}

\bibitem[{{Mapelli} {et~al.}(2021){Mapelli}, {Dall'Amico}, {Bouffanais},
  {Giacobbo}, {Arca Sedda}, {Artale}, {Ballone}, {Di Carlo}, {Iorio},
  {Santoliquido}, \& {Torniamenti}}]{Mapelli2021}
{Mapelli}, M., {Dall'Amico}, M., {Bouffanais}, Y., {et~al.} 2021, \mnras, 505,
  339, \dodoi{10.1093/mnras/stab1334}

\bibitem[{{Marchant} \& {Moriya}(2020)}]{Marchant-2020A&A...640L..18M}
{Marchant}, P., \& {Moriya}, T.~J. 2020, \aap, 640, L18,
  \dodoi{10.1051/0004-6361/202038902}

\bibitem[{Mills \& Fairhurst(2021)}]{Mills:2020thr}
Mills, C., \& Fairhurst, S. 2021, Phys. Rev. D, 103, 024042,
  \dodoi{10.1103/PhysRevD.103.024042}

\bibitem[{Mills {et~al.}(2018)Mills, Tiwari, \&
  Fairhurst}]{mills_localization_2018}
Mills, C., Tiwari, V., \& Fairhurst, S. 2018, Physical Review D, 97, 104064,
  \dodoi{10.1103/PhysRevD.97.104064}

\bibitem[{{Ng} {et~al.}(2022){Ng}, {Chen}, {Goncharov}, {Dupletsa},
  {Borhanian}, {Branchesi}, {Harms}, {Maggiore}, {Sathyaprakash}, \&
  {Vitale}}]{Ng2022}
{Ng}, K. K.~Y., {Chen}, S., {Goncharov}, B., {et~al.} 2022, \apjl, 931, L12,
  \dodoi{10.3847/2041-8213/ac6bea}

\bibitem[{{Ng} {et~al.}(2023){Ng}, {Goncharov}, {Chen}, {Borhanian},
  {Dupletsa}, {Franciolini}, {Branchesi}, {Harms}, {Maggiore}, {Riotto},
  {Sathyaprakash}, \& {Vitale}}]{Ng2023PhRvD.107b4041N}
{Ng}, K. K.~Y., {Goncharov}, B., {Chen}, S., {et~al.} 2023, \prd, 107, 024041,
  \dodoi{10.1103/PhysRevD.107.024041}

\bibitem[{{Nitz} \& {Capano}(2021)}]{Nitz2021}
{Nitz}, A.~H., \& {Capano}, C.~D. 2021, \apjl, 907, L9,
  \dodoi{10.3847/2041-8213/abccc5}

\bibitem[{{Pezzulli} {et~al.}(2016){Pezzulli}, {Valiante}, \&
  {Schneider}}]{Pezzulli16}
{Pezzulli}, E., {Valiante}, R., \& {Schneider}, R. 2016, \mnras, 458, 3047,
  \dodoi{10.1093/mnras/stw505}

\bibitem[{Pratten {et~al.}(2020)}]{Pratten:2020ceb}
Pratten, G., {et~al.} 2020.
\newblock \doarXiv{2004.06503}

\bibitem[{{Punturo} {et~al.}(2014){Punturo}, {L{\"u}ck}, \&
  {Beker}}]{Punturo2014}
{Punturo}, M., {L{\"u}ck}, H., \& {Beker}, M. 2014, in Astrophysics and Space
  Science Library, Vol. 404, Advanced Interferometers and the Search for
  Gravitational Waves, ed. M.~{Bassan}, 333,
  \dodoi{10.1007/978-3-319-03792-9_13}

\bibitem[{{Renzo} {et~al.}(2020){Renzo}, {Cantiello}, {Metzger}, \&
  {Jiang}}]{Renzo2020ApJ...904L..13R}
{Renzo}, M., {Cantiello}, M., {Metzger}, B.~D., \& {Jiang}, Y.~F. 2020, \apjl,
  904, L13, \dodoi{10.3847/2041-8213/abc6a6}

\bibitem[{{Ricarte} \& {Natarajan}(2018)}]{Ricarte2018}
{Ricarte}, A., \& {Natarajan}, P. 2018, \mnras, 481, 3278,
  \dodoi{10.1093/mnras/sty2448}

\bibitem[{{Roupas} \& {Kazanas}(2019)}]{Kazanas2019A&A...632L...8R}
{Roupas}, Z., \& {Kazanas}, D. 2019, \aap, 632, L8,
  \dodoi{10.1051/0004-6361/201937002}

\bibitem[{{Safarzadeh} \& {Haiman}(2020)}]{Zoltan2020ApJ...903L..21S}
{Safarzadeh}, M., \& {Haiman}, Z. 2020, \apjl, 903, L21,
  \dodoi{10.3847/2041-8213/abc253}

\bibitem[{{Schneider} {et~al.}(2023){Schneider}, {Valiante}, {Trinca},
  {Graziani}, {Volonteri}, \& {Maiolino}}]{schneider2023}
{Schneider}, R., {Valiante}, R., {Trinca}, A., {et~al.} 2023, arXiv e-prints,
  arXiv:2305.12504, \dodoi{10.48550/arXiv.2305.12504}

\bibitem[{Singer {et~al.}(2014)Singer, Price, Farr, Urban, Pankow,
  {et~al.}}]{Singer:2014qca}
Singer, L.~P., Price, L.~R., Farr, B., {et~al.} 2014.
\newblock \doarXiv{1404.5623}

\bibitem[{{Spera} \& {Mapelli}(2017)}]{Spera17}
{Spera}, M., \& {Mapelli}, M. 2017, ArXiv e-prints.
\newblock \doarXiv{1706.06109}

\bibitem[{{Stacy} \& {Bromm}(2013)}]{Stacy2013}
{Stacy}, A., \& {Bromm}, V. 2013, \mnras, 433, 1094,
  \dodoi{10.1093/mnras/stt789}

\bibitem[{{Sugimura} {et~al.}(2023){Sugimura}, {Matsumoto}, {Hosokawa},
  {Hirano}, \& {Omukai}}]{Sugimura2023}
{Sugimura}, K., {Matsumoto}, T., {Hosokawa}, T., {Hirano}, S., \& {Omukai}, K.
  2023, arXiv e-prints, arXiv:2307.15108, \dodoi{10.48550/arXiv.2307.15108}

\bibitem[{{Tanikawa} {et~al.}(2021){Tanikawa}, {Kinugawa}, {Yoshida},
  {Hijikawa}, \& {Umeda}}]{Tanikawa2021MNRAS.505.2170T}
{Tanikawa}, A., {Kinugawa}, T., {Yoshida}, T., {Hijikawa}, K., \& {Umeda}, H.
  2021, \mnras, 505, 2170, \dodoi{10.1093/mnras/stab1421}

\bibitem[{{The LIGO Scientific Collaboration} {et~al.}(2021{\natexlab{a}}){The
  LIGO Scientific Collaboration}, {the Virgo Collaboration}, {Abbott},
  {Abbott}, {Acernese}, {Ackley}, {Adams}, {Adhikari}, {Adhikari}, {Adya},
  {Affeldt}, {Agarwal}, {Agathos}, {Agatsuma}, {Aggarwal}, {Aguiar}, {Aiello},
  {Ain}, {Ajith}, {Albanesi}, {Allocca}, {Altin}, {Amato}, {Anand}, {Anand},
  {Ananyeva}, {Anderson}, {Anderson}, {Andrade}, {Andres}, {Andri{\'c}},
  {Angelova}, {Ansoldi}, {Antelis}, {Antier}, {Appert}, {Arai}, {Araya},
  {Areeda}, {Ar{\`e}ne}, {Arnaud}, {Aronson}, {Arun}, {Asali}, {Ashton},
  {Assiduo}, {Aston}, {Astone}, {Aubin}, {Austin}, {Babak}, {Badaracco},
  {Bader}, {Badger}, {Bae}, {Baer}, {Bagnasco}, {Bai}, {Baird}, {Ball},
  {Ballardin}, {Ballmer}, {Balsamo}, {Baltus}, {Banagiri}, {Bankar},
  {Barayoga}, {Barbieri}, {Barish}, {Barker}, {Barneo}, {Barone}, {Barr},
  {Barsotti}, {Barsuglia}, {Barta}, {Bartlett}, {Barton}, {Bartos}, {Bassiri},
  {Basti}, {Bawaj}, {Bayley}, {Baylor}, {Bazzan}, {B{\'e}csy}, {Bedakihale},
  {Bejger}, {Belahcene}, {Benedetto}, {Beniwal}, {Bennett}, {Bentley},
  {BenYaala}, {Bergamin}, {Berger}, {Bernuzzi}, {Berry}, {Bersanetti},
  {Bertolini}, {Betzwieser}, {Beveridge}, {Bhandare}, {Bhardwaj},
  {Bhattacharjee}, {Bhaumik}, {Bilenko}, {Billingsley}, {Bini}, {Birney},
  {Birnholtz}, {Biscans}, {Bischi}, {Biscoveanu}, {Bisht}, {Biswas}, {Bitossi},
  {Bizouard}, {Blackburn}, {Blair}, {Blair}, {Blair}, {Bobba}, {Bode}, {Boer},
  {Bogaert}, {Boldrini}, {Bonavena}, {Bondu}, {Bonilla}, {Bonnand}, {Booker},
  {Boom}, {Bork}, {Boschi}, {Bose}, {Bose}, {Bossilkov}, {Boudart},
  {Bouffanais}, {Bozzi}, {Bradaschia}, {Brady}, {Bramley}, {Branch},
  {Branchesi}, {Brau}, {Breschi}, {Briant}, {Briggs}, {Brillet}, {Brinkmann},
  {Brockill}, {Brooks}, {Brooks}, {Brown}, {Brunett}, {Bruno}, {Bruntz},
  {Bryant}, {Bulik}, {Bulten}, {Buonanno}, {Buscicchio}, {Buskulic}, {Buy},
  {Byer}, {Cadonati}, {Cagnoli}, {Cahillane}, {Calder{\'o}n Bustillo},
  {Callaghan}, {Callister}, {Calloni}, {Cameron}, {Camp}, {Canepa},
  {Canevarolo}, {Cannavacciuolo}, {Cannon}, {Cao}, {Capote}, {Carapella},
  {Carbognani}, {Carlin}, {Carney}, {Carpinelli}, {Carrillo}, {Carullo},
  {Carver}, {Casanueva Diaz}, {Casentini}, {Castaldi}, {Caudill},
  {Cavagli{\`a}}, {Cavalier}, {Cavalieri}, {Ceasar}, {Cella},
  {Cerd{\'a}-Dur{\'a}n}, {Cesarini}, {Chaibi}, {Chakravarti}, {Chalathadka
  Subrahmanya}, {Champion}, {Chan}, {Chan}, {Chan}, {Chan}, {Chandra},
  {Chanial}, {Chao}, {Charlton}, {Chase}, {Chassande-Mottin}, {Chatterjee},
  {Chatterjee}, {Chatterjee}, {Chattopadhyay}, {Chaturvedi}, {Chaty},
  {Chatziioannou}, {Chen}, {Chen}, {Chen}, {Chen}, {Chen}, {Cheng}, {Cheong},
  {Cheung}, {Chia}, {Chiadini}, {Chiarini}, {Chierici}, {Chincarini},
  {Chiofalo}, {Chiummo}, {Cho}, {Cho}, {Choudhary}, {Choudhary}, {Christensen},
  {Chu}, {Chua}, {Chung}, {Ciani}, {Ciecielag}, {Cie{\'s}lar}, {Cifaldi},
  {Ciobanu}, {Ciolfi}, {Cipriano}, {Cirone}, {Clara}, {Clark}, {Clark},
  {Clarke}, {Clearwater}, {Clesse}, {Cleva}, {Coccia}, {Codazzo}, {Cohadon},
  {Cohen}, {Cohen}, {Colleoni}, {Collette}, {Colombo}, {Colpi}, {Compton},
  {Constancio}, {Conti}, {Cooper}, {Corban}, {Corbitt}, {Cordero-Carri{\'o}n},
  {Corezzi}, {Corley}, {Cornish}, {Corre}, {Corsi}, {Cortese}, {Costa},
  {Cotesta}, {Coughlin}, {Coulon}, {Countryman}, {Cousins}, {Couvares},
  {Coward}, {Cowart}, {Coyne}, {Coyne}, {Creighton}, {Creighton}, {Criswell},
  {Croquette}, {Crowder}, {Cudell}, {Cullen}, {Cumming}, {Cummings},
  {Cunningham}, {Cuoco}, {Cury{\l}o}, {Dabadie}, {Dal Canton}, {Dall'Osso},
  {D{\'a}lya}, {Dana}, {DaneshgaranBajastani}, {D'Angelo}, {Danila},
  {Danilishin}, {D'Antonio}, {Danzmann}, {Darsow-Fromm}, {Dasgupta}, {Datrier},
  {Datta}, {Dattilo}, {Dave}, {Davier}, {Davies}, {Davis}, {Davis}, {Daw},
  {Dean}, {DeBra}, {Deenadayalan}, {Degallaix}, {De Laurentis},
  {Del{\'e}glise}, {Del Favero}, {De Lillo}, {De Lillo}, {Del Pozzo},
  {DeMarchi}, {De Matteis}, {D'Emilio}, {Demos}, {Dent}, {Depasse}, {De
  Pietri}, {De Rosa}, {De Rossi}, {DeSalvo}, {De Simone}, {Dhurandhar},
  {D{\'\i}az}, {Diaz-Ortiz}, {Didio}, {Dietrich}, {Di Fiore}, {Di Fronzo}, {Di
  Giorgio}, {Di Giovanni}, {Di Giovanni}, {Di Girolamo}, {Di Lieto}, {Ding},
  {Di Pace}, {Di Palma}, {Di Renzo}, {Divakarla}, {Divyajyoti}, {Dmitriev},
  {Doctor}, {D'Onofrio}, {Donovan}, {Dooley}, {Doravari}, {Dorrington},
  {Drago}, {Driggers}, {Drori}, {Ducoin}, {Dupej}, {Durante}, {D'Urso},
  {Duverne}, {Dwyer}, {Eassa}, {Easter}, {Ebersold}, {Eckhardt}, {Eddolls},
  {Edelman}, {Edo}, {Edy}, {Effler}, {Eichholz}, {Eikenberry}, {Eisenmann},
  {Eisenstein}, {Ejlli}, {Engelby}, {Errico}, {Essick}, {Estell{\'e}s},
  {Estevez}, {Etienne}, {Etzel}, {Evans}, {Evans}, {Ewing}, {Fafone}, {Fair},
  {Fairhurst}, {Fanning}, {Farah}, {Farinon}, {Farr}, {Farr}, {Farrow},
  {Fauchon-Jones}, {Favaro}, {Favata}, {Fays}, {Fazio}, {Feicht}, {Fejer},
  {Fenyvesi}, {Ferguson}, {Fernandez-Galiana}, {Ferrante}, {Ferreira},
  {Fidecaro}, {Figura}, {Fiori}, {Fishbach}, {Fisher}, {Fittipaldi}, {Fiumara},
  {Flaminio}, {Floden}, {Fong}, {Font}, {Fornal}, {Forsyth}, {Franke},
  {Frasca}, {Frasconi}, {Frederick}, {Freed}, {Frei}, {Freise}, {Frey},
  {Fritschel}, {Frolov}, {Fronz{\'e}}, {Fulda}, {Fyffe}, {Gabbard}, {Gabella},
  {Gadre}, {Gair}, {Gais}, {Galaudage}, {Gamba}, {Ganapathy}, {Ganguly},
  {Gaonkar}, {Garaventa}, {Garc{\'\i}a}, {Garc{\'\i}a-N{\'u}{\~n}ez},
  {Garc{\'\i}a-Quir{\'o}s}, {Garufi}, {Gateley}, {Gaudio}, {Gayathri}, {Gemme},
  {Gennai}, {George}, {George}, {Gerberding}, {Gergely}, {Gewecke}, {Ghonge},
  {Ghosh}, {Ghosh}, {Ghosh}, {Ghosh}, {Giacomazzo}, {Giacoppo}, {Giaime},
  {Giardina}, {Gibson}, {Gier}, {Giesler}, {Giri}, {Gissi}, {Glanzer},
  {Gleckl}, {Godwin}, {Goetz}, {Goetz}, {Gohlke}, {Goncharov}, {Gonz{\'a}lez},
  {Gopakumar}, {Gosselin}, {Gouaty}, {Gould}, {Grace}, {Grado}, {Granata},
  {Granata}, {Grant}, {Gras}, {Grassia}, {Gray}, {Gray}, {Greco}, {Green},
  {Green}, {Gretarsson}, {Gretarsson}, {Griffith}, {Griffiths}, {Griggs},
  {Grignani}, {Grimaldi}, {Grimm}, {Grote}, {Grunewald}, {Gruning}, {Guerra},
  {Guidi}, {Guimaraes}, {Guix{\'e}}, {Gulati}, {Guo}, {Guo}, {Gupta}, {Gupta},
  {Gupta}, {Gustafson}, {Gustafson}, {Guzman}, {Haegel}, {Halim}, {Hall},
  {Hamilton}, {Hammond}, {Haney}, {Hanks}, {Hanna}, {Hannam}, {Hannuksela},
  {Hansen}, {Hansen}, {Hanson}, {Harder}, {Hardwick}, {Haris}, {Harms},
  {Harry}, {Harry}, {Hartwig}, {Haskell}, {Hasskew}, {Haster}, {Haughian},
  {Hayes}, {Healy}, {Heidmann}, {Heidt}, {Heintze}, {Heinze}, {Heinzel},
  {Heitmann}, {Hellman}, {Hello}, {Helmling-Cornell}, {Hemming}, {Hendry},
  {Heng}, {Hennes}, {Hennig}, {Hennig}, {Hernandez}, {Hernandez Vivanco},
  {Heurs}, {Hild}, {Hill}, {Hines}, {Hochheim}, {Hofman}, {Hohmann}, {Holcomb},
  {Holland}, {Holley-Bockelmann}, {Hollows}, {Holmes}, {Holt}, {Holz},
  {Hopkins}, {Hough}, {Hourihane}, {Howell}, {Hoy}, {Hoyland}, {Hreibi}, {Hsu},
  {Huang}, {H{\"u}bner}, {Huddart}, {Hughey}, {Hui}, {Husa}, {Huttner},
  {Huxford}, {Huynh-Dinh}, {Idzkowski}, {Iess}, {Ingram}, {Isi}, {Isleif},
  {Iyer}, {JaberianHamedan}, {Jacqmin}, {Jadhav}, {Jadhav}, {James}, {Jan},
  {Jani}, {Janquart}, {Janssens}, {Janthalur}, {Jaranowski}, {Jariwala},
  {Jaume}, {Jenkins}, {Jenner}, {Jeunon}, {Jia}, {Johns}, {Johnson-McDaniel},
  {Jones}, {Jones}, {Jones}, {Jones}, {Jones}, {Jonker}, {Ju}, {Junker},
  {Juste}, {Kalaghatgi}, {Kalogera}, {Kamai}, {Kandhasamy}, {Kang}, {Kanner},
  {Kao}, {Kapadia}, {Kapasi}, {Karat}, {Karathanasis}, {Karki}, {Kashyap},
  {Kasprzack}, {Kastaun}, {Katsanevas}, {Katsavounidis}, {Katzman}, {Kaur},
  {Kawabe}, {K{\'e}f{\'e}lian}, {Keitel}, {Key}, {Khadka}, {Khalili}, {Khan},
  {Khazanov}, {Khetan}, {Khursheed}, {Kijbunchoo}, {Kim}, {Kim}, {Kim}, {Kim},
  {Kim}, {Kimball}, {Kinley-Hanlon}, {Kirchhoff}, {Kissel}, {Kleybolte},
  {Klimenko}, {Knee}, {Knowles}, {Knyazev}, {Koch}, {Koekoek}, {Koley},
  {Kolitsidou}, {Kolstein}, {Komori}, {Kondrashov}, {Kontos}, {Koper},
  {Korobko}, {Kovalam}, {Kozak}, {Kringel}, {Krishnendu}, {Kr{\'o}lak},
  {Kuehn}, {Kuei}, {Kuijer}, {Kumar}, {Kumar}, {Kumar}, {Kumar}, {Kuns},
  {Kuwahara}, {Lagabbe}, {Laghi}, {Lalande}, {Lam}, {Lamberts}, {Landry},
  {Lane}, {Lang}, {Lange}, {Lantz}, {La Rosa}, {Lartaux-Vollard}, {Lasky},
  {Laxen}, {Lazzarini}, {Lazzaro}, {Leaci}, {Leavey}, {Lecoeuche}, {Lee},
  {Lee}, {Lee}, {Lee}, {Lehmann}, {Lema{\^\i}tre}, {Leroy}, {Letendre},
  {Levesque}, {Levin}, {Leviton}, {Leyde}, {Li}, {Li}, {Li}, {Li}, {Li},
  {Linde}, {Linker}, {Linley}, {Littenberg}, {Liu}, {Liu}, {Liu}, {Llamas},
  {Llorens-Monteagudo}, {Lo}, {Lockwood}, {London}, {Longo}, {Lopez}, {Lopez
  Portilla}, {Lorenzini}, {Loriette}, {Lormand}, {Losurdo}, {Lott}, {Lough},
  {Lousto}, {Lovelace}, {Lucaccioni}, {L{\"u}ck}, {Lumaca}, {Lundgren},
  {Lynam}, {Macas}, {MacInnis}, {Macleod}, {MacMillan}, {Macquet}, {Maga{\~n}a
  Hernandez}, {Magazz{\`u}}, {Magee}, {Maggiore}, {Magnozzi}, {Mahesh},
  {Majorana}, {Makarem}, {Maksimovic}, {Maliakal}, {Malik}, {Man}, {Mandic},
  {Mangano}, {Mango}, {Mansell}, {Manske}, {Mantovani}, {Mapelli},
  {Marchesoni}, {Marion}, {Mark}, {M{\'a}rka}, {M{\'a}rka}, {Markakis},
  {Markosyan}, {Markowitz}, {Maros}, {Marquina}, {Marsat}, {Martelli},
  {Martin}, {Martin}, {Martinez}, {Martinez}, {Martinez}, {Martinovic},
  {Martynov}, {Marx}, {Masalehdan}, {Mason}, {Massera}, {Masserot},
  {Massinger}, {Masso-Reid}, {Mastrogiovanni}, {Matas}, {Mateu-Lucena},
  {Matichard}, {Matiushechkina}, {Mavalvala}, {McCann}, {McCarthy},
  {McClelland}, {McClincy}, {McCormick}, {McCuller}, {McGhee}, {McGuire},
  {McIsaac}, {McIver}, {McRae}, {McWilliams}, {Meacher}, {Mehmet}, {Mehta},
  {Meijer}, {Melatos}, {Melchor}, {Mendell}, {Menendez-Vazquez}, {Menoni},
  {Mercer}, {Mereni}, {Merfeld}, {Merilh}, {Merritt}, {Merzougui}, {Meshkov},
  {Messenger}, {Messick}, {Meyers}, {Meylahn}, {Mhaske}, {Miani}, {Miao},
  {Michaloliakos}, {Michel}, {Middleton}, {Milano}, {Miller}, {Miller},
  {Miller}, {Millhouse}, {Mills}, {Milotti}, {Minazzoli}, {Minenkov}, {Mir},
  {Miravet-Ten{\'e}s}, {Mishra}, {Mishra}, {Mistry}, {Mitra}, {Mitrofanov},
  {Mitselmakher}, {Mittleman}, {Mo}, {Moguel}, {Mogushi}, {Mohapatra},
  {Mohite}, {Molina}, {Molina-Ruiz}, {Mondin}, {Montani}, {Moore}, {Moraru},
  {Morawski}, {More}, {Moreno}, {Moreno}, {Morisaki}, {Mours}, {Mow-Lowry},
  {Mozzon}, {Muciaccia}, {Mukherjee}, {Mukherjee}, {Mukherjee}, {Mukherjee},
  {Mukherjee}, {Mukund}, {Mullavey}, {Munch}, {Mu{\~n}iz}, {Murray},
  {Musenich}, {Muusse}, {Nadji}, {Nagar}, {Napolano}, {Nardecchia},
  {Naticchioni}, {Nayak}, {Nayak}, {Neil}, {Neilson}, {Nelemans}, {Nelson},
  {Nery}, {Neubauer}, {Neunzert}, {Ng}, {Ng}, {Nguyen}, {Nguyen}, {Nguyen},
  {Nichols}, {Nissanke}, {Nitoglia}, {Nocera}, {Norman}, {North}, {Nuttall},
  {Oberling}, {O'Brien}, {O'Dell}, {Oelker}, {Oganesyan}, {Oh}, {Oh}, {Ohme},
  {Ohta}, {Okada}, {Olivetto}, {Oram}, {O'Reilly}, {Ormiston}, {Ormsby},
  {Ortega}, {O'Shaughnessy}, {O'Shea}, {Ossokine}, {Osthelder}, {Ottaway},
  {Overmier}, {Pace}, {Pagano}, {Page}, {Pagliaroli}, {Pai}, {Pai}, {Palamos},
  {Palashov}, {Palomba}, {Pan}, {Panda}, {Pang}, {Pankow}, {Pannarale}, {Pant},
  {Panther}, {Paoletti}, {Paoli}, {Paolone}, {Park}, {Parker}, {Pascucci},
  {Pasqualetti}, {Passaquieti}, {Passuello}, {Patel}, {Pathak}, {Patricelli},
  {Patron}, {Patrone}, {Paul}, {Payne}, {Pedraza}, {Pegoraro}, {Pele}, {Penn},
  {Perego}, {Pereira}, {Pereira}, {Perez}, {P{\'e}rigois}, {Perkins},
  {Perreca}, {Perri{\`e}s}, {Petermann}, {Petterson}, {Pfeiffer}, {Pham},
  {Phukon}, {Piccinni}, {Pichot}, {Piendibene}, {Piergiovanni}, {Pierini},
  {Pierro}, {Pillant}, {Pillas}, {Pilo}, {Pinard}, {Pinto}, {Pinto},
  {Piotrzkowski}, {Pirello}, {Pitkin}, {Placidi}, {Planas}, {Plastino},
  {Pluchar}, {Poggiani}, {Polini}, {Pong}, {Ponrathnam}, {Popolizio}, {Porter},
  {Poulton}, {Powell}, {Pracchia}, {Pradier}, {Prajapati}, {Prasai},
  {Prasanna}, {Pratten}, {Principe}, {Prodi}, {Prokhorov}, {Prosposito},
  {Prudenzi}, {Puecher}, {Punturo}, {Puosi}, {Puppo}, {P{\"u}rrer}, {Qi},
  {Quetschke}, {Quitzow-James}, {Raab}, {Raaijmakers}, {Radkins}, {Radulesco},
  {Raffai}, {Rail}, {Raja}, {Rajan}, {Ramirez}, {Ramirez}, {Ramos-Buades},
  {Rana}, {Rapagnani}, {Rapol}, {Ray}, {Raymond}, {Raza}, {Razzano}, {Read},
  {Rees}, {Regimbau}, {Rei}, {Reid}, {Reid}, {Reitze}, {Relton}, {Renzini},
  {Rettegno}, {Reza}, {Rezac}, {Ricci}, {Richards}, {Richardson}, {Richardson},
  {Riemenschneider}, {Riles}, {Rinaldi}, {Rink}, {Rizzo}, {Robertson}, {Robie},
  {Robinet}, {Rocchi}, {Rodriguez}, {Rolland}, {Rollins}, {Romanelli},
  {Romano}, {Romel}, {Romero-Rodr{\'\i}guez}, {Romero-Shaw}, {Romie},
  {Ronchini}, {Rosa}, {Rose}, {Rosell}, {Rosi{\'n}ska}, {Ross}, {Rowan},
  {Rowlinson}, {Roy}, {Roy}, {Roy}, {Rozza}, {Ruggi}, {Ruiz-Rocha}, {Ryan},
  {Sachdev}, {Sadecki}, {Sadiq}, {Sakellariadou}, {Salafia}, {Salconi},
  {Saleem}, {Salemi}, {Samajdar}, {Sanchez}, {Sanchez}, {Sanchez},
  {Sanchis-Gual}, {Sanders}, {Sanuy}, {Saravanan}, {Sarin}, {Sassolas},
  {Satari}, {Sauter}, {Savage}, {Sawant}, {Sawant}, {Sayah}, {Schaetzl},
  {Scheel}, {Scheuer}, {Schiworski}, {Schmidt}, {Schmidt}, {Schnabel},
  {Schneewind}, {Schofield}, {Sch{\"o}nbeck}, {Schulte}, {Schutz}, {Schwartz},
  {Scott}, {Scott}, {Seglar-Arroyo}, {Sellers}, {Sengupta}, {Sentenac}, {Seo},
  {Sequino}, {Sergeev}, {Setyawati}, {Shaffer}, {Shahriar}, {Shams}, {Sharma},
  {Sharma}, {Shawhan}, {Shcheblanov}, {Shikauchi}, {Shoemaker}, {Shoemaker},
  {ShyamSundar}, {Sieniawska}, {Sigg}, {Singer}, {Singh}, {Singh}, {Singha},
  {Sintes}, {Sipala}, {Skliris}, {Slagmolen}, {Slaven-Blair}, {Smetana},
  {Smith}, {Smith}, {Soldateschi}, {Somala}, {Son}, {Soni}, {Soni}, {Sordini},
  {Sorrentino}, {Sorrentino}, {Soulard}, {Souradeep}, {Sowell}, {Spagnuolo},
  {Spencer}, {Spera}, {Srinivasan}, {Srivastava}, {Srivastava}, {Staats},
  {Stachie}, {Steer}, {Steinhoff}, {Steinlechner}, {Steinlechner}, {Stevenson},
  {Stops}, {Stover}, {Strain}, {Strang}, {Stratta}, {Strunk}, {Sturani},
  {Stuver}, {Sudhagar}, {Sudhir}, {Suh}, {Summerscales}, {Sun}, {Sun}, {Sunil},
  {Sur}, {Suresh}, {Sutton}, {Swinkels}, {Szczepa{\'n}czyk}, {Szewczyk},
  {Tacca}, {Tait}, {Talbot}, {Talbot}, {Tanasijczuk}, {Tanner}, {Tao}, {Tao},
  {Tapia San Mart{\'\i}n}, {Taranto}, {Tasson}, {Tenorio}, {Terhune},
  {Terkowski}, {Thirugnanasambandam}, {Thomas}, {Thomas}, {Thomas}, {Thompson},
  {Thondapu}, {Thorne}, {Thrane}, {Tiwari}, {Tiwari}, {Tiwari}, {Toivonen},
  {Toland}, {Tolley}, {Tonelli}, {Torres-Forn{\'e}}, {Torrie}, {Tosta e Melo},
  {T{\"o}yr{\"a}}, {Trapananti}, {Travasso}, {Traylor}, {Trevor}, {Tringali},
  {Tripathee}, {Troiano}, {Trovato}, {Trozzo}, {Trudeau}, {Tsai}, {Tsai},
  {Tsang}, {Tse}, {Tso}, {Tsukada}, {Tsuna}, {Tsutsui}, {Turbang}, {Turconi},
  {Ubhi}, {Udall}, {Ueno}, {Unnikrishnan}, {Urban}, {Utina}, {Vahlbruch},
  {Vajente}, {Vajpeyi}, {Valdes}, {Valentini}, {Valsan}, {van Bakel}, {van
  Beuzekom}, {van den Brand}, {Van Den Broeck}, {Vander-Hyde}, {van der
  Schaaf}, {van Heijningen}, {Vanosky}, {van Remortel}, {Vardaro}, {Vargas},
  {Varma}, {Vas{\'u}th}, {Vecchio}, {Vedovato}, {Veitch}, {Veitch},
  {Venneberg}, {Venugopalan}, {Verkindt}, {Verma}, {Verma}, {Veske}, {Vetrano},
  {Vicer{\'e}}, {Vidyant}, {Viets}, {Vijaykumar}, {Villa-Ortega}, {Vinet},
  {Virtuoso}, {Vitale}, {Vo}, {Vocca}, {von Reis}, {von Wrangel}, {Vorvick},
  {Vyatchanin}, {Wade}, {Wade}, {Wagner}, {Walet}, {Walker}, {Wallace},
  {Wallace}, {Walsh}, {Wang}, {Wang}, {Ward}, {Warner}, {Was}, {Washington},
  {Watchi}, {Weaver}, {Webster}, {Weinert}, {Weinstein}, {Weiss}, {Weller},
  {Weller}, {Wellmann}, {Wen}, {We{\ss}els}, {Wette}, {Whelan}, {White},
  {Whiting}, {Whittle}, {Wilken}, {Williams}, {Williams}, {Williamson},
  {Willis}, {Willke}, {Wilson}, {Winkler}, {Wipf}, {Wlodarczyk}, {Woan},
  {Woehler}, {Wofford}, {Wong}, {Wu}, {Wysocki}, {Xiao}, {Yamamoto}, {Yang},
  {Yang}, {Yang}, {Yang}, {Yap}, {Yeeles}, {Yelikar}, {Ying}, {Yoo}, {Yu},
  {Yu}, {Zadro{\.z}ny}, {Zanolin}, {Zelenova}, {Zendri}, {Zevin}, {Zhang},
  {Zhang}, {Zhang}, {Zhang}, {Zhao}, {Zhao}, {Zhao}, {Zhou}, {Zhou}, {Zhu},
  {Zimmerman}, {Zlochower}, {Zucker}, \& {Zweizig}}]{GWTC-2.1}
{The LIGO Scientific Collaboration}, {the Virgo Collaboration}, {Abbott}, R.,
  {et~al.} 2021{\natexlab{a}}, arXiv e-prints, arXiv:2108.01045.
\newblock \doarXiv{2108.01045}

\bibitem[{{The LIGO Scientific Collaboration} {et~al.}(2021{\natexlab{b}}){The
  LIGO Scientific Collaboration}, {the Virgo Collaboration}, {the KAGRA
  Collaboration}, {Abbott}, {Abbott}, {Acernese}, {Ackley}, {Adams},
  {Adhikari}, {Adhikari}, {Adya}, {Affeldt}, {Agarwal}, {Agathos}, {Agatsuma},
  {Aggarwal}, {Aguiar}, {Aiello}, {Ain}, {Ajith}, {Akcay}, {Akutsu},
  {Albanesi}, {Allocca}, {Altin}, {Amato}, {Anand}, {Anand}, {Ananyeva},
  {Anderson}, {Anderson}, {Ando}, {Andrade}, {Andres}, {Andri{\'c}},
  {Angelova}, {Ansoldi}, {Antelis}, {Antier}, {Appert}, {Arai}, {Arai}, {Arai},
  {Araki}, {Araya}, {Araya}, {Areeda}, {Ar{\`e}ne}, {Aritomi}, {Arnaud},
  {Arogeti}, {Aronson}, {Arun}, {Asada}, {Asali}, {Ashton}, {Aso}, {Assiduo},
  {Aston}, {Astone}, {Aubin}, {Austin}, {Babak}, {Badaracco}, {Bader},
  {Badger}, {Bae}, {Bae}, {Baer}, {Bagnasco}, {Bai}, {Baiotti}, {Baird},
  {Bajpai}, {Ball}, {Ballardin}, {Ballmer}, {Balsamo}, {Baltus}, {Banagiri},
  {Bankar}, {Barayoga}, {Barbieri}, {Barish}, {Barker}, {Barneo}, {Barone},
  {Barr}, {Barsotti}, {Barsuglia}, {Barta}, {Bartlett}, {Barton}, {Bartos},
  {Bassiri}, {Basti}, {Bawaj}, {Bayley}, {Baylor}, {Bazzan}, {B{\'e}csy},
  {Bedakihale}, {Bejger}, {Belahcene}, {Benedetto}, {Beniwal}, {Bennett},
  {Bentley}, {BenYaala}, {Bergamin}, {Berger}, {Bernuzzi}, {Berry},
  {Bersanetti}, {Bertolini}, {Betzwieser}, {Beveridge}, {Bhandare}, {Bhardwaj},
  {Bhattacharjee}, {Bhaumik}, {Bilenko}, {Billingsley}, {Bini}, {Birney},
  {Birnholtz}, {Biscans}, {Bischi}, {Biscoveanu}, {Bisht}, {Biswas}, {Bitossi},
  {Bizouard}, {Blackburn}, {Blair}, {Blair}, {Blair}, {Bobba}, {Bode}, {Boer},
  {Bogaert}, {Boldrini}, {Bonavena}, {Bondu}, {Bonilla}, {Bonnand}, {Booker},
  {Boom}, {Bork}, {Boschi}, {Bose}, {Bose}, {Bossilkov}, {Boudart},
  {Bouffanais}, {Bozzi}, {Bradaschia}, {Brady}, {Bramley}, {Branch},
  {Branchesi}, {Brandt}, {Brau}, {Breschi}, {Briant}, {Briggs}, {Brillet},
  {Brinkmann}, {Brockill}, {Brooks}, {Brooks}, {Brown}, {Brunett}, {Bruno},
  {Bruntz}, {Bryant}, {Bulik}, {Bulten}, {Buonanno}, {Buscicchio}, {Buskulic},
  {Buy}, {Byer}, {Cabourn Davies}, {Cadonati}, {Cagnoli}, {Cahillane},
  {Calder{\'o}n Bustillo}, {Callaghan}, {Callister}, {Calloni}, {Cameron},
  {Camp}, {Canepa}, {Canevarolo}, {Cannavacciuolo}, {Cannon}, {Cao}, {Cao},
  {Capocasa}, {Capote}, {Carapella}, {Carbognani}, {Carlin}, {Carney},
  {Carpinelli}, {Carrillo}, {Carullo}, {Carver}, {Casanueva Diaz}, {Casentini},
  {Castaldi}, {Caudill}, {Cavagli{\`a}}, {Cavalier}, {Cavalieri}, {Ceasar},
  {Cella}, {Cerd{\'a}-Dur{\'a}n}, {Cesarini}, {Chaibi}, {Chakravarti},
  {Chalathadka Subrahmanya}, {Champion}, {Chan}, {Chan}, {Chan}, {Chan},
  {Chan}, {Chandra}, {Chanial}, {Chao}, {Chapman-Bird}, {Charlton}, {Chase},
  {Chassande-Mottin}, {Chatterjee}, {Chatterjee}, {Chatterjee}, {Chaturvedi},
  {Chaty}, {Chatziioannou}, {Chen}, {Chen}, {Chen}, {Chen}, {Chen}, {Chen},
  {Chen}, {Chen}, {Cheng}, {Cheong}, {Cheung}, {Chia}, {Chiadini}, {Chiang},
  {Chiarini}, {Chierici}, {Chincarini}, {Chiofalo}, {Chiummo}, {Cho}, {Cho},
  {Choudhary}, {Choudhary}, {Christensen}, {Chu}, {Chu}, {Chu}, {Chua},
  {Chung}, {Ciani}, {Ciecielag}, {Cie{\'s}lar}, {Cifaldi}, {Ciobanu}, {Ciolfi},
  {Cipriano}, {Cirone}, {Clara}, {Clark}, {Clark}, {Clarke}, {Clearwater},
  {Clesse}, {Cleva}, {Coccia}, {Codazzo}, {Cohadon}, {Cohen}, {Cohen},
  {Colleoni}, {Collette}, {Colombo}, {Colpi}, {Compton}, {Constancio}, {Conti},
  {Cooper}, {Corban}, {Corbitt}, {Cordero-Carri{\'o}n}, {Corezzi}, {Corley},
  {Cornish}, {Corre}, {Corsi}, {Cortese}, {Costa}, {Cotesta}, {Coughlin},
  {Coulon}, {Countryman}, {Cousins}, {Couvares}, {Coward}, {Cowart}, {Coyne},
  {Coyne}, {Creighton}, {Creighton}, {Criswell}, {Croquette}, {Crowder},
  {Cudell}, {Cullen}, {Cumming}, {Cummings}, {Cunningham}, {Cuoco},
  {Cury{\l}o}, {Dabadie}, {Dal Canton}, {Dall'Osso}, {D{\'a}lya}, {Dana},
  {DaneshgaranBajastani}, {D'Angelo}, {Danila}, {Danilishin}, {D'Antonio},
  {Danzmann}, {Darsow-Fromm}, {Dasgupta}, {Datrier}, {Datta}, {Dattilo},
  {Dave}, {Davier}, {Davis}, {Davis}, {Daw}, {de Alarc{\'o}n}, {Dean}, {DeBra},
  {Deenadayalan}, {Degallaix}, {De Laurentis}, {Del{\'e}glise}, {Del Favero},
  {De Lillo}, {De Lillo}, {Del Pozzo}, {DeMarchi}, {De Matteis}, {D'Emilio},
  {Demos}, {Dent}, {Depasse}, {De Pietri}, {De Rosa}, {De Rossi}, {DeSalvo},
  {De Simone}, {Dhurandhar}, {D{\'\i}az}, {Diaz-Ortiz}, {Didio}, {Dietrich},
  {Di Fiore}, {Di Fronzo}, {Di Giorgio}, {Di Giovanni}, {Di Giovanni}, {Di
  Girolamo}, {Di Lieto}, {Ding}, {Di Pace}, {Di Palma}, {Di Renzo},
  {Divakarla}, {Dmitriev}, {Doctor}, {D'Onofrio}, {Donovan}, {Dooley},
  {Doravari}, {Dorrington}, {Drago}, {Driggers}, {Drori}, {Ducoin}, {Dupej},
  {Durante}, {D'Urso}, {Duverne}, {Dwyer}, {Eassa}, {Easter}, {Ebersold},
  {Eckhardt}, {Eddolls}, {Edelman}, {Edo}, {Edy}, {Effler}, {Eguchi},
  {Eichholz}, {Eikenberry}, {Eisenmann}, {Eisenstein}, {Ejlli}, {Engelby},
  {Enomoto}, {Errico}, {Essick}, {Estell{\'e}s}, {Estevez}, {Etienne}, {Etzel},
  {Evans}, {Evans}, {Ewing}, {Fafone}, {Fair}, {Fairhurst}, {Farah}, {Farinon},
  {Farr}, {Farr}, {Farrow}, {Fauchon-Jones}, {Favaro}, {Favata}, {Fays},
  {Fazio}, {Feicht}, {Fejer}, {Fenyvesi}, {Ferguson}, {Fernandez-Galiana},
  {Ferrante}, {Ferreira}, {Fidecaro}, {Figura}, {Fiori}, {Fishbach}, {Fisher},
  {Fittipaldi}, {Fiumara}, {Flaminio}, {Floden}, {Fong}, {Font}, {Fornal},
  {Forsyth}, {Franke}, {Frasca}, {Frasconi}, {Frederick}, {Freed}, {Frei},
  {Freise}, {Frey}, {Fritschel}, {Frolov}, {Fronz{\'e}}, {Fujii}, {Fujikawa},
  {Fukunaga}, {Fukushima}, {Fulda}, {Fyffe}, {Gabbard}, {Gabella}, {Gadre},
  {Gair}, {Gais}, {Galaudage}, {Gamba}, {Ganapathy}, {Ganguly}, {Gao},
  {Gaonkar}, {Garaventa}, {Garc{\'\i}a}, {Garc{\'\i}a-N{\'u}{\~n}ez},
  {Garc{\'\i}a-Quir{\'o}s}, {Garufi}, {Gateley}, {Gaudio}, {Gayathri}, {Ge},
  {Gemme}, {Gennai}, {George}, {George}, {Gerberding}, {Gergely}, {Gewecke},
  {Ghonge}, {Ghosh}, {Ghosh}, {Ghosh}, {Ghosh}, {Giacomazzo}, {Giacoppo},
  {Giaime}, {Giardina}, {Gibson}, {Gier}, {Giesler}, {Giri}, {Gissi},
  {Glanzer}, {Gleckl}, {Godwin}, {Goetz}, {Goetz}, {Gohlke}, {Golomb},
  {Goncharov}, {Gonz{\'a}lez}, {Gopakumar}, {Gosselin}, {Gouaty}, {Gould},
  {Grace}, {Grado}, {Granata}, {Granata}, {Grant}, {Gras}, {Grassia}, {Gray},
  {Gray}, {Greco}, {Green}, {Green}, {Gretarsson}, {Gretarsson}, {Griffith},
  {Griffiths}, {Griggs}, {Grignani}, {Grimaldi}, {Grimm}, {Grote}, {Grunewald},
  {Gruning}, {Guerra}, {Guidi}, {Guimaraes}, {Guix{\'e}}, {Gulati}, {Guo},
  {Guo}, {Gupta}, {Gupta}, {Gupta}, {Gustafson}, {Gustafson}, {Guzman}, {Ha},
  {Haegel}, {Hagiwara}, {Haino}, {Halim}, {Hall}, {Hamilton}, {Hammond}, {Han},
  {Haney}, {Hanks}, {Hanna}, {Hannam}, {Hannuksela}, {Hansen}, {Hansen},
  {Hanson}, {Harder}, {Hardwick}, {Haris}, {Harms}, {Harry}, {Harry},
  {Hartwig}, {Hasegawa}, {Haskell}, {Hasskew}, {Haster}, {Hattori}, {Haughian},
  {Hayakawa}, {Hayama}, {Hayes}, {Healy}, {Heidmann}, {Heidt}, {Heintze},
  {Heinze}, {Heinzel}, {Heitmann}, {Hellman}, {Hello}, {Helmling-Cornell},
  {Hemming}, {Hendry}, {Heng}, {Hennes}, {Hennig}, {Hennig}, {Hernandez},
  {Hernandez Vivanco}, {Heurs}, {Hild}, {Hill}, {Himemoto}, {Hines},
  {Hiranuma}, {Hirata}, {Hirose}, {Hochheim}, {Hofman}, {Hohmann}, {Holcomb},
  {Holland}, {Holley-Bockelmann}, {Hollows}, {Holmes}, {Holt}, {Holz}, {Hong},
  {Hopkins}, {Hough}, {Hourihane}, {Howell}, {Hoy}, {Hoyland}, {Hreibi},
  {Hsieh}, {Hsu}, {Huang}, {Huang}, {Huang}, {Huang}, {Huang}, {Huang},
  {H{\"u}bner}, {Huddart}, {Hughey}, {Hui}, {Hui}, {Husa}, {Huttner},
  {Huxford}, {Huynh-Dinh}, {Ide}, {Idzkowski}, {Iess}, {Ikenoue}, {Imam},
  {Inayoshi}, {Ingram}, {Inoue}, {Ioka}, {Isi}, {Isleif}, {Ito}, {Itoh},
  {Iyer}, {Izumi}, {JaberianHamedan}, {Jacqmin}, {Jadhav}, {Jadhav}, {James},
  {Jan}, {Jani}, {Janquart}, {Janssens}, {Janthalur}, {Jaranowski}, {Jariwala},
  {Jaume}, {Jenkins}, {Jenner}, {Jeon}, {Jeunon}, {Jia}, {Jin}, {Johns},
  {Johnson-McDaniel}, {Jones}, {Jones}, {Jones}, {Jones}, {Jones}, {Jonker},
  {Ju}, {Jung}, {Jung}, {Junker}, {Juste}, {Kaihotsu}, {Kajita}, {Kakizaki},
  {Kalaghatgi}, {Kalogera}, {Kamai}, {Kamiizumi}, {Kanda}, {Kandhasamy},
  {Kang}, {Kanner}, {Kao}, {Kapadia}, {Kapasi}, {Karat}, {Karathanasis},
  {Karki}, {Kashyap}, {Kasprzack}, {Kastaun}, {Katsanevas}, {Katsavounidis},
  {Katzman}, {Kaur}, {Kawabe}, {Kawaguchi}, {Kawai}, {Kawasaki},
  {K{\'e}f{\'e}lian}, {Keitel}, {Key}, {Khadka}, {Khalili}, {Khan}, {Khazanov},
  {Khetan}, {Khursheed}, {Kijbunchoo}, {Kim}, {Kim}, {Kim}, {Kim}, {Kim},
  {Kim}, {Kimball}, {Kimura}, {Kinley-Hanlon}, {Kirchhoff}, {Kissel}, {Kita},
  {Kitazawa}, {Kleybolte}, {Klimenko}, {Knee}, {Knowles}, {Knyazev}, {Koch},
  {Koekoek}, {Kojima}, {Kokeyama}, {Koley}, {Kolitsidou}, {Kolstein}, {Komori},
  {Kondrashov}, {Kong}, {Kontos}, {Koper}, {Korobko}, {Kotake}, {Kovalam},
  {Kozak}, {Kozakai}, {Kozu}, {Kringel}, {Krishnendu}, {Kr{\'o}lak}, {Kuehn},
  {Kuei}, {Kuijer}, {Kulkarni}, {Kumar}, {Kumar}, {Kumar}, {Kumar}, {Kume},
  {Kuns}, {Kuo}, {Kuo}, {Kuromiya}, {Kuroyanagi}, {Kusayanagi}, {Kuwahara},
  {Kwak}, {Lagabbe}, {Laghi}, {Lalande}, {Lam}, {Lamberts}, {Landry}, {Lane},
  {Lang}, {Lange}, {Lantz}, {La Rosa}, {Lartaux-Vollard}, {Lasky}, {Laxen},
  {Lazzarini}, {Lazzaro}, {Leaci}, {Leavey}, {Lecoeuche}, {Lee}, {Lee}, {Lee},
  {Lee}, {Lee}, {Lee}, {Lehmann}, {Lema{\^\i}tre}, {Leonardi}, {Leroy},
  {Letendre}, {Levesque}, {Levin}, {Leviton}, {Leyde}, {Li}, {Li}, {Li}, {Li},
  {Li}, {Li}, {Lin}, {Lin}, {Lin}, {Lin}, {Lin}, {Linde}, {Linker}, {Linley},
  {Littenberg}, {Liu}, {Liu}, {Liu}, {Liu}, {Llamas}, {Llorens-Monteagudo},
  {Lo}, {Lockwood}, {Loh}, {London}, {Longo}, {Lopez}, {Lopez Portilla},
  {Lorenzini}, {Loriette}, {Lormand}, {Losurdo}, {Lott}, {Lough}, {Lousto},
  {Lovelace}, {Lucaccioni}, {L{\"u}ck}, {Lumaca}, {Lundgren}, {Luo}, {Lynam},
  {Macas}, {MacInnis}, {Macleod}, {MacMillan}, {Macquet}, {Maga{\~n}a
  Hernandez}, {Magazz{\`u}}, {Magee}, {Maggiore}, {Magnozzi}, {Mahesh},
  {Majorana}, {Makarem}, {Maksimovic}, {Maliakal}, {Malik}, {Man}, {Mandic},
  {Mangano}, {Mango}, {Mansell}, {Manske}, {Mantovani}, {Mapelli},
  {Marchesoni}, {Marchio}, {Marion}, {Mark}, {M{\'a}rka}, {M{\'a}rka},
  {Markakis}, {Markosyan}, {Markowitz}, {Maros}, {Marquina}, {Marsat},
  {Martelli}, {Martin}, {Martin}, {Martinez}, {Martinez}, {Martinez},
  {Martinovic}, {Martynov}, {Marx}, {Masalehdan}, {Mason}, {Massera},
  {Masserot}, {Massinger}, {Masso-Reid}, {Mastrogiovanni}, {Matas},
  {Mateu-Lucena}, {Matichard}, {Matiushechkina}, {Mavalvala}, {McCann},
  {McCarthy}, {McClelland}, {McClincy}, {McCormick}, {McCuller}, {McGhee},
  {McGuire}, {McIsaac}, {McIver}, {McRae}, {McWilliams}, {Meacher}, {Mehmet},
  {Mehta}, {Meijer}, {Melatos}, {Melchor}, {Mendell}, {Menendez-Vazquez},
  {Menoni}, {Mercer}, {Mereni}, {Merfeld}, {Merilh}, {Merritt}, {Merzougui},
  {Meshkov}, {Messenger}, {Messick}, {Meyers}, {Meylahn}, {Mhaske}, {Miani},
  {Miao}, {Michaloliakos}, {Michel}, {Michimura}, {Middleton}, {Milano},
  {Miller}, {Miller}, {Miller}, {Millhouse}, {Mills}, {Milotti}, {Minazzoli},
  {Minenkov}, {Mio}, {Mir}, {Miravet-Ten{\'e}s}, {Mishra}, {Mishra}, {Mistry},
  {Mitra}, {Mitrofanov}, {Mitselmakher}, {Mittleman}, {Miyakawa}, {Miyamoto},
  {Miyazaki}, {Miyo}, {Miyoki}, {Mo}, {Modafferi}, {Moguel}, {Mogushi},
  {Mohapatra}, {Mohite}, {Molina}, {Molina-Ruiz}, {Mondin}, {Montani}, {Moore},
  {Moraru}, {Morawski}, {More}, {Moreno}, {Moreno}, {Mori}, {Morisaki},
  {Moriwaki}, {Morr{\'a}s}, {Mours}, {Mow-Lowry}, {Mozzon}, {Muciaccia},
  {Mukherjee}, {Mukherjee}, {Mukherjee}, {Mukherjee}, {Mukherjee}, {Mukund},
  {Mullavey}, {Munch}, {Mu{\~n}iz}, {Murray}, {Musenich}, {Muusse}, {Nadji},
  {Nagano}, {Nagano}, {Nagar}, {Nakamura}, {Nakano}, {Nakano}, {Nakashima},
  {Nakayama}, {Napolano}, {Nardecchia}, {Narikawa}, {Naticchioni}, {Nayak},
  {Nayak}, {Negishi}, {Neil}, {Neilson}, {Nelemans}, {Nelson}, {Nery},
  {Neubauer}, {Neunzert}, {Ng}, {Ng}, {Nguyen}, {Nguyen}, {Nguyen}, {Nguyen
  Quynh}, {Ni}, {Nichols}, {Nishizawa}, {Nissanke}, {Nitoglia}, {Nocera},
  {Norman}, {North}, {Nozaki}, {Nu{\~n}o Siles}, {Nuttall}, {Oberling},
  {O'Brien}, {Obuchi}, {O'Dell}, {Oelker}, {Ogaki}, {Oganesyan}, {Oh}, {Oh},
  {Oh}, {Ohashi}, {Ohishi}, {Ohkawa}, {Ohme}, {Ohta}, {Okada}, {Okutani},
  {Okutomi}, {Olivetto}, {Oohara}, {Ooi}, {Oram}, {O'Reilly}, {Ormiston},
  {Ormsby}, {Ortega}, {O'Shaughnessy}, {O'Shea}, {Oshino}, {Ossokine},
  {Osthelder}, {Otabe}, {Ottaway}, {Overmier}, {Pace}, {Pagano}, {Page},
  {Pagliaroli}, {Pai}, {Pai}, {Palamos}, {Palashov}, {Palomba}, {Pan}, {Pan},
  {Panda}, {Pang}, {Pang}, {Pankow}, {Pannarale}, {Pant}, {Panther},
  {Paoletti}, {Paoli}, {Paolone}, {Parisi}, {Park}, {Park}, {Parker},
  {Pascucci}, {Pasqualetti}, {Passaquieti}, {Passuello}, {Patel}, {Pathak},
  {Patricelli}, {Patron}, {Paul}, {Payne}, {Pedraza}, {Pegoraro}, {Pele},
  {Pe{\~n}a Arellano}, {Penn}, {Perego}, {Pereira}, {Pereira}, {Perez},
  {P{\'e}rigois}, {Perkins}, {Perreca}, {Perri{\`e}s}, {Petermann},
  {Petterson}, {Pfeiffer}, {Pham}, {Phukon}, {Piccinni}, {Pichot},
  {Piendibene}, {Piergiovanni}, {Pierini}, {Pierro}, {Pillant}, {Pillas},
  {Pilo}, {Pinard}, {Pinto}, {Pinto}, {Piotrzkowski}, {Piotrzkowski},
  {Pirello}, {Pitkin}, {Placidi}, {Planas}, {Plastino}, {Pluchar}, {Poggiani},
  {Polini}, {Pong}, {Ponrathnam}, {Popolizio}, {Porter}, {Poulton}, {Powell},
  {Pracchia}, {Pradier}, {Prajapati}, {Prasai}, {Prasanna}, {Pratten},
  {Principe}, {Prodi}, {Prokhorov}, {Prosposito}, {Prudenzi}, {Puecher},
  {Punturo}, {Puosi}, {Puppo}, {P{\"u}rrer}, {Qi}, {Quetschke},
  {Quitzow-James}, {Qutob}, {Raab}, {Raaijmakers}, {Radkins}, {Radulesco},
  {Raffai}, {Rail}, {Raja}, {Rajan}, {Ramirez}, {Ramirez}, {Ramos-Buades},
  {Rana}, {Rapagnani}, {Rapol}, {Ray}, {Raymond}, {Raza}, {Razzano}, {Read},
  {Rees}, {Regimbau}, {Rei}, {Reid}, {Reid}, {Reitze}, {Relton}, {Renzini},
  {Rettegno}, {Reza}, {Rezac}, {Ricci}, {Richards}, {Richardson}, {Richardson},
  {Riemenschneider}, {Riles}, {Rinaldi}, {Rink}, {Rizzo}, {Robertson}, {Robie},
  {Robinet}, {Rocchi}, {Rodriguez}, {Rolland}, {Rollins}, {Romanelli},
  {Romano}, {Romel}, {Romero-Rodr{\'\i}guez}, {Romero-Shaw}, {Romie},
  {Ronchini}, {Rosa}, {Rose}, {Rosi{\'n}ska}, {Ross}, {Rowan}, {Rowlinson},
  {Roy}, {Roy}, {Roy}, {Rozza}, {Ruggi}, {Ruiz-Rocha}, {Ryan}, {Sachdev},
  {Sadecki}, {Sadiq}, {Sago}, {Saito}, {Saito}, {Sakai}, {Sakai},
  {Sakellariadou}, {Sakuno}, {Salafia}, {Salconi}, {Saleem}, {Salemi},
  {Samajdar}, {Sanchez}, {Sanchez}, {Sanchez}, {Sanchis-Gual}, {Sanders},
  {Sanuy}, {Saravanan}, {Sarin}, {Sassolas}, {Satari}, {Sathyaprakash}, {Sato},
  {Sato}, {Sauter}, {Savage}, {Sawada}, {Sawant}, {Sawant}, {Sayah},
  {Schaetzl}, {Scheel}, {Scheuer}, {Schiworski}, {Schmidt}, {Schmidt},
  {Schnabel}, {Schneewind}, {Schofield}, {Sch{\"o}nbeck}, {Schulte}, {Schutz},
  {Schwartz}, {Scott}, {Scott}, {Seglar-Arroyo}, {Sekiguchi}, {Sekiguchi},
  {Sellers}, {Sengupta}, {Sentenac}, {Seo}, {Sequino}, {Sergeev}, {Setyawati},
  {Shaffer}, {Shahriar}, {Shams}, {Shao}, {Sharma}, {Sharma}, {Shawhan},
  {Shcheblanov}, {Shibagaki}, {Shikauchi}, {Shimizu}, {Shimoda}, {Shimode},
  {Shinkai}, {Shishido}, {Shoda}, {Shoemaker}, {Shoemaker}, {ShyamSundar},
  {Sieniawska}, {Sigg}, {Singer}, {Singh}, {Singh}, {Singha}, {Sintes},
  {Sipala}, {Skliris}, {Slagmolen}, {Slaven-Blair}, {Smetana}, {Smith},
  {Smith}, {Soldateschi}, {Somala}, {Somiya}, {Son}, {Soni}, {Soni}, {Sordini},
  {Sorrentino}, {Sorrentino}, {Sotani}, {Soulard}, {Souradeep}, {Sowell},
  {Spagnuolo}, {Spencer}, {Spera}, {Srinivasan}, {Srivastava}, {Srivastava},
  {Staats}, {Stachie}, {Steer}, {Steinhoff}, {Steinlechner}, {Steinlechner},
  {Stevenson}, {Stops}, {Stover}, {Strain}, {Strang}, {Stratta}, {Strunk},
  {Sturani}, {Stuver}, {Sudhagar}, {Sudhir}, {Sugimoto}, {Suh}, {Sullivan},
  {Sullivan}, {Summerscales}, {Sun}, {Sun}, {Sunil}, {Sur}, {Suresh}, {Sutton},
  {Suzuki}, {Suzuki}, {Swinkels}, {Szczepa{\'n}czyk}, {Szewczyk}, {Tacca},
  {Tagoshi}, {Tait}, {Takahashi}, {Takahashi}, {Takamori}, {Takano}, {Takeda},
  {Takeda}, {Talbot}, {Talbot}, {Tanaka}, {Tanaka}, {Tanaka}, {Tanaka},
  {Tanaka}, {Tanasijczuk}, {Tanioka}, {Tanner}, {Tao}, {Tao}, {Tapia San
  Mart{\'\i}n}, {Taranto}, {Tasson}, {Telada}, {Tenorio}, {Terhune},
  {Terkowski}, {Thirugnanasambandam}, {Thomas}, {Thomas}, {Thomas}, {Thompson},
  {Thondapu}, {Thorne}, {Thrane}, {Tiwari}, {Tiwari}, {Tiwari}, {Toivonen},
  {Toland}, {Tolley}, {Tomaru}, {Tomigami}, {Tomura}, {Tonelli},
  {Torres-Forn{\'e}}, {Torrie}, {Tosta e Melo}, {T{\"o}yr{\"a}}, {Trapananti},
  {Travasso}, {Traylor}, {Trevor}, {Tringali}, {Tripathee}, {Troiano},
  {Trovato}, {Trozzo}, {Trudeau}, {Tsai}, {Tsai}, {Tsang}, {Tsang}, {Tsao},
  {Tse}, {Tso}, {Tsubono}, {Tsuchida}, {Tsukada}, {Tsuna}, {Tsutsui},
  {Tsuzuki}, {Turbang}, {Turconi}, {Tuyenbayev}, {Ubhi}, {Uchikata},
  {Uchiyama}, {Udall}, {Ueda}, {Uehara}, {Ueno}, {Ueshima}, {Unnikrishnan},
  {Uraguchi}, {Urban}, {Ushiba}, {Utina}, {Vahlbruch}, {Vajente}, {Vajpeyi},
  {Valdes}, {Valentini}, {Valsan}, {van Bakel}, {van Beuzekom}, {van den
  Brand}, {Van Den Broeck}, {Vander-Hyde}, {van der Schaaf}, {van Heijningen},
  {Vanosky}, {van Putten}, {van Remortel}, {Vardaro}, {Vargas}, {Varma},
  {Vas{\'u}th}, {Vecchio}, {Vedovato}, {Veitch}, {Veitch}, {Venneberg},
  {Venugopalan}, {Verkindt}, {Verma}, {Verma}, {Veske}, {Vetrano},
  {Vicer{\'e}}, {Vidyant}, {Viets}, {Vijaykumar}, {Villa-Ortega}, {Vinet},
  {Virtuoso}, {Vitale}, {Vo}, {Vocca}, {von Reis}, {von Wrangel}, {Vorvick},
  {Vyatchanin}, {Wade}, {Wade}, {Wagner}, {Walet}, {Walker}, {Wallace},
  {Wallace}, {Walsh}, {Wang}, {Wang}, {Wang}, {Ward}, {Warner}, {Was},
  {Washimi}, {Washington}, {Watchi}, {Weaver}, {Webster}, {Weinert},
  {Weinstein}, {Weiss}, {Weller}, {Weller}, {Wellmann}, {Wen}, {We{\ss}els},
  {Wette}, {Whelan}, {White}, {Whiting}, {Whittle}, {Wilken}, {Williams},
  {Williams}, {Williams}, {Williamson}, {Willis}, {Willke}, {Wilson},
  {Winkler}, {Wipf}, {Wlodarczyk}, {Woan}, {Woehler}, {Wofford}, {Wong}, {Wu},
  {Wu}, {Wu}, {Wu}, {Wysocki}, {Xiao}, {Xu}, {Yamada}, {Yamamoto}, {Yamamoto},
  {Yamamoto}, {Yamamoto}, {Yamashita}, {Yamazaki}, {Yang}, {Yang}, {Yang},
  {Yang}, {Yang}, {Yap}, {Yeeles}, {Yelikar}, {Ying}, {Yokogawa}, {Yokoyama},
  {Yokozawa}, {Yoo}, {Yoshioka}, {Yu}, {Yu}, {Yuzurihara}, {Zadro{\.z}ny},
  {Zanolin}, {Zeidler}, {Zelenova}, {Zendri}, {Zevin}, {Zhan}, {Zhang},
  {Zhang}, {Zhang}, {Zhang}, {Zhang}, {Zhao}, {Zhao}, {Zhao}, {Zhao}, {Zheng},
  {Zhou}, {Zhou}, {Zhu}, {Zhu}, {Zimmerman}, {Zlochower}, {Zucker}, \&
  {Zweizig}}]{GWTC-3}
{The LIGO Scientific Collaboration}, {the Virgo Collaboration}, {the KAGRA
  Collaboration}, {et~al.} 2021{\natexlab{b}}, arXiv e-prints,
  arXiv:2111.03606.
\newblock \doarXiv{2111.03606}

\bibitem[{{Trinca} {et~al.}(2022){Trinca}, {Schneider}, {Maiolino}, {Valiante},
  {Graziani}, \& {Volonteri}}]{Trinca2022}
{Trinca}, A., {Schneider}, R., {Maiolino}, R., {et~al.} 2022, arXiv e-prints,
  arXiv:2211.01389.
\newblock \doarXiv{2211.01389}

\bibitem[{Usman {et~al.}(2019)Usman, Mills, \& Fairhurst}]{Usman:2018imj}
Usman, S.~A., Mills, J.~C., \& Fairhurst, S. 2019, Astrophys. J., 877, 82,
  \dodoi{10.3847/1538-4357/ab0b3e}

\bibitem[{{Valiante} {et~al.}(2017){Valiante}, {Agarwal}, {Habouzit}, \&
  {Pezzulli}}]{Valiante17review}
{Valiante}, R., {Agarwal}, B., {Habouzit}, M., \& {Pezzulli}, E. 2017, \pasa,
  34, e031, \dodoi{10.1017/pasa.2017.25}

\bibitem[{{Valiante} {et~al.}(2021){Valiante}, {Colpi}, {Schneider},
  {Mangiagli}, {Bonetti}, {Cerini}, {Fairhurst}, {Haardt}, {Mills}, \&
  {Sesana}}]{Valiante-colpi2021}
{Valiante}, R., {Colpi}, M., {Schneider}, R., {et~al.} 2021, \mnras, 500, 4095,
  \dodoi{10.1093/mnras/staa3395}

\bibitem[{{van Son} {et~al.}(2020){van Son}, {De Mink}, {Broekgaarden},
  {Renzo}, {Justham}, {Laplace}, {Mor{\'a}n-Fraile}, {Hendriks}, \&
  {Farmer}}]{vanSon2020}
{van Son}, L.~A.~C., {De Mink}, S.~E., {Broekgaarden}, F.~S., {et~al.} 2020,
  \apj, 897, 100, \dodoi{10.3847/1538-4357/ab9809}

\bibitem[{Veitch {et~al.}(2015)Veitch, Raymond, Farr,
  {et~al.}}]{Veitch:2014wba}
Veitch, J., Raymond, V., Farr, B., {et~al.} 2015, Phys. Rev. D, D91, 042003,
  \dodoi{10.1103/PhysRevD.91.042003}

\bibitem[{{Vink} {et~al.}(2021){Vink}, {Higgins}, {Sander}, \&
  {Sabhahit}}]{Vink2021MNRAS.504..146V}
{Vink}, J.~S., {Higgins}, E.~R., {Sander}, A. A.~C., \& {Sabhahit}, G.~N. 2021,
  \mnras, 504, 146, \dodoi{10.1093/mnras/stab842}

\bibitem[{{Vitale}(2016)}]{Vitale2016PhRvD..94l1501V}
{Vitale}, S. 2016, \prd, 94, 121501, \dodoi{10.1103/PhysRevD.94.121501}

\bibitem[{{Volonteri} {et~al.}(2021){Volonteri}, {Habouzit}, \&
  {Colpi}}]{SeedNAT2021}
{Volonteri}, M., {Habouzit}, M., \& {Colpi}, M. 2021, Nature Reviews Physics,
  3, 732, \dodoi{10.1038/s42254-021-00364-9}

\bibitem[{{Volonteri} \& {Rees}(2005)}]{Volonteri05}
{Volonteri}, M., \& {Rees}, M.~J. 2005, \apj, 633, 624, \dodoi{10.1086/466521}

\bibitem[{{Volonteri} {et~al.}(2015){Volonteri}, {Silk}, \&
  {Dubus}}]{Volonteri2015c}
{Volonteri}, M., {Silk}, J., \& {Dubus}, G. 2015, \apj, 804, 148,
  \dodoi{10.1088/0004-637X/804/2/148}

\bibitem[{{Woosley} \& {Heger}(2021)}]{Woosley2021}
{Woosley}, S.~E., \& {Heger}, A. 2021, \apjl, 912, L31,
  \dodoi{10.3847/2041-8213/abf2c4}

\end{thebibliography}

\end{document}